\documentclass[12pt, a4paper, letterpaper , twoside , openright]{Thesis} 
\usepackage{wrapfig}
\usepackage{lscape}
\usepackage{rotating}
\usepackage{graphicx}
\usepackage{caption}
\usepackage{amsmath}
\usepackage{cuted}
\usepackage{amsmath}
\usepackage{mathtools}

\usepackage{floatflt}
\usepackage{wrapfig}
\usepackage{graphicx}
\usepackage{stfloats}
\usepackage{array}
\usepackage{enumitem}
\usepackage[openbib]{currvita}
\usepackage{epsfig}
\usepackage{multirow}
\usepackage{algorithm}
\usepackage{wrapfig}
\usepackage{algorithmicx}
\usepackage{algpseudocode}
\algrenewcommand\algorithmicrequire{\textbf{Input:}}
\algrenewcommand\algorithmicensure{\textbf{Output:}}
\usepackage{amssymb}
\usepackage{amsthm}
\usepackage{xcolor}
\usepackage{float}
\usepackage{subfloat}
\usepackage{xspace}

\usepackage{url}
\usepackage{doi}
\usepackage[spacing,kerning]{microtype}
\usepackage{pdflscape}
\usepackage[section]{placeins}
\usepackage{placeins}
\hypersetup{
	bookmarks=false,      
	unicode=false,          
	pdftoolbar=true,        
	pdfmenubar=true,        
	pdffitwindow=false,    
	pdfstartview={FitH},    
	pdftitle={My title},    
	pdfauthor={Author},    
	pdfsubject={Subject},   
	pdfcreator={Creator},   
	pdfproducer={Producer}, 
	pdfkeywords={keyword1} {key2} {key3}, 
	pdfnewwindow=true,      
	colorlinks=true,      
	linkcolor=blue, 
	citecolor=blue,        
	filecolor=magenta,     
	urlcolor=blue         
}
\usepackage{epstopdf}

\algnewcommand\algorithmicinput{\textbf{INPUT:}}
\algnewcommand\INPUT{\item[\algorithmicinput]}

\algnewcommand\algorithmicoutput{\textbf{OUTPUT:}}
\algnewcommand\OUTPUT{\item[\algorithmicoutput]}

\theoremstyle{plain}
\newtheorem{thm}{Theorem}[section]
\newtheorem{lem}[thm]{Lemma}
\newtheorem{prop}[thm]{Proposition}

\theoremstyle{definition}
\newtheorem{defn}{Definition}[section]

\usepackage[english]{babel}
\theoremstyle{remark}

\usepackage[none]{hyphenat}
\dimen\footins=2 cm
\usepackage{mathptmx}      
\emergencystretch=\maxdimen
\usepackage{varioref}

\graphicspath{{Pictures/}} 
\usepackage[square, numbers]{natbib} 

\hypersetup{urlcolor=black, colorlinks=true} 	
\title{\ttitle} 

\begin{document}
\makeatletter
\renewcommand*{\NAT@nmfmt}[1]{\textsc{#1}}
\makeatother

\frontmatter 

\setstretch{1.6} 

\fancyhead{} 
\rhead{\thepage} 
\lhead{} 

\pagestyle{fancy} 

\newcommand{\HRule}{\rule{\linewidth}{0.5mm}} 

% PDF meta-data
\hypersetup{pdftitle={\ttitle}}
\hypersetup{pdfsubject=\subjectname}
\hypersetup{pdfauthor=\authornames}
\hypersetup{pdfkeywords=\keywordnames}

%----------------------------------------------------------------------------------------
%	TITLE PAGE
%----------------------------------------------------------------------------------------

\begin{titlepage}
\begin{center}

\HRule \\[0.4cm] % Horizontal line
{\huge \bfseries \ttitle}\\[0.4cm] % Thesis title
\HRule \\[1cm] % Horizontal line
 
\large \textit{Thesis submitted in partial fulfillment \\  for the Award of  Degree\\ \degreename }\\[0.2cm] 
\textit{by}\\[0.15cm]

\href{https://orcid.org/0000-0002-7810-0859}{\authornames}

\vfill
\graphicspath{ {./Figures/} }
\begin{figure}[hb]
  \centering
  \includegraphics[width=0.3\linewidth]{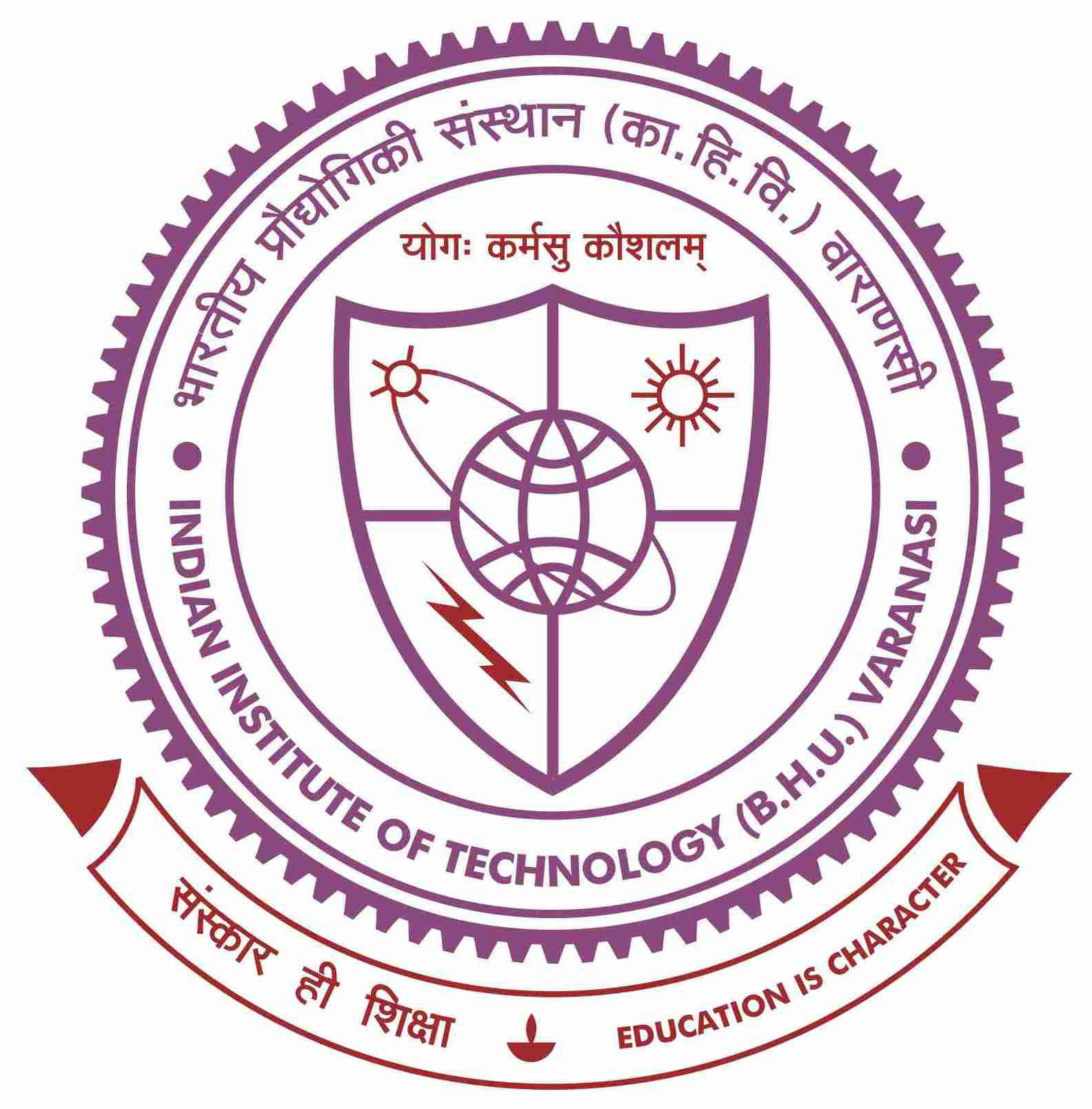}
\end{figure}

\DEPTNAME\\ % Research group name and department name
\textsc{ \UNIVNAME}\\[1cm] % University name
%\large \today\\[1.5cm] % Date
\large \noindent{\textbf{2019} }

\end{center}

\end{titlepage}

%----------------------------------------------------------------------------------------
%	DEDICATION
%----------------------------------------------------------------------------------------

\clearpage % Start a new page
\thispagestyle{empty} 

\begin{figure}
  \centering
  \includegraphics[width=0.8\textwidth]{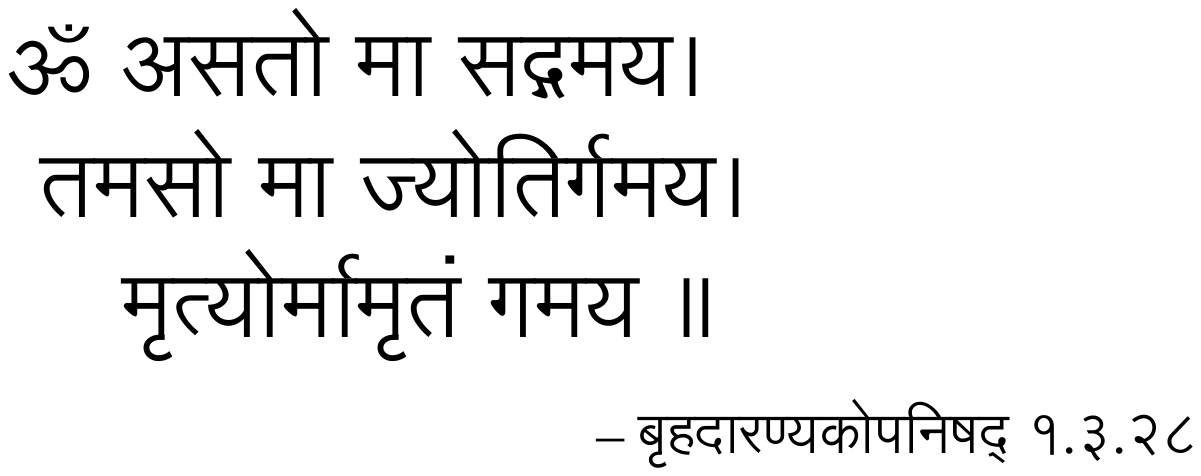}
\end{figure}

%-----------------------------------------------------------------------------------
%-----------------------------------------------------------------------------------

\clearpage % Start a new page

%----------------------------------------------------------------------------------------
%	ACKNOWLEDGEMENTS
%----------------------------------------------------------------------------------------

\setstretch{1.3} % Reset the line-spacing to 1.3 for body text (if it has changed)

\acknowledgements{\addtocontents{toc}{\vspace{1em}} % Add a gap in the Contents, for aesthetics

%I would like to take this.\\  

Firstly, I would like to express my sincere gratitude to my thesis supervisor Dr. R.S. Singh for his extended support and guidance during my research work.  It would not have been possible to successfully carry out this research work without his encouragement and support.

I wish to express my thanks to Dr. R. Srivastava, Head, Computer Science and Engineering Department, for his valuable support in carrying out the research. I would like to thanks to all the faculty members and staffs of the department. I would also like to thank my Research Program Evaluation Committee, Dr. S.K. Singh, Dr. B. Biswas, Dr. A.J. Gupta, for their valuable suggestion for my work.

I would like to take this opportunity to thank everyone who has helped me and contributed directly or indirectly to complete my research and thesis. 

I would like to thanks all my family members who always supported me for my work. I especially thank my parents, who have continuously given me encouragement and support throughout my life.

Lastly, I am grateful to \textit{Paramātmā} for his blessings and giving me the strength and courage to follow the research path. 

\vspace*{2.5cm} \hspace*{8 cm}{\large - \authornames}
\vfill{}}

%\clearpage % Start a new page
% Added to resolve the issue of blank page coming before Copyright Transfer Certificate on 2019-07-10 
\cleardoublepage
%----------------------------------------------------------------------------------------
%	LIST OF CONTENTS/FIGURES/TABLES PAGES
%----------------------------------------------------------------------------------------

\pagestyle{fancy} % The page style headers have been "empty" all this time, now use the "fancy" headers as defined before to bring them back

\lhead{\emph{Contents}} % Set the left side page header to "Contents"
\tableofcontents % Write out the Table of Contents

\lhead{\emph{List of Figures}} % Set the left side page header to "List of Figures"
\listoffigures % Write out the List of Figures

\lhead{\emph{List of Tables}} % Set the left side page header to "List of Tables"
\listoftables % Write out the List of Tables

%----------------------------------------------------------------------------------------
%	ABBREVIATIONS
%----------------------------------------------------------------------------------------

\clearpage % Start a new page

\setstretch{1.5} % Set the line spacing to 1.5, this makes the following tables easier to read

\lhead{\emph{Abbreviations}} % Set the left side page header to "Abbreviations"
\listofsymbols{ll} % Include a list of Abbreviations (a table of two columns)
{
\textbf{GM} & \textbf{G}raph \textbf{M}atching \\
\textbf{GED} & \textbf{G}raph \textbf{E}dit \textbf{D}istance \\
\textbf{ARG} & \textbf{A}ttributed \textbf{R}elational \textbf{G}raph \\
\textbf{PR} & \textbf{P}attern \textbf{R}ecognition \\
\textbf{NC} & \textbf{N}ode \textbf{C}ontraction \\
\textbf{VD} & \textbf{V}ertec \textbf{D}istance \\
\textbf{ED} & \textbf{E}dge \textbf{D}istance \\
\textbf{EDM}& \textbf{E}dge \textbf{D}istance \textbf{M}etric \\
\textbf{GD} & \textbf{G}raph \textbf{D}istance \\
\textbf{GDM} & \textbf{G}raph \textbf{D}istance \textbf{M}etric \\
\textbf{WGM} & \textbf{W}eighted \textbf{G}raph \textbf{M}atching \\
\textbf{HGED} & \textbf{H}omeomorphic \textbf{G}raph \textbf{E}dit \textbf{D}istance 
%\textbf{Acronym} & \textbf{W}hat (it) \textbf{S}tands \textbf{F}or \\
}

%----------------------------------------------------------------------------------------
%	SYMBOLS
%----------------------------------------------------------------------------------------

\clearpage % Start a new page

\lhead{\emph{Symbols}} % Set the left side page header to "Symbols"

\listofnomenclature{lll} % Include a list of Symbols (a two column table)
{
$G$ & Graph \\
$V$ & Vertex set \\
$E$ & Edge set \\
$O$ & Big-O \\
$\mu$ & Node labeling function \\
$\nu$ & Edge labeling function \\
$L_V$ & Node label set \\
$L_E$ & Edge label set \\
$k$-$NC$ & $k$-degree node contraction \\
$k^*$-$NC$ & $k^*$-degree node contraction \\
$k^*$-$GED$ & $k^*$-graph edit distance \\
$r$-$NC$ & $r$-centrality node contraction \\
$r$-$GED$ & $r$-centrality graph edit distance \\
$E_{ij}^{A}$ & Angular component of Edge Distance \\
$E_{ij}^{L}$ & Length component of Edge Distance \\
$E_{ij}^{P}$ & Position component of Edge Distance Metric
% Symbol & Name & Unit \\

}

\clearpage % Start a new page

%----------------------------------------------------------------------------------------
%	ABSTRACT PAGE
%----------------------------------------------------------------------------------------

%\addtotoc{Preface} % Add the "Abstract" page entry to the Contents
\addtotoc{Abstract}
\Abstract{\addtocontents{toc}{\vspace{1em}} % Add a gap in the Contents, for aesthetics

%Enter Abstract Content here...
%}
The graph is one of the most widely used mathematical structure in engineering and science because of its representational power and inherent ability to demonstrate the relationship between objects. Graph matching is the process of finding the similarity between the two graphs. It has a wide range of applications of object identification in various graph-based representations. Graph matching is broadly classified into two types, exact and error-tolerant graph matching. While exact graph matching requires a strict correspondence between nodes and edges of the two graphs the error-tolerant matching allows some flexibility to measure the similarity between two graphs from a broader perspective. Due to non-availability of the efficient graph matching solutions, various approximation and suboptimal algorithms have been proposed.

The objective of this work is to introduce the novel graph matching techniques using the representational power of the graph and apply it to structural pattern recognition applications. We present an extensive survey of various exact and inexact graph matching techniques. A category of graph matching algorithms is presented, which reduces the graph size by removing the less important nodes using some measure of relevance. Graph matching using the concept of homeomorphism is presented, it uses path contraction to remove the nodes with degree two from all simple paths of input graphs in which every node except first and last have degree two.

We present an approach to error-tolerant graph matching using node contraction where the given graph is transformed into another graph by contracting smaller degree nodes. Node contraction is the process of removing a node and its associated edges provided that the node is not an articulation point. It leads to a reduction in search space required to perform graph matching. We use this scheme to extend the notion of graph edit distance, which can be used as a trade-off between execution time and accuracy requirements of various graph matching applications. Experimental results show that the algorithm achieves efficiency without disturbing the topology of graphs too much. 

We describe an approach to graph matching by utilizing the various node centrality information which reduces the graph size by removing a fraction of nodes from both graphs based on a given centrality measure. Depending on the structure and properties of various graph datasets, we can choose the appropriate centrality measure to reduce the size of the graphs. Experiments show that different centrality criteria lead to a different saving in computation time and classification ratio. Depending on the application requirements, a suitable centrality measure can be selected to achieve the best performance. 

The graph matching problem is inherently linked to geometry and topology of graphs. We introduce a novel approach to measure graph similarity using geometric graphs. We define the vertex distance between two geometric graphs using the position of their vertices and show it to be a metric over the set of all graphs with vertices only. We define edge distance between two graphs based on the angular orientation, length and position of the edges. Then we combine the notion of vertex distance and edge distance to define the graph distance between two geometric graphs and show it to be a metric. We describe a geometric graph isomorphism algorithm using the above concept of graph similarity to perform exact graph matching. Finally, we use the proposed graph similarity framework to perform error-tolerant graph matching. The experimental results show that this graph matching approach is promising to graph dataset in which every node has a coordinate position in a two-dimensional plane.

\setstretch{1.3} % Return the line spacing back to 1.3

\addtocontents{toc}{\vspace{2em}} % Add a gap in the Contents, for aesthetics

%----------------------------------------------------------------------------------------
%	THESIS CONTENT - CHAPTERS
%----------------------------------------------------------------------------------------

\mainmatter % Begin numeric (1,2,3...) page numbering

\pagestyle{fancy} % Return the page headers back to the "fancy" style

% Include the chapters of the thesis as separate files from the Chapters folder
% Uncomment the lines as you write the chapters

% Chapter Template

\chapter{Introduction} % Main chapter title

\label{Chapter1} % Change X to a consecutive number; for referencing this chapter elsewhere, use \ref{ChapterX}

\lhead{Chapter 1. \emph{Introduction}} % Change X to a consecutive number; this is for the header on each page - perhaps a shortened title

%----------------------------------------------------------------------------------------
%	SECTION 1
%----------------------------------------------------------------------------------------

\section{Graph Matching}
The graph is one of the most prominent mathematical structure in computer science. It is defined as a set of nodes and edges, where each edge is a connection between a pair of nodes. Due to its inherent ability to demonstrate the structural representation between objects, it is used in a wide range of applications in science and engineering.

For example, in biological applications, graphs are used as biological networks where a node represents biological units like cells, protein, neuron, etc., and an edge represents the connection between them. In chemical applications graphs are used as a molecular graph, where nodes represent atoms or molecules, and edge denotes valence or bond between chemical units. In computer science, graphs are ubiquitous and are used in almost every area of computing. For example, in the operating system, graphs are used to characterize resource allocation graph in which process and resources are designated by nodes. An edge from resource to process denotes that resource is allocated to process, whereas an edge from process to resource indicates that the process has requested for the corresponding resource. In computer network graphs are used to find the shortest path for routing the data packets over the communication network. In software engineering control flow graph is used to find the complexity of a program and, to envisage the dependency and association between the different components of a software project. In this thesis, we use graphs for matching of structured data.

Graph matching is the process of computing the similarity between two graphs. It is a generalization of tree matching, which itself is a generalization of string matching. While the string is a linear structure, implying that at each successive step only one element like a symbol can be appended to an existing element; on the other hand trees and graphs can be connected to more than one data element. This increase in the representation power from strings to trees and trees to graphs also leads to the increased computational complexity of matching algorithms as shown in Figure \ref{fig:string-tree-graph}. While polynomial time algorithms for string and tree matching are known but no polynomial time algorithm is known for graph matching.

Depending on the nature of matching, graph matching is broadly classified into exact and inexact matching. Exact graph matching requires a strict one-to-one correspondence between vertices and edges of two graphs. It is like graph isomorphism problem, where a bijection is required between the vertices of one graph to another one such that for every edge in the first graph there exists a corresponding edge in the second graph which connects the same set of nodes. Exact graph matching even though theoretically important, may not be useful in real-world applications, where input data often get modified due to the presence of noise and error in storage and transmission process. To overcome the effect of noise and distortion on input data, inexact graph matching is used, which finds a similarity score between two graphs. Inexact graph matching is also known as error-tolerant graph matching as it can accommodate some tolerance to noise or error that may have incurred during the processing of data.    

A major application of graph matching is in structural pattern recognition. Structural pattern recognition utilizes the underlying structure of the object to perform the various pattern recognition tasks. The ability to identify patterns is one of the fundamental characteristics of human beings and up to certain extents whole of living beings. Pattern recognition consists of analyzing and classifying patterns based on their characteristics and features. Every person deals with a large number of pattern recognition problems in day to day life. Examples include recognition of a relative from a group of persons, identifying a particular book from a book self, identification of streets and friend's home, etc. Because of the complex cognition system of the human brain, the task of recognizing different object seems to be intuitive and very simple, but the same tasks using a machine or computer can be very complicated. Pattern recognition system develops algorithms to perform such tasks automatically using a machine.

Depending on the nature of the underlying problem, pattern recognition can be categorized into statistical and structural pattern recognition. Statistical pattern recognition uses feature vectors to represent different patterns. Use of feature vectors allows standard mathematical techniques, which apply to vector space, are also applicable in statistical pattern recognition domain. Therefore the use of feature vectors leads to many efficient algorithms in statistical pattern recognition. The limitation of the use of feature vectors of fixed dimension is that it can not be used for pattern having structures like strings, trees and graphs (Figure \ref{fig:string-tree-graph}). In such situations, structural pattern recognition offers an alternative to statistical pattern recognition by recognizing patterns using graph-based representations instead of feature vectors as shown in Figure \ref{fig:structural-vs-statistical-PR}. The major benefit of using graph is that its representational ability is higher than that of feature vectors. Here we can observe a classic trade-off between using vector and graph. While the efficiency of mathematical operations using vector are high, but their representational power is low, on the other hand, representational power of graphs are high, but the efficiency of various mathematical operations is low. Due to unavailability of efficient polynomial time solutions to graph matching problem, various approximation and suboptimal algorithms have been proposed recently.

%-------------------------------------
%       Add More Contents Here
%-------------------------------------

\begin{figure}[!t]
\centering
 \includegraphics[scale=.4]{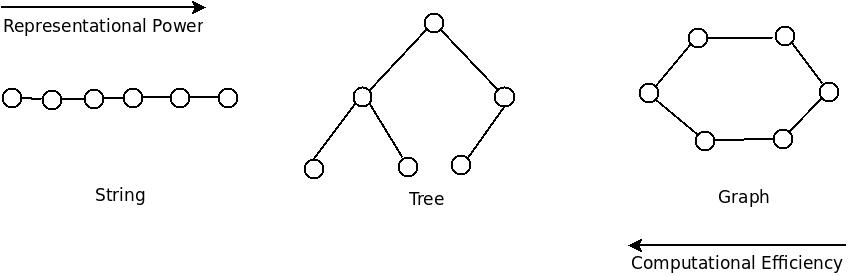}
 \caption{Generalization from string to tree and tree to graph}
 \label{fig:string-tree-graph}
\end{figure}

\begin{figure}[!t]
\centering
 \includegraphics[scale=.4]{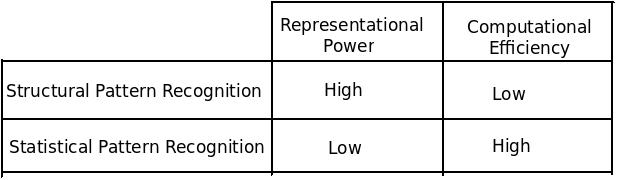}
 \caption{structural versus statistical pattern recognition}
 \label{fig:structural-vs-statistical-PR}
\end{figure}
%----------------------------------------------------------------------------------------
%	SECTION 2
%----------------------------------------------------------------------------------------

\section{Contribution}
In this section, a brief description of the contribution of this thesis is given. 

\begin{itemize}
\item Graph matching using the concept of homeomorphism is introduced. It uses path contraction to remove the nodes with degree two from all simple paths of input graphs in which every node except first and last have degree two. Since path contraction is reverse of subdivision operation, the resulting graphs are homeomorphic and topologically equivalent. We use this concept to define homeomorphic graph edit distance.

\item Error-tolerant graph matching using node contraction is presented, in which input graphs are transformed by contracting smaller degree nodes. Node contraction is the process of deleting a node and its linked edges provided that the node is not a cut vertex. This approach is used to present the notion of extended graph edit distance. The proposed framework can be used as a trade-off between time versus accuracy considerations. Experimental results show that the algorithm attains efficiency without altering the topology of the graphs drastically.

\item A class of graph matching algorithms is proposed, which reduces the graph size by removing the less relevant nodes or edges using some measure of centrality. It reduces the search space by contracting the nodes with least centrality values. Four different centrality measures: degree, betweenness, eigenvector and PageRank are used for comparison. Node contraction can be considered as a special case of this approach, where centrality measure is degree centrality. Depending on the structure and properties of various graph datasets, we can choose the appropriate centrality measure to reduce the size of the graphs. The proposed algorithm is implemented on the letter and molecules dataset and results show that different centrality criteria can be selected to save computation time during graph matching.

\item A novel approach to measure graph similarity between geometric graphs is introduced. Vertex distance between two geometric graphs is defined using the coordinate positions of the vertices. Edge distance between two geometric graphs is defined using the angular orientation, length and position of edges and the resulting distance is shown to be a metric. Finally, graph distance is defined using the linear combination of vertex distance and edge distance. The resulting graph distance notion is shown to be a metric over the set of all geometric graphs.

\item Exact and error-tolerant geometric graph matching is described using the notion of the proposed geometric graph similarity metric. A geometric graph isomorphism algorithm is presented to check the isomorphism between two geometric graphs. Prior to checking geometric graph isomorphism, this uses graph alignment algorithm to perform the geometric transformation on a graph so that its reference coordinates are aligned to the input graph. Error-tolerant graph matching using the geometric graph similarity is presented. The input graph size is made equal by appending additional vertices and edges of appropriate values. Weighting parameters are used to combine the vertex distance and the three components of edge distance to compute the final graph distance metric. Proposed graph matching algorithm is both error-tolerant as well as computed in polynomial time. Experimental evaluation shows that the proposed geometric graph matching framework is promising to graph dataset having two-dimensional coordinates for each vertex.
\end{itemize}

\section{Organization of the Thesis}

This thesis is organized as follows. The present chapter provides an introduction to graph matching and a brief overview of the proposed work. 

Chapter 2, provides a survey of exact and error-tolerant graph matching. It also describes the definitions and basic concepts used in exact and inexact graph matching. 

Chapter 3, introduces graph matching using various extensions to graph edit distance. It describes the homeomorphic graph edit distance, which uses path contraction as a preprocessing step. It also discusses the error-tolerant graph matching using node contraction and presents the extended graph edit distance. 

Chapter 4, presents graph matching utilizing node centrality information to ignore the less relevant nodes and introduces $r$-centrality graph edit distance. Here $r$ is the fraction of nodes to be contracted from the graph based on given centrality criteria. It uses degree, betweenness, eigenvector and PageRank centrality measures for computing the centrality of nodes in the graphs. 

Chapter 5, introduces an intuitive approach to measure the similarity between two geometric graphs. It defines vertex distance using the position of vertices of the two graphs; then it defines edge distance using the alignment, length and position features of edges. Finally, it combines both vertex and edge distance to compute graph distance. It also presents algorithms for exact and error-tolerant graph matching using the proposed geometric graph similarity framework. 

Finally, chapter 6, provides the concluding remarks.  

% Chapter Template

\chapter{Graph Matching: A Survey} % Main chapter title

\label{Chapter2} % Change X to a consecutive number; for referencing this chapter elsewhere, use \ref{ChapterX}

\lhead{Chapter 2. \emph{Graph Matching: A Survey}} % Change X to a consecutive number; this is for the header on each page - perhaps a shortened title

%-----------Chapter 2---------------------

\section{Introduction}

%----------------Add main survey references here--------------------------------
Graph matching is the process of computing the similarity between two graphs. It is broadly classified into exact and inexact graph matching. While exact graph matching has more theoretical implications in computer science, on the other side inexact graph matching has more practical implications in computer science and pattern recognition. For example, an optimal solution to graph isomorphism problem which is an exact graph matching problem will lead to resolution of its complexity class, which is currently neither known to be in class $P$, nor in $NP$-complete. On the other hand, the subgraph isomorphism problem is known to be in $NP$-complete and therefore, no efficient polynomial time algorithm is available.

In this chapter, a review of various exact, approximate and error-tolerant graph matching techniques is provided. Conte \textit{et al.} \citep{Conteetal2004} in 2004 describe an extensive survey of different exact and inexact graph matching techniques used in pattern recognition. In 2014 Foggia \textit{et al.} \citep{Foggiaetal2014} provide a more recent survey of graph matching and learning techniques used in pattern recognition. The 1998 paper by Bunke \citep{Bunke1998} presents a formal framework and algorithm for graph matching. In 2010 Gao \textit{et al.} \citep{Gaoetal2010} describe a survey of the various algorithm for graph edit distance. Livi and Rizzi \citep{LiviRizzi2013} in 2013 provide a review of graph matching problem focusing methodological and algorithmic results.

This chapter is organized as follows. Section 2.2, presents basic definitions and concepts used in exact and error-tolerant graph matching. Section 2.3, describes a survey of exact graph matching techniques. Section 2.4, presents a review of error-tolerant graph matching methods. Finally, section 2.5 describes the summary.

\section{Basic Concepts and Definitions}

In this section, we review the basic concepts and definitions used in exact and inexact graph matching \citep{AggarwalWang2010,NeuhausBunke2007,Riesen2015,Rosen2017}.

\begin{defn}[Graph] 
 A graph $G$ is defined as $G= (V,E,\mu,\nu)$, where 
 \setlist{nolistsep}
 \begin{itemize}[noitemsep]
  \item $V$ is the set of vertices or nodes,
  \item $E$ is the set of edges or links,
  \item $\mu: V \rightarrow L_V$ is a function that assigns a node label set $l_v \in L_V$ to each vertex $v \in V$,
  \item $\nu: E \rightarrow L_E$ is a function that assigns an edge label set $l_e \in L_E$ to each edge in $E$,
  \item $L_V$ and $L_E$ are node label set and edge label set respectively.
 \end{itemize} 
% A graph $G$ is defined as $G= (V,E,\mu,\nu)$, where $V$ is the set of vertices or nodes, $E$ is the set of edges or links, $\mu: V \rightarrow L_V$ is a function that allocates a node label alphabet $l_v \in L_V$ to each vertex $v \in V$, $\nu: E \rightarrow L_E$ is a function that allocates an edge label alphabet $l_e \in L_E$ to each edge in $E$. Here, $L_V$ and $L_E$ are node label set and edge label set respectively. In this paper, we use $G_i$ to represent $(V_i, E_i, \mu_i,\nu_i)$. 
\end{defn}

When the nodes or edges of a graph have labels, then the graph is called a labeled graph. If $L_V = L_E = \emptyset$ then $G$ is called the unlabeled graph. Figure \ref{fig:graph-examples} shows examples of labeled and unlabeled graphs. $|G|$ denotes the number of nodes in a graph $G$. 

\begin{defn}[Subgraph] 
 Let $G_1 = (V_1, E_1, \mu_1,\nu_1)$ and $G_2 = (V_2,E_2,\mu_2,\nu_2)$ be two graphs. Graph $G_1$ is said to be a subgraph of graph $G_2$, if
 \setlist{nolistsep}
 \begin{itemize}[noitemsep]
  \item $V_1 \subseteq V_2$,
  \item $E_1 \subseteq E_2$,
  \item For every node $u \in G_1$, we have $\mu_1(u)=\mu_2(u)$,
  \item similarly, for every edge $e \in G_1$, we have $\nu_1(e)=\nu_2(e)$.
  \end{itemize} 
% A graph $G_1$ is said to be a subgraph of graph $G_2$, if $V_1 \subseteq V_2$;  $E_1 \subseteq E_2$; for every node $u \in G_1$, we have $\mu_1(u)=\mu_2(u)$; similarly, for every edge $e \in G_1$, we have $\nu_1(e)=\nu_2(e)$.  
\end{defn}

\begin{figure}
\begin{minipage}[b]{.3\linewidth}
\centering
\includegraphics{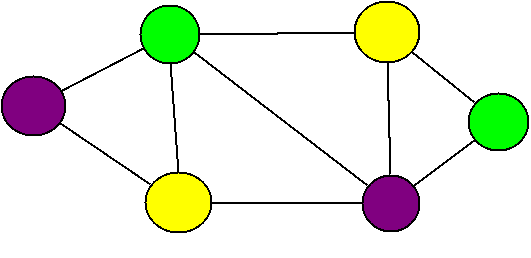}
\subcaption{Undirected labeled}\label{fig:1a}
\end{minipage}%
\hfill
\begin{minipage}[b]{.3\linewidth}
\centering
\includegraphics{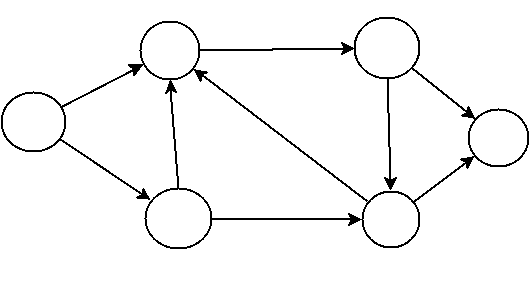} 
\subcaption{Directed unlabeled}\label{fig:1b}
\end{minipage}
\hfill
\begin{minipage}[b]{.3\linewidth}
\centering
\includegraphics{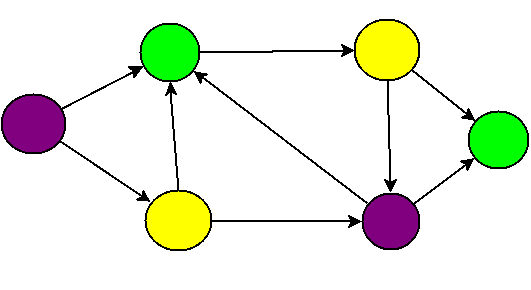} 
\subcaption{Directed labeled}\label{fig:1b}
\end{minipage}
\caption{Graph Examples}\label{fig:graph-examples}
\end{figure}

\begin{defn}[Graph Isomorphism] 
 Let $G_1 = (V_1, E_1, \mu_1,\nu_1)$ and $G_2 = (V_2,E_2,\mu_2,\nu_2)$ be two graphs. A graph isomorphism between $G_1$ and $G_2$ is a function $f: V_1 \rightarrow V_2$ such that 
 \setlist{nolistsep}
 \begin{itemize}[noitemsep]
  \item $\mu_1(u) = \mu_2(f(u)), \forall u \in V_1$,
  \item $\forall e_1 = (u, v) \in E_1, \exists e_2 = (f(u), f(v)) \in E_2$ such that $\nu_1(e_1) = \nu_2(e_2)$,
  \item $\forall e_2 = (u, v) \in E_2, \exists e_1 = (f^{-1}(u), f^{-1}(v)) \in E_1$ such that $\nu_1(e_1) = \nu_2(e_2)$,
 \end{itemize} 
\end{defn}

In other words, graph isomorphism between two graphs $G_1$ and $G_2$ is a bijection between every vertex $u \in G_1$ to a unique vertex $v \in G_2$, such that their labels and connecting edges are preserved.

\begin{defn}[Subgraph Isomorphism] 
 Let $G_1$ and $G_2$ be two graphs. A mapping $f:V_1 \rightarrow V_2$ from $G_1$ to $G_2$ is called as subgraph isomorphism if there is a graph isomorphism between $G_1$ and a subgraph of $G_2$. 
\end{defn}

\begin{defn}[Maximum Common Subgraph] 
 Let $G_1$ and $G_2$ be two graphs. A graph $G$ is said to be a common subgraph of $G_1$ and $G_2$, when $G$ is subgraph isomorphic to both $G_1$ and $G_2$. $G$ is called as Maximum Common Subgraph (mcs) of $G_1$ and $G_2$, when there does not exists any larger size graph than $G$ which is common subgraph to both $G_1$ and $G_2$.  
\end{defn}

\begin{defn}[Minimum Common Supergraph] 
 Let $G_1$ and $G_2$ be two graphs. A graph $G$ is said to be a common supergraph of $G_1$ and $G_2$, when $G_1$ and $G_2$ are subgraph isomorphic to $G$. $G$ is called as Minimum Common Supergraph (MCS) of $G_1$ and $G_2$, when there does not exists any smaller size graph than $G$ which is common supergraph to both $G_1$ and $G_2$.  
\end{defn}

A sequence of edit operations that transform a graph $G_1$ into $G_2$ is called an edit path between $G_1$ and $G_2$. A common set of edit operations include insertions, deletions, and substitutions of nodes and edges. The notation $\epsilon \rightarrow u$ is used to represent the insertion of a vertex $u$, $u \rightarrow \epsilon$ represents deletion of vertex $u$, and $u \rightarrow v$ denotes substitution of a vertex $u$ by vertex $v$. Similarly, $\epsilon \rightarrow e$ is used to represent the insertion of an edge $e$, $e \rightarrow \epsilon$ represents deletion of the edge $e$, and $e \rightarrow f$ denotes substitution of an edge $e$ by edge $f$.

\begin{defn}[Graph Edit Distance] 
 The Graph Edit Distance (GED) between two graphs $G_1$ and $G_2$ is defined by 
% \begin{equation}
$$ GED(G_1, G_2)= min_{(e_1,...,e_k) \in \varphi (G_1, G_2)} \sum_{i=1}^{k} c(e_i) \label{eq} $$
% \end{equation}
Here $\varphi (G_1, G_2)$ indicates the set of edit distance paths converting $G_1$ to $G_2$, and $c(e_i)$ is the cost function associated with every edit operation $e_i$. 
\end{defn}

The graph edit distance between two graphs is the minimum number of modifications in terms of edit operations needed to transform one graph into another one. Graph edit distance is a generalization of tree edit distance, which in turn is a generalization of string edit distance. While the edit operations in string edit distance consist of insertion, deletion and substitution of alphabets only, the set of edit operations in tree edit distance as well as graph edit distance consists of insertion, deletion and substitution of nodes and edges.

% Explain edit costs

\section{Exact Graph Matching}

%----------------Intro of Exact GM and theoretical algorithms with complexity-------------------------------
Exact graph matching is a one-to-one correspondence between vertices and edges of two graphs. It is also known as the graph isomorphism problem. We may note that “graph matching” is different from “matching” used in graph theory. Graph matching is a comparison between two graphs; on the other hand, matching in a graph is a set of edges such that no two edges share the same vertex. In other words, matching is defined for a single graph, whereas graph matching is defined between two graphs. Matching in a graph is a subgraph so that each node of the subgraph can have either zero or one edge incident to it.

In graph theory, an isomorphism of two graphs is a one-to-one correspondence (bijection) between the vertex set of both graphs. If an isomorphism exists between two graphs, then we say the graphs to be isomorphic. The problem of determining whether two graphs are isomorphic is called a graph isomorphism problem. As mentioned earlier, the graph isomorphism problem is one of the few problems in computational complexity theory that belongs to $NP$; however, it is still not known whether it is solvable in polynomial time or $NP$-complete. Perhaps it is only one of the two problems given by Garey and Johnson in \citep{GareyJohnson1979} 1979, whose complexity is still unresolved. The other problem is the equally famous so-called integer factorization problem. In fact, the graph isomorphism problem is suspected to be in class $NP$-intermediate. In complexity theory, class $NP$-intermediate consists of those problems that are in class $NP$ but neither know to be in class $P$ nor $NP$-complete. Ladner \citep{Ladner1975} in 1975 demonstrated that if $P \neq NP$ then $NP$-intermediate is not empty. In other words, if class $P$ is not equal to class $NP$, then there will exist some problems that will be in class $NP$-intermediate. It can also be shown that if an NP-intermediate problem exists, then $P \neq NP$ and $P = NP$, then $NP$-intermediate is empty.

Currently, the fastest accepted theoretical algorithm for graph isomorphism problem is given by Babai and Luks \citep{BabaiLuks1983} in 1983 with running time $2^{O(\sqrt{n \log n})}$ for graphs with $n$ vertices. In 2015, Babai proposed a quasipolynomial time algorithm for graph isomorphism problem having running time $2^{O({(\log n)}^c)}$ for some fixed $c>0$. While the general graph isomorphism problem's complexity is not resolved, polynomial-time algorithms are presented for special classes of graphs. For example, Luks \citep{Luks1982} in 1982 describe a polynomial algorithm of graph isomorphism problem for bounded valence graphs. Zemlyachenko \textit{et al.} \citep{Zemlyachenkoetal1985} in 1985 describe a creative compilation of work related to the graph isomorphism problem. Kobler \textit{et al.} \citep{Kobler1993} in 1993 summarize the results of the graph isomorphism problem and its structural complexity.

Subgraph isomorphism problem is a generalization of the graph isomorphism problem. In subgraph isomorphism, the task is to determine whether a graph contains a subgraph that is isomorphic to another graph. In other words, the graph isomorphism problem is a special case of a subgraph isomorphism problem. Since the subgraph isomorphism problem is a generalization of the Hamiltonian cycle problem, therefore it is NP-complete \citep{Cook1971}. As subgraph isomorphism is $NP$-complete and has no polynomial time efficient solutions available, various approximation and suboptimal algorithms have been proposed.
In the remainder of this section, we will review some of the more practical algorithms for exact graph matching.

\subsection{Tree Search-Based Techniques}

Tree search-based methods are among the first techniques used for graph matching task. In the 1968 paper, Hart \textit{et al.} \citep{Hartetal1968} describe $A^*$ search algorithm to find an optimal path between two nodes of a graph having the minimum cost. In order to select the best node to expand the search path, it uses an evaluation function $f(n)$ which is the sum of two component $g(n)$ and $h(n)$. Here $g(n)$ is the actual cost of an optimal search path from source node $s$ to $n$ and $h(n)$ is the actual cost of an optimal search path from $n$ to a target node $t$. Authors show that by using a suitable choice of $f(n)$ the $A^*$ search algorithm always find an optimal path from a source node to a target node. The reader may refer to Nilsson's book \citep{Nilsson1980} for further details and $A^*$ search algorithm variants.
% The technique of $A^*$ search algorithm is used to implement graph edit distance computation between two graphs.
% --------ADD MORE-----------------

Ullmann \citep{Ullmann1976} in 1976 describes an algorithm for subgraph isomorphism which can also be used to find graph isomorphism as well as monomorphism. The author proposed a refinement procedure to inferentially eliminate the successor nodes which are not consistent with the previous matching. Ullmann \citep{Ullmann2011} substantially updates the above paper in 2011 by computing subgraph isomorphism using binary constraint satisfaction. Specifically, the author proposed a new bit-vector algorithm for binary constraint satisfaction, which depends more on the search and less on domain editing. In their 1980 paper, Ghahraman \textit{et al.} \citep{Ghahraman1980} proposed a graph monomorphism algorithm using netgraph computed from the Cartesian product of the vertices of two graphs to prune the search space.  

Cordella \textit{et al.} \citep{Cordella1998} in 1998 describe the VF algorithm, which performs graph matching using state-space representation. In this technique, each state represents a partial solution to the graph matching. The transformation from one state to other denotes inclusion of a pair of matched nodes. They use a set of criteria to prune the states of partial matching, which does not lead to required graph matching. Cordella \textit{et al.} \citep{Cordella1999} in 1999, using experiments, show that the VF algorithm performs significantly better than Ullmann's algorithm. In 2001, the same authors \citep{Cordella2001} proposed the VF2 algorithm, which is a modified version of the VF algorithm. VF2 is particularly suitable for large graphs as it reduces memory requirement to linear with respect to the number of vertices in the graph. The 2002 paper by Larrosa and Valiente \citep{LarrosaValiente2002} provides a formulation of the graph isomorphism problem within constraint satisfaction framework. The authors also proposed a new algorithm to take advantage of the problem structure to enhance the look-ahead procedure. 

McGregor \citep{McGregor1982}, in 1982, describes a backtrack search algorithm for the maximal common subgraph problem. The 2001 paper by Koch \citep{Koch2001} describes the enumeration of all connected maximal common subgraph between two graphs by transforming the maximal common subgraph problem into the clique problem. Bunke \textit{et al.} \citep{Bunkeetal2002} in 2002 describe algorithms for maximum common subgraph detection and their performance evaluation.

The 2007 paper by Konc and Janezic \citep{KoncJanezic2007} presents an improvement to the approximate coloring algorithm and uses it to compute the maximum clique problem. A computed maximum clique can be used to find the maximum common subgraph. The proposed improvement is based on the idea that the tightest upper bounds can be calculated close to the root of the recursion tree of the branch-and-bound algorithm. Zampelli \textit{et al.} \citep{Zampellietal2010} in 2010 proposed a constraint programming approach to subgraph isomorphism problem by using a global constraint and an associated filtering algorithm. The 2010 paper by Solnon \citep{Solnon2010} describes a new filtering algorithm based on local all different constraints and shows that this filtering prunes more branches than other existing filtering and therefore, it is more efficient.

\subsection{Algebraic Graph Theory Based Techniques}
One of the important algorithm for exact graph matching and graph isomorphism proposed by McKay \citep{McKay1981} in 1981, known as nauty, is based on group theory. This paper describes an efficient algorithm for computing the generators for the automorphism group of the graphs. The automorphism group can be used to find canonical labeling of the input graphs. The canonical labeling determines the set of isomorphic graphs by comparing the adjacency matrices of canonical form for corresponding graphs. One drawback of the above algorithm is that the computation of canonical labeling can be exponential in the worst case. The 2003 paper by De Santo \textit{et al.} \citep{DeSantoetal2003} has shown that the nauty program may take more computation time than some other algorithms like VF2. This paper describes an experimental comparison of four exact graph matching algorithms, namely: Ullmann’s algorithm, the algorithm by Schmidt and Druffel, the VF2 algorithm and Nauty. To perform the benchmarking activity, the authors have also built the domain of a large database of graphs.

Darga \textit{et al.} \citep{Dargaetal2004} in 2004 introduced a new symmetry detection tool, saucy, which is an implementation of the automorphism group using a sparse data structure. They demonstrate saucy to be more efficient than nauty for several practical graph datasets. The 2008 paper by Darga \textit{et al.} \citep{Dargaetal2008}, improved the saucy program and described a symmetry discovery algorithm by exploiting the sparsity present in input as well as output. McKay and Piperno \citep{McKayPiperno2014} in 2014, improved the nauty program and introduced a novel technique called Traces that is more efficient than most of the other existing graph isomorphism tools.

\subsection{Special Type of Graphs}

Although subgraph isomorphism is NP-complete, and graph isomorphism problem is neither known to be P nor NP-complete, polynomial time algorithms for the special class of graphs are known. Aho \textit{et al.} \citep{Ahoetal1974} in 1974 describe an algorithm for tree isomorphism in polynomial time. %--------------------- Add details of tree isomorphism --------------------
Hoffmann and O’Donnel \citep{HoffmannODonnel1982} in 1982 described several algorithms for tree pattern matching and comparison with other methods. This paper also provides a time and space complexity analysis of these algorithms. The 2005 paper by Bille \citep{Bille2005} provides an extensive survey of tree edit distance, alignment distance and inclusion problem. Demaine \textit{et al.} in 2009 present an optimal decomposition algorithm for tree edit distance having worst-case complexity of $O(n^3)$ time, where $n$ is the size of the trees. The 2011 paper by Akutsu \textit{et al.} \citep{Akutsuetal2011} proposed exact algorithms for computing the tree edit distance for unordered trees. 

Hopcroft and Wong \citep{HopcroftWong1974} in 1974 describe a linear time algorithm for finding isomorphism between two planar graphs . This algorithm improved the solution of planar graph isomorphism from $O(n.logn)$ \citep{HopcroftTarjan1973} to linear time. The proposed algorithm can be modified to partition a collection of planar graphs into equivalence class of isomorphic planar graphs in linear time.
Luks \citep{Luks1982} in 1982 describes a polynomial-time algorithm of the graph isomorphism problem for bounded valence graph. The author has performed a polynomial-time reduction from the graph isomorphism problem of bounded valence to the color automorphism problem of groups having the composition factor of bounded order. The color automorphism problem for a group consists of evaluating generators for the subgroup of the group that stabilizes the color classes, and this problem is solvable in polynomial time. Dickinson \textit{et al.} \citep{Dickinsonetal2004} in 2004, presented a method to find graph isomorphism for a class of graphs with unique node labels in polynomial time. They have shown that subgraph isomorphism, maximum common subgraph and graph edit distance can be computed in quadratic time. The 2007 paper by Kuramochi and Karypis \citep{KuramochiKarypis2007} has proposed an algorithm to compute graph isomorphism between geometric graphs in polynomial time.

\subsection{Other Techniques}

Messmer and Bunke \citep{MessmerBunke1998}, in 1998, proposed a new technique to graph and subgraph isomorphism detection from a sample graph to a large dataset of model graphs. They use a preprocessing step to create a decision tree for model graphs, which can be used to detect subgraph isomorphism from the candidate input graph to model graphs in polynomial time. The drawback of this approach is that the preprocessing step take exponential computation time and decision tree construction take exponential space with respect to the number of nodes in the graphs. Weber \textit{et al.} \citep{Weberetal2011}, in 2011, described an extension of the above approach to reduce the storage amount and indexing time for graphs significantly. Messmer and Bunke \citep{MessmerBunke2000}, in 2000, proposed a new technique for finding subgraph isomorphism between an input graph and a large database of pre-processed graphs. In this method, subgraphs which occur frequently are represented only once to save the execution time to find them in the input graph.  

Irniger and Bunke \citep{IrnigerBunke2004}, in 2004, presented a decision tree structures for graph database filtering. In this paper, they represent graph utilizing feature vectors, and then they build a decision tree to index the graph database using these feature vectors. In 2005, Gori \textit{et al.} \citep{Gorietal2005} described a general framework for exact and approximate graph matching using a random walk based on the notion of Markovian graph spectral analysis. This paper presents a polynomial time algorithm for graph isomorphism problem for Markovian spectrally distinguishable graphs, which are a set of graphs that do not easily reduce to other graphs. Using experiments, the authors have shown that their proposed algorithm is remarkably more efficient than other existing techniques. This idea is also demonstrated to be effective for approximate graph matching.

Dahm \textit{et al.} \citep{Dahmetal2012}, in 2012, describe an improvement for subgraph isomorphism detection by eliminating the nodes using their topological features. For large graphs, this technique removes the nodes from the second graph which are not compatible with the nodes from the first graph based on the topological features of its nodes, before testing them for subgraph isomorphism.

\section{Error-Tolerant Graph Matching}

One of the significant limitations of exact graph matching is its strict constraints imposed for matching. Because of its stringent conditions, exact matching is often not suitable for real-world applications. For example, exact matching can only identify whether two graphs are exactly similar or not; but it can not tell how much they are similar. In other words, exact matching does not explore similarity space between exactly similar graphs, which are isomorphic, and dissimilar graphs, which are non-isomorphic.

In many real-world applications, input graph data may get modified due to the presence of noise during the storage and acquisition process. The graph data may also get deformed during the processing step of extracting graphs from objects. In such circumstances, exact matching cannot be used to find a matching or similarity between two graphs. Due to the above reasons, error-tolerant graph matching is used to perform general matching between two graphs apart from checking graph and subgraph isomorphism. In this sense, error-tolerant matching is a generalization of exact matching, which can tell how much two graphs are similar when they are not exactly similar. 

Error-tolerant matching is also known as an inexact matching because it finds an approximate matching between two graphs in a reasonable time when finding the optimal solution is often computationally expansive and intractable. Therefore error-tolerant matching is useful even when the possibility of distortion in input data is not present, as it returns an approximate but efficient matching when finding the best solution is not feasible.

Error-tolerant matching is also known as error-correcting graph matching because it can be used to correct errors in the form of deformation, that may have been occurred due to the presence of noise or distortions during the processing of graph. Many error-tolerant graph matching techniques are based on different kind of errors with their associated cost, and the task is to find a matching with minimum error. For example, a class of techniques for error-tolerant matching is based on graph edit operations. The standard edit operations on a graph can be insertion, deletion and substitution of nodes and edges, with each edit operation has its associated cost. Then the graph edit cost can be defined as the minimum cost of a set of edit operations that transform one graph to the other one.

%------------------------Graph distance and Graph edit distance definition can be added here--------------------

In the remainder of this section, we survey the important techniques of error-tolerant graph matching proposed in the literature. 

\subsection{Tree Search-Based Techniques}

Tsai and Fu \citep{TsaiFu1979} in 1979 proposed an error-correcting isomorphism of attributed relational graph for matching deformed patterns using an ordered search tree based algorithm. In this paper, the authors use a combination of structural technique by representing patterns as a relational graph, and statistical technique by using deformation probabilities for error-tolerant matching. Same authors in their 1983 paper \citep{TsaiFu1983}, extended their work to error-correcting subgraph isomorphism by formulating it as a state-space tree search problem. They also described an ordered search algorithm for computing optimal error-tolerant subgraph isomorphism. 

The 1980 paper \citep{Ghahraman1980} by Ghahraman \textit{et al.} describes an algorithm for optimal graph monomorphism problem by specifying graph morphisms using subgraphs of Cartesian product graph, and utilizing a branch and bound algorithm. Shapiro and Haralick \citep{ShapiroHaralick1981} in 1981, proposed relational homomorphism between two structural description using efficient tree-search techniques. The 1983 paper by Sanfeliu and Fu \citep{SanfeliuFu1983} proposed a distance measure for a non-hierarchical attributed relational graph using the recognition of nodes. It finds the minimum number of modification needed to transform one graph into another. To determine the cost of recognition of nodes, the main features of vertices are expressed by a different cost function, which are used to find the similarity between the nodes.

Eshera and Fu \citep{EsheraFu1984} in 1984, proposed a new method for computing global distance measure between attributed graphs for image analysis. It works by decomposition of attributed relational graphs into a basic attributed relational graph, which is a graph in the form of one level tree. Then, state space representation is generated from these basic attributed relational graphs to find an optimal matching between the sets of basic attributed relational graphs.

The 1996 paper \citep{Cordellaetal1996} by Cordella \textit{et al.} describes an efficient algorithm for the error-tolerant matching of ARG graphs using a defined set of syntactic and semantic transformations. This paper defines transformation contextually with reference to a prototype, which is applicable when the sample graph matches the prototype. The same authors \citep{Cordellaetal1998} in 1998, extend their work to perform matching using state space representation to exhibit a partial matching between two graphs. Serratosa \textit{et al.} \citep{Serratosaetal1999} in 1999, proposed an efficient algorithm to find a suboptimal similarity measure between Function Described Graph (FDG) and ARG using tree search, where FDG itself is depicted as the ensemble of ARG's.

Berretti \textit{et al.} \citep{Berrettietal2000} in 2000, describe an efficient solution for indexing and matching of ARG's using  heuristic based on $A^*$ search. The 2001 paper \citep{Berrettietal2001} by the same authors presents an efficient solution for subgraph error-correcting isomorphism problem using a look-ahead technique for computing object distances. Gregory and Kittler  \citep{GregoryKittlerl2002} in 2002, propose color image retrieval using graph search techniques utilizing matching through graph edit operations and optimal search methods.

Sanfeliu \textit{et al.} \citep{Sanfeliuetal2002} in 2002 extends their work of function-described graphs and explains a representation of a set of attributed graphs, denominated function-described graphs and a distance measure for matching attributed graphs with function-described graphs. Authors explains graph-based representation and techniques for image segmentation and object recognition. The 2004 paper by Sanfeliu \textit{et al.} \citep{Sanfeliuetal2004} discuss the object learning and recognition using the so called second-order random graphs, which includes both marginal probability as well as second order joint probability functions of graph elements. Serratosa \textit{et al.} \citep{Serratosaetal2002} in 2002 describe the supervised as well as the unsupervised synthesis of function-described graphs from a set of graphs. The 2003 paper by Serratosa \textit{et al.} \citep{Serratosaetal2003} describes distance measure and efficient matching algorithm between attributed graphs and function-described graphs. Authors also use function-described graphs for modelling and matching objects represented by attributed graphs. The paper by Cook \textit{et al.} \citep{Cookeetal2003} in 2003 describes a structural search engine that utilizes the hyperlink structure of the web and the textual information to find the websites of interest. Hidovic and Pelilo \citep{HidovicPelilo2004} in 2004 proposed distance measures for attributed graphs, which are based on the maximal similarity common subgraph of two graphs.

\subsection{Continuous Optimization}
Graph matching is an inherently discrete optimization problem. A recent class of graph matching techniques is based on transforming the discrete optimization problem into a continuous one and then solving this continuous optimization problem using the various optimization algorithms. Finally, convert the appropriate solution of this continuous optimization problem to the discrete domain.

A novel class of error-tolerant graph matching techniques is based on relaxation labeling. The idea is to assign each node of the first graph a label from a set of labels so that it matches a node of the other graph. For each node of the one graph, there is a vector of probabilities based on a normal probability distribution for assigning it to the nodes of the other graph. During the initial step, these probabilities are computed using node and edge attributes and other information about graphs. It is then updated in successive iterative steps until the desired labeling is achieved. Fischler and Elschlager \citep{FischlerElschlager1973} in 1973, proposed relaxation labeling by providing a combined descriptive scheme and decision metric. Authors also described an algorithm using an approach similar to dynamic programming to reduce a huge amount of computational time. Rosenfeld \textit{et al.} \citep{Rosenfeldetal1976} in 1976 described scene labeling by extending discrete relaxation to probabilistic relaxation and performing ambiguity reduction among the objects in a scene using iterated relaxation operations. 

Relaxation labeling technique has been further improved and extended in several works. The 1989 paper \citep{KittlerHancock1989} by Kittler and Hancock try to resolve the issue of combining evidence in probabilistic relaxation in a systematic way by providing a specification that does not need the use of a support function. Above authors \citep{HancockKittler1990} in 1990, describe discrete relaxation to perform the maximum a posteriori probability estimation of globally consistent label assignment.
Christmas \textit{et al.} \citep{Christmasetal1995} in 1995, proposed probabilistic relaxation for graph matching using features extracted from 2D images and expressed the matching problem in the Bayesian framework for label assignment. Rangarajan and Mjolsness \citep{RangarajanMjolsness1996} in 1996 proposed a Lagrangian relaxation network for graph matching. The idea is to find a permutation matrix that makes nodes of two graphs into a correspondence. Wilson and Hancock \citep{WilsonHancock1997} in 1997, proposed an error-tolerant matching using Bayesian consistency measure by extending the probabilistic framework. The 1999 paper \citep{HuetHancock1999} by Huet and Hancock extends the above work by describing a Bayesian matching for large image libraries using edge consistency as well as node attribute similarity. Myers \textit{et al.} \citep{Myersetal2000} in 2000, proposed Bayesian graph edit distance for performing error-tolerant matching of corrupted graphs.
Torsello and Hancock \citep{TorselloHancock2003} in 2003, used relaxation labeling to perform approximate tree edit distance having uniform edit cost. Kostin \textit{et al.} \citep{Kostinetal2005} in 2005 describe an extension using the probabilistic relaxation algorithm given by \citep{Christmasetal1995}. The 2007 paper by Chevalier \textit{et al.} \citep{Chevalieretal2007} presented an improvement over relaxation labeling techniques for region adjacency graphs, which varies with time. Sanroma et al. \citep{Sanromaetal2012} in 2012 proposed a graph matching technique to solve the point-set correspondence problem using the Expectation-Maximization algorithm. 

\subsection{Spectral Methods}

A class of graph matching techniques is based on the spectral method of algebraic graph theory. The spectral method is based on the fact that adjacency matrices of graphs remain unchanged during node rearrangement and therefore for similar graphs, adjacency matrices will have the same eigendecomposition. Computing eigenvalue and eigenvector are well-known problems, which can be solved efficiently in polynomial time. Therefore many graph matching techniques use the spectral techniques by evaluating eigenvalues and eigenvector of matrices. 

Umeyama \citep{Umeyama1988} in 1988, proposed the matching between two weighted graphs, having weight for each edge, using eigendecomposition of the adjacency matrices. The author shows that near optimal matching can be achieved If the graphs are close to each other. One of the restriction of the proposed method is that the input graphs should be of the same size. The paper \citep{CarcassoniHancock2001} by Carcassoni and Hancock in 2001, proposed an eigendecomposition based technique to weighted graph matching using a hierarchical approach. They use a probabilistic approach to restrict the individual matching using the matching between the pairwise clusters. In 2004, Caelli and Kosinov \citep{CaelliKosinov2004} presented graph eigen-space based method for error-tolerant graph matching using eigen-subspace projection and vertex clustering techniques for graphs having an equal and different number of nodes. Robles-Kelly and Hancock \citep{KellyHancock2005} in 2005, proposed graph spectral seriation technique to transform the adjacency matrix of a graph into a string sequence and then apply efficient string matching methods to it. In 2005, Shokoufandeh \textit{et al.} \citep{Shokoufandehetal2005} described a framework for indexing hierarchical image structures using the spectral characterization of a directed acyclic graph. Cour \textit{et al.} \citep{Couretal2007} in 2007, proposed balanced graph matching using spectral relaxation technique for an approximate solution to graph matching problem. The authors also present a normalization technique to enhance the accuracy of the existing graph matching methods. Duchenne \textit{et al.} \citep{Duchenneetal2011} in 2011, proposed a tensor-based algorithm for hypergraph matching by maximizing the multilinear objective function through the generalization of spectral techniques.

\subsection{Artificial Neural Networks}

A category of graph matching methods using artificial neural network has also been proposed. Suganthan \textit{et al.} \citep{Suganthanetal1995} in 1995, applied the Hopfield neural network to homomorphic graph matching by optimizing an input energy criterion. The 1997 paper \citep{SperdutiStarita1997} by Sperduti and Starita proposed supervised neural network for graph classification using the concept of generalized recursive structure.  In 1998, Frasconi \textit{et al.} \citep{Frasconietal1998} presented a general framework for processing of structural information by unifying artificial neural network and belief network models, where the connection between data variables are represented by directed acyclic graphs. Barsi \citep{Barsi2003} in 2003, described the generalization of self-organizing feature map by replacing the regular neuron grid by the undirected graph. Jain and Wysotzki \citep{JainWysotzki2005} in 2005, proposed a neural network based technique to error-tolerant graph matching problem by reducing the search space using neural refinement procedure and subsequently performing energy minimization process. The 2009 paper \citep{Scarsellietal2009}, by Scarselli \textit{et al.}, presented a graph neural network model by extending the existing neural network models to support the data represented in the structural domain.

\subsection{Graph Edit Distance}

Graph edit distance is one of the most flexible and widely used technique for error-tolerant graph matching. Fu and Bhargava \citep{FuBhargava1973} in 1973, proposed a method to represent a pattern by a tree instead of strings. Authors describe a tree system consisting of grammar, transformation and mapping on tree and tree automata; and apply it to the structural pattern recognition. The 1983 paper \citep{SanfeliuFu1983} by Sanfeliu and Fu presented a distance measure for finding the minimum number of alterations needed to convert one graph to another one using insertion, deletion, and substitution of nodes and edges. Bunke and Allerman \citep{BunkeAllerman1983} in 1983 described an error-tolerant graph matching of attributed graphs for structural pattern recognition using state space search with heuristic information. Authors define the cost function of edit operations so that the graph distance satisfies the properties of metrics under particular conditions. 

Graph edit distance applies to a wide range of applications, as specific edit cost functions can be defined for different applications. One of the limitations of graph edit distance is that it is computationally expensive, and its worst-case complexity is exponential with the number of nodes in input graphs as Zeng \textit{et al.} \citep{Zeng2009} demonstrated graph edit distance to be an NP-hard problem. To overcome this issue, a large number of approximate and suboptimal algorithms for error-tolerant graph matching has been proposed. In 2004, Neuhaus and Bunke \citep{NeuhausBunke2004} presented an approximate and error-tolerant graph matching for planar graphs by successively matching subgraphs to locally optimize structural similarity until it obtains a complete edit path. To find the candidate path the authors use the concept of the neighborhood of a node, to match the subgraph constructed from the neighborhood of the nodes of the first graph to subgraphs created from the neighborhood of the various nodes of the other graph. Neuhaus \textit{et al.} \citep{Neuhausetal2006} in 2006, presented techniques for suboptimal computation of graph edit distance using the fact that $A^*$ search-based graph edit distance traverse a large area of search space which may not be pertinent to particular classification problems. Authors proposed two variants of $A^*$, namely $A^*$-beamsearch and $A^*$-pathlength. $A^*$-beamsearch process only a fixed number of nodes called beam-width at any given level of the search tree rather than exploring all the successor nodes of the search tree. $A^*$-pathlength uses an extra weight factor to select the longer partial edit path over the short ones. Riesen and Bunke \citep{RiesenBunke2009} in 2009, proposed suboptimal graph edit distance computation using bipartite graph matching, which considers only local instead of global edge structure in the course of the optimization process. Their method uses bipartite optimization process to map vertices and its local structure of the first graph to vertices and its local structure of the second graph. Sole-Ribalta \textit{et al.} \citep{Sole-Ribaltaetal2012} in 2012, described the various properties and applications of the graph edit distance cost. Ferrer \textit{et al.} \citep{Ferreretal2015} in 2015, further enhance the distance accuracy of the above bipartite graph matching framework by exploiting the order of the assignment computed by the approximation algorithm. In 2015, Riesen \textit{et al.} \citep{Riesenetal2015} described the estimation of exact graph edit distance using the regression analysis on the lower and upper bounds of bipartite approximation. Riesen and Bunke \citep{RiesenBunke2015} in 2015 used various search strategies like iterative search, greedy search, and beam search, etc., to improve the approximation of bipartite graph edit distance computation. Riesen \citep{Riesen2015} in 2015, describes the various algorithm for structural pattern recognition using graph edit distance. Dwivedi and Singh \citep{DwivediSingh2017} in 2017, proposed homeomorphic graph edit distance, which performs path contraction on the two inputs graphs before computing graph edit distance. Same authors in 2018, extend the above approach to describe error-tolerant graph matching using node contraction \citep{DwivediSingh2018b}. 
%\citep{SorlinSolnon2005}

\subsection{Special Type of Graphs}

Error-tolerant graph matching algorithms for special kind of graphs have also been proposed. Lu \citep{Lu1983} in 1983 presents an algorithm to compute best matching of two trees using node splitting and merging. The 1992 paper by Zhang \textit{et al.} \citep{Zhangetal1992} provides an efficient polynomial time algorithm for computing the editing distance between unordered labeled trees. Shasha \textit{et al.} \citep{Shashaetal1994} in 1994, described efficient algorithms and heuristics, leading to approximate solutions of tree edit distance. Bernard \textit{et al.} \citep{Bernardetal2008} in 2008 describes probabilistic models of tree edit distance based on expectation maximization algorithm. In 1997 Oflazer \citep{Oflazer1997} describes an error-tolerant tree matching algorithm using branch and bound technique. Rocha and Pavlidis in \citep{RochaPavlidis1994} presented efficient error-correcting homomorphism for planar graphs. Valiente \citep{Valiente2001} in 2001, described a tree distance which can be computed efficiently in polynomial time. Dwivedi and Singh \citep{DwivediSingh2019} in 2019, presented an efficient error-tolerant graph matching algorithm for geometric graphs. Other inexact graph matching for a particular class of graphs is described in \citep{WangAbe1995}, \citep{Llados2001}.

\subsection{Graph Kernels and Embeddings}

Kernel methods enable us to apply statistical pattern recognition to graph domain. Different type of graph kernels used in graph matching is described by Neuhaus and Bunke \citep{NeuhausBunke2007} and Gartner \citep{Gartner2008}. Neuhaus \textit{et al.} \citep{Neuhausetal2009} described various kernels for error-tolerant graph classification. Kashima \textit{et al.} \citep{Kashimaetal2003} in 2003, proposed marginalized kernels between labeled graphs based on an infinite dimensional feature space.
In \citep{BorgwardtKriegel2005}, the authors present graph kernels based on the shortest path, which is positive definite and computed in polynomial time. Lafferty and Lebanon \citep{LaffertyLebanon2005} in 2005, described a class of kernel known as diffusion kernel for statistical learning which utilizes the geometric structures of statistical models. In \citep{Haussler1999}, the Haussler presents convolution kernels for discrete structures which generalizes the class of radial basis kernels. Gazere \textit{et al.} \citep{Gazereetal2012} in 2012, describe two new graph kernels for regression and classification problem. The first kernel, which is Laplacian kernel, is based on the notion of edit distance, whereas the second kernel, known as treelet kernel is based on subtrees enumeration. Strug \citep{Strug2011} in 2011, described a kernel for the hierarchical graph using a combination of tree and graph kernel. In the 2013 paper, Bai and Hancock \citep{BaiHancock2013} proposed graph kernels based on Jensen-Shannon kernel, which is non-extensive information theoretic kernel. Riesen and Bunke \citep{RiesenBunke2010} in 2010, describe graph embedding and various classification and clustering techniques for graphs based on vector space embedding.

\subsection{Other Methods}

Now, we briefly review other techniques for error-tolerant graph matching proposed in literature. In \citep{LuoHancock2001}, the authors describe the error-tolerant graph matching using an expectation maximization algorithm. Gold and Rangarajan \citep{GoldRangarajan1996} in 1996, proposed graduated assignment algorithm for graph matching. The 2011 paper by Chang and Kimia \citep{ChangKimia2011} extends the graduated assignment graph matching algorithm for hypergraphs. Liu \textit{et al.} \citep{Liuetal1995} proposed a modified genetic algorithm for solving weighted graph matching problem. Khoo and Suganthan \citep{KhooSuganthan2001} in 2001, describe a genetic algorithm based optimization for multiple relational graph mapping. Wang \textit{et al.} \citep{Wangetal1994} present an approximate graph matching algorithm for labeled graphs. Mori \textit{et al.} \citep{Morietal2014} in 2014 describe the significance of global features for online characters recognition. Wu \textit{et al.} \citep{Wuetal2015} in 2015 presented object tracking benchmark and effective approaches for robust tracking. The 2019 paper by Cilia  \textit{et al.} \citep{Ciliaetal2019} proposed a feature ranking based approach with a greedy search strategy to perform handwritten character recognition. Bouillon \textit{et al.} \citep{Bouillonetal2019} in 2019 presented an improvement to handwritten historical document analysis using a novel preprocessing step for enhancing the input image. Some other techniques are described in \citep{Jagotaetal2000,JainObermayer2011,SantacruzSerratosa2018,Solnon2019}.

Recent trends in graph-based representation for pattern recognition is provided by Brun \textit{et al.} \citep{Brunetal2020}. The 2012 paper by Bunke and Riesen \citep{BunkeRiesen2012} survey the attempts made towards unifying structural and statistical pattern recognition. Hancock and Wilson \citep{HancockWilson2012} present a review of recent work on the pattern analysis with graphs. Jain \textit{et al.} \citep{Jainetal2016} in 2016 provide a comprehensive survey of biometric research techniques of the last five decades, including its accomplishments, challenges, and opportunities. Santosh \citep{Santosh2018} in 2018 describes various statistical, structural and syntactic approaches for graphics recognition.

\section{Summary}
This chapter presented basic concepts and definition used in graph matching. It also described a survey of various graph matching techniques and algorithms. We discussed methods for both exact as well as error-tolerant graph matching techniques. A brief survey of graph edit distance is also presented. Different type of graph matching techniques may be suitable for different applications depending on their requirement. For example, exact graph matching is suited for applications which require rigorous matching such as graph isomorphism. On the other hand, for the applications which are affected by the presence of noise or distortion inexact matching may be perfect. We also described exact and inexact graph matching techniques for special types of graphs. 

In the survey on graph matching, one issue we found is that the graph matching techniques are computationally much expansive. We have addressed this issue in chapters 3, 4 by proposing several approximate solutions. In chapter 3, we present an approach to graph matching using node contraction, which reduces the search space based on the degree of nodes. Chapter 4, describes graph matching derived from a given fraction of nodes of both graphs using some centrality measure. It can be used to perform matching under a time constraint.  

Another issue we found in graph matching is the lack of an adequate similarity measure. Towards this direction, we proposed a similarity measure in chapter 5, to find the similarity between two geometric graphs. This framework requires that every node of the graph dataset should have its associated coordinate point. Given any graph, it can be embedded in a two-dimensional plane to get the coordinates of its vertices. We applied this similarity framework to perform exact and error-tolerant graph matching.
 
% Chapter Template

\chapter{Graph Matching using Extensions to Graph Edit Distance}
%\chapter{Error-Tolerant Graph Matching Using Homeomorphism} % Main chapter title

\label{Chapter3} % Change X to a consecutive number; for referencing this chapter elsewhere, use \ref{ChapterX}

\lhead{Chapter 3. \emph{Graph Matching using Extensions to Graph Edit Distance}} % Change X to a consecutive number; this is for the header on each page - perhaps a shortened title

%----------------------------------------------------------------------------------------
%	SECTION 1
%----------------------------------------------------------------------------------------

\section{Introduction}

%----------------Introduction Chapter 3-----------------------------------

Graph edit distance is one of the important and most widely used technique for graph matching. The basic concept of graph edit distance is to transform one graph to the other one using the minimum number of deformations. These distortions are defined as edit operations consisting of insertion, deletion, and substitution of nodes and edges. Due to the flexibility of this distortion model and its associated cost, graph edit distance applies to a wide range of applications. The main advantage of the graph edit distance is that it can be applied to any arbitrary graph labeled or unlabeled and its ability to handle any type of structural distortion. On the other hand, the major disadvantage of graph edit distance is that it is computationally expensive, and it takes exponential time with respect to the number of nodes in input graphs. In this chapter, we describe the extensions of graph edit distance, which can be used as a trade-off between computation time and accuracy of various graph matching applications.
 
We describe an approach to error-tolerant graph matching using graph homeomorphism to find the structural similarity between two graphs \citep{DwivediSingh2017}. In this technique, we use path contraction to reduce the number of nodes in the input graphs, while preserving the topological equivalence of these graphs. We also present an approach to error-tolerant graph matching using node contraction, in which a given graph is transformed into another graph by contracting smaller degree nodes \citep{DwivediSingh2018b}. The above approaches lead to a reduction in the search space needed to compute graph edit distances between transformed graphs.

This chapter is organized as follows. Section 3.2 describes preliminaries and section 3.3 presents homeomorphic graph edit distance. Error-tolerant graph matching using extended graph edit distance is introduced in section 3.4. Experimental evaluation of the above techniques is shown in section 3.5. Finally, section 3.6 summarizes this chapter. 
 
\section{Preliminaries}
In this section, we review the basic definitions and concepts related to graph matching and the proposed work.

%Two graphs are homeomorphic, if one of them can be obtained by subdividing the edges of another graph by inserting new nodes along the edges. 
Let $G_1$ and $G_2$ be two graphs. A bijective mapping $f:V_1' \rightarrow V_2'$ from $G_1$ to $G_2$ is called an \textit{error-tolerant graph matching}, if $V_1' \subseteq V_1$ and $V_2' \subseteq V_2$ \citep{Bunke1998}.

A common set of edit operations involves insertion of vertices or edges, deletion of vertices or edges, and substitution of vertices or edges. Insertion of a vertex $u$ is denoted by $\epsilon \rightarrow u$, deletion of vertex $u$ by $u \rightarrow \epsilon$, and substitution of a vertex $u$ by vertex $v$ is denoted by $u \rightarrow v$. We use an additional edit operation for path contraction of a simple path in which every vertex except first and last have degree two. We denote this path contraction by $(u_1,...,u_n) \rightarrow (u_1, u_n)$. We use path contraction to remove each node with degree two, while maintaining the topological equivalence of the graphs.

Let $G=(V, E)$ be a graph having an edge $e=(u, v) \in E$. Subdivision of the edge $e$ produces another graph $G'=(V', E')$ such that $V'=V \cup \{w\}$, and $E' = E \setminus \{(u, v )\} \cup \{(u, w), (w, v)\}$, where $w$ is an additional vertex and $(u, w)$, $(w, v)$ are additional edges. A subdivision of a graph $G$ is another graph $G'$ obtained by performing the subdivision operation on one or more edges in $G$. Two graph $G_1$ and $G_2$ are homeomorphic or topologically equivalent, if both graphs $G_1$ and $G_2$ are a subdivision of some graph $G$. 

Since subdivision of $G$, changes the count of vertices of degree two only, hence for two graphs to be homeomorphic they must have the same number of vertices of each degree except two.

\section{Homeomorphic Graph Edit Distance}\label{sec:Homeomorphic Graph Edit Distance}

Two graphs are said to be homeomorphic, when one of the graph can be transformed to the other one after performing subdivision operations on the edges by inserting the additional nodes along its edges. 

Let $G=(V, E)$ be a graph with $V$ and $E$ be vertex set and edge respectively and let $e=(u, v) \in E$. Subdivision operation on the edge $E$ generates another graph $G'=(V', E')$ so that $V'=V \cup \{w\}$, and $E' = E \setminus \{(u, v )\} \cup \{(u, w), (w, v)\}$, here $w$ is a new vertex and $(u, w)$, $(w, v)$ are new edges.

%------------------------------------------------------------------------------------

Homeomorphic graph edit distance is defined as the minimum count of edit operations required to convert a graph $G_1$ to make it homeomorphic to graph $G_2$. The elementary idea behind the homeomorphic graph edit distance is that if two graphs are homeomorphic or topologically equivalent then they must have the same number of vertices of each degree, except degree two vertices. Therefore we can remove each vertex of degree 2, and again get a homeomorphic graph.

\begin{defn}
%\textbf{Definition 1.}
 Let $G_i = (V_i, E_i, \mu_i,\nu_i)$ for $i=1,2$ be two graphs, the homeomorphic graph edit distance (HGED) between $G_1$ and $G_2$ is defined by
 $$ HGED(G_1, G_2) = GED(G_1', G_2') $$ 
 \begin{equation}
 = min_{(e_1,...,e_k) \in \varphi (G_1', G_2')} \sum_{i=1}^{k} c(e_i) \label{eq}
 \end{equation}
 Where $G_1'$ and $G_2'$ are graphs obtained from $G_1$ and $G_2$ respectively by performing path contraction from each vertex,  $\varphi (G_1', G_2')$ denotes the set of edit distance paths converting $G_1'$ into $G_2'$, and $c(e_i)$ is the cost function associated with every edit operation $e_i$.
\end{defn}
 
Specifically, we perform a path contraction of all simple paths $(u_1,...,u_n)$ in a graph, such that all vertices along this path have degree two (for $i=2$ to $n-1$, $Deg(u_i)=2$) except first $(u_1)$, and last vertex $(u_n)$. These path contractions will save the huge processing of several vertices and edges, which are required in the computation of graph edit distance. 

\begin{prop} \label{prop:homeomorphic}
%\textbf{Proposition 1.}
 Let $G = (V, E, \mu,\nu)$ be a graph. If $G' = (V', E', \mu',\nu')$ be the graph obtained by performing path contraction $(u_1,...,u_n) \rightarrow (u_1, u_n)$ by removing or smoothing the vertices $u_2,...,u_{(n-1)}$ where $Deg(u_i)=2$ for $i=2$ to $n-1$, then $G$ and $G'$ are homeomorphic.
\end{prop}
 
The process of contraction removes the 2-degree nodes only, and other nodes remain unaffected. Since smoothing reverses the operation of the subdivision, therefore $G$ and $G'$ are homeomorphic.

\subsection{Homeomorphic Edit Cost}
Homeomorphic edit cost functions are based on Euclidean distance measure, which allocates constant cost to insertion, deletion of nodes and edges. Substitution cost of nodes, edges and paths are assigned proportional to their Euclidean distance. 

For two graphs $G_1$ and $G_2$, we define homeomorphic graph edit cost function for all nodes $u \in V_1$, $v \in V_2$ and for all edges $e \in E_1$, $e' \in E_2$ by \\
$c(u \rightarrow \epsilon)= x_{node}$ \\
$c(\epsilon \rightarrow v)= x_{node}$ \\
$c(u \rightarrow v)= y_{node}.|| \mu_1(u) - \mu_2(v)||$ \\
$c(e \rightarrow \epsilon)= x_{edge}$ \\
$c(\epsilon \rightarrow e')= x_{edge}$ \\
$c(e \rightarrow e')= y_{edge}. || \nu_1(e) - \nu_2(e')||$ \\
%$c((u_1,...,u_n) \rightarrow (u_1, u_n))= z_{path} $ \\
$c((u_1,...,u_n) \rightarrow (u_1, u_n))= z_{path}.|| \mu_1(u_1) - \mu_1(u_n)||$ \\
Where $x_{node}$, $x_{edge}$, $y_{node}$, $y_{edge}$, and $z_{path}$ are non-negative parameters, and $c(u \rightarrow \epsilon)$ is the cost of deletion of node $u$, $c(\epsilon \rightarrow v)$ is the cost of insertion of node $v$, $c(u \rightarrow v)$ is the cost of substitution of node $u$ by node $v$, $c(e \rightarrow \epsilon)$ is the cost of deletion of edge $e$, $c(\epsilon \rightarrow e')$ is the cost of insertion of edge $e'$, and $c((u_1,...,u_n) \rightarrow (u_1, u_n))$ is the cost of path contraction from $(u_1,...,u_n)$ to $(u_1, u_n)$. We observe that the substitution cost defined above is proportionate to Euclidean metric of the corresponding vertex and edge labels. The above graph edit function is similar to the edit cost of graph edit distance with the introduction of an additional operation $(u_1,...,u_n) \rightarrow (u_1, u_n)$ and a parameter $z_{path}$. 

\begin{prop}
%\textbf{Proposition 2.}
Path substitution $c((u_1,...,u_n) \rightarrow (u_1, u_n))$ can save up to $(2n-2)$ edit operations.
\end{prop}
%\begin{proof}
Here, we can observe that $z_{path}=(n-1)x_{edge} + (n-2)x_{node} + x_{edge} = n.x_{edge} + (n-2)x_{node}$, and hence we save up to $(2n-2)$ operations for every path contraction.
%\end{proof}

An example for homeomorphic edit path computation from graph $G$ to graph $H$ is shown in Figure \ref{fig:edit-path}. First, $G$ is converted to $G'$ and $H$ is converted to $H'$ using path contraction, and finally $G'$ is transformed to $H'$ using four edit operations, these are node insertion followed by edge insertion and again node insertion followed by edge insertion.

\begin{figure}[!t]
\centering
 \includegraphics[scale=.4]{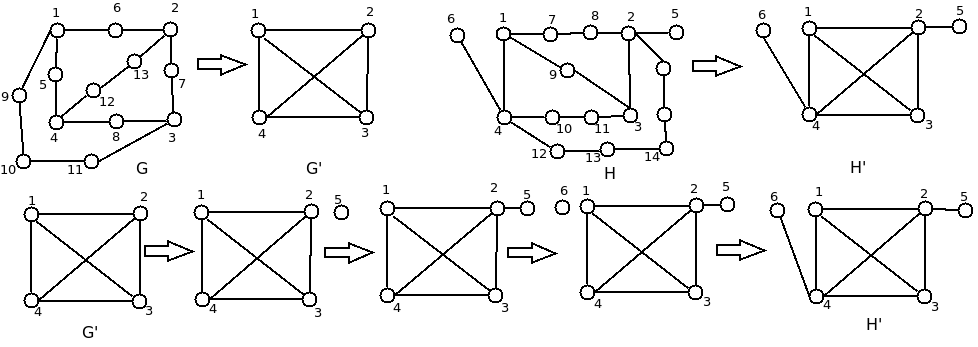}
 \caption{Homeomorphic edit path from $G$ to $H$}
 \label{fig:edit-path}
\end{figure}

% \begin{figure}
% \centerline{\includegraphics[width=1.00\textwidth]{Chapters/figures/Dia4.png}}
% \caption{Homeomorphic edit path from $G$ to $H$}
% \end{figure} 
 
\subsection{Algorithm}

\begin{algorithm}
\caption{\bf :  Homeomorphic-Graph-Edit-Distance $(G_1,G_2)$} \label{algorithm:homeomorphic-ged}
\begin{algorithmic}[1]
\Require Two Graphs $G_1$, $G_2$, where $G_i = (V_i, E_i, \mu_i,\nu_i)$ for $i=1,2$ 
                       where $V_1 = \{u_1,...,u_n\}$ and $V_2 = \{v_1,...,v_m\}$
\Ensure A min. cost homeomorphic GED between $G_1$ and $G_2$
  
   \For {each $(u_i \in V_1)$ }   
   \If {(there is a path $(u_i, u_{i+1},...,u_{i+k})$ such that 
   $deg(u_{i+1})=deg(u_{i+2})=...=deg(u_{i+k-1})=2$)}
   \State $(u_i, u_{i+1},...,u_{i+k}) \rightarrow (u_i, u_{i+k}) $ 
   \State $V_1 \leftarrow V_1 \setminus \{u_{i+1},..., u_{i+k-1} \}$ 
   \EndIf
   \EndFor
   \For {each $(v_j \in V_2)$ }
   \If {(there is a path $(v_j, v_{j+1},...,v_{j+k})$ such that  
   $deg(v_{j+1})=deg(v_{j+2})=...=deg(v_{j+k-1})=2$)}
   \State $(v_j, v_{j+1},...,v_{j+k}) \rightarrow (v_j, v_{j+k}) $ 
   \State $V_2 \leftarrow V_2 \setminus \{v_{j+1},..., v_{j+k-1} \}$ 
   \EndIf
   \EndFor
   \State Update $G_1, G_2, n \leftarrow n', m \leftarrow m'$
   \State $A \leftarrow \emptyset$
   \For {each $(v_j \in V_2)$}
   \State $A \leftarrow A \cup \{ u_1 \rightarrow v_j \}$ 
   \EndFor
   \State $A \leftarrow A \cup \{ u_1 \rightarrow \epsilon \}$ 
   \While { $(1)$ }
   \State Prune $A$ using optimizing techniques 
   \State Compute min. cost graph edit path $C_{min}$ from $A$ 
   \If {($C_{min}$ is a complete graph edit path)}
   \Return $C_{min}$
   \Else
       \If {(all nodes $(u_i \in V_1)$ are processed)}
        \For {each unprocessed $(v_j \in V_2)$}
         \State $C_{min} \leftarrow C_{min} \cup \{ \epsilon \rightarrow v_j \} $ 
        \EndFor
        \State $A \leftarrow A \cup \{ C_{min} \}$
        \Else
       \For {(each unprocessed node $(u_i \in V_1)$)}
         \For {(each $(v_j \in V_2)$)}
         \State $C_{min} \leftarrow C_{min} \cup \{ u_i \rightarrow v_j \} \cup \{ u_i \rightarrow \epsilon \}$ 
         \EndFor
       \EndFor
       \State $A \leftarrow A \cup \{ C_{min} \}$
   \EndIf
   \EndIf
   \EndWhile
    
\end{algorithmic}
\end{algorithm} 

The computation of error-tolerant graph matching using graph homeomorphism is described in Algorithm \ref{algorithm:homeomorphic-ged}. The input to the Homeomorphic-Graph-Edit-Distance algorithm is two graphs $G_1$ and $G_2$, where $G_i = (V_i, E_i, \mu_i,\nu_i)$ for $i=1,2$. The graphs $G_1$ and $G_2$ have $n$ and $m$ vertices respectively, and output of the algorithm is a minimum cost homeomorphic graph edit distance between $G_1$ and $G_2$. A brief description of this algorithm is as follows. The algorithm proceeds by performing path contraction on both graph $G_1$ and $G_2$. For each vertex of graph $G_1$, it searches for simple paths in which all intermediate vertices have degree two, and then it updates this path by removing all such intermediate vertices and insert an edge between first and last vertices. After performing same path contraction operations on $G_2$, both graphs and their parameters are updated. The algorithm then initializes an empty set $A$. Simultaneous substitution of first vertex $u_1$ of $G_1$ with every other vertex of $G_2$ is inserted in $A$ along with the deletion of $u_1$. After that, \textit{while} loop is executed until we get a minimum cost edit distance $C_{min}$, which is also a complete graph edit path. In \textit{while} loop, pruning on $A$ is performed to reduce the search space using various optimizing techniques and heuristic methods. If $C_{min}$ is the complete edit path, i.e., the set of graph edit path is identical to input graph, then algorithm returns this value. Otherwise, if all nodes of $G_1$ are processed and some nodes of $G_2$ are not processed then these unprocessed nodes are inserted in $C_{min}$, and $A$ is updated in the inner loop of \textit{if} part. Finally, in the inner loop of the \textit{else} part, every unprocessed node of $G_1$ is substituted by all nodes of second graph $G_2$ and these substitutions are inserted in $C_{min}$ along with the deletion of unprocessed nodes of $G_1$ and $A$ is updated.

The correctness of Homeomorphic-Graph-Edit-Distance algorithm can be established using Proposition \ref{prop:homeomorphic}. The algorithm reduces the number of vertices in input graph, which lowers the search space required for further processing of edit distance computation. Various pruning strategies can be used to decrease the total execution time. Like, we can keep a fixed number of vertices in the set $A$ at any given time and whenever there is an update in $A$ only that much minimum cost fix entries are retained out of all available edit path entries.

\section{Extended Graph Edit Distance}  

% From Basic concepts and motivation (start)
The usefulness of error-tolerant graph matching is on the premise that it can accommodate errors that may have acquired in the input graph due to the presence of noise and distortions during the processing and retrieval steps. For a dense graph, if the distortion occurs with the addition (or deletion) of a node or an edge followed by successive additions (or deletion) of nodes or edges, then to perform the graph matching it may be reasonable to ignore the smallest degree nodes, if we want to do the matching using as small number of nodes of graph as possible. For example, suppose we have a graph $G$ as given in Figure \ref{fig:node-contraction}, and due to distortion and presence of noise, additional nodes $(u, v, w)$ got added, shown in the figure to be connected with dotted edges in graph $H$. After removing or contracting the 1-degree, 2-degree and 3-degree nodes, we get the same graph as $G$.  

\begin{figure}[!t]
\centering
\includegraphics[scale=0.1]{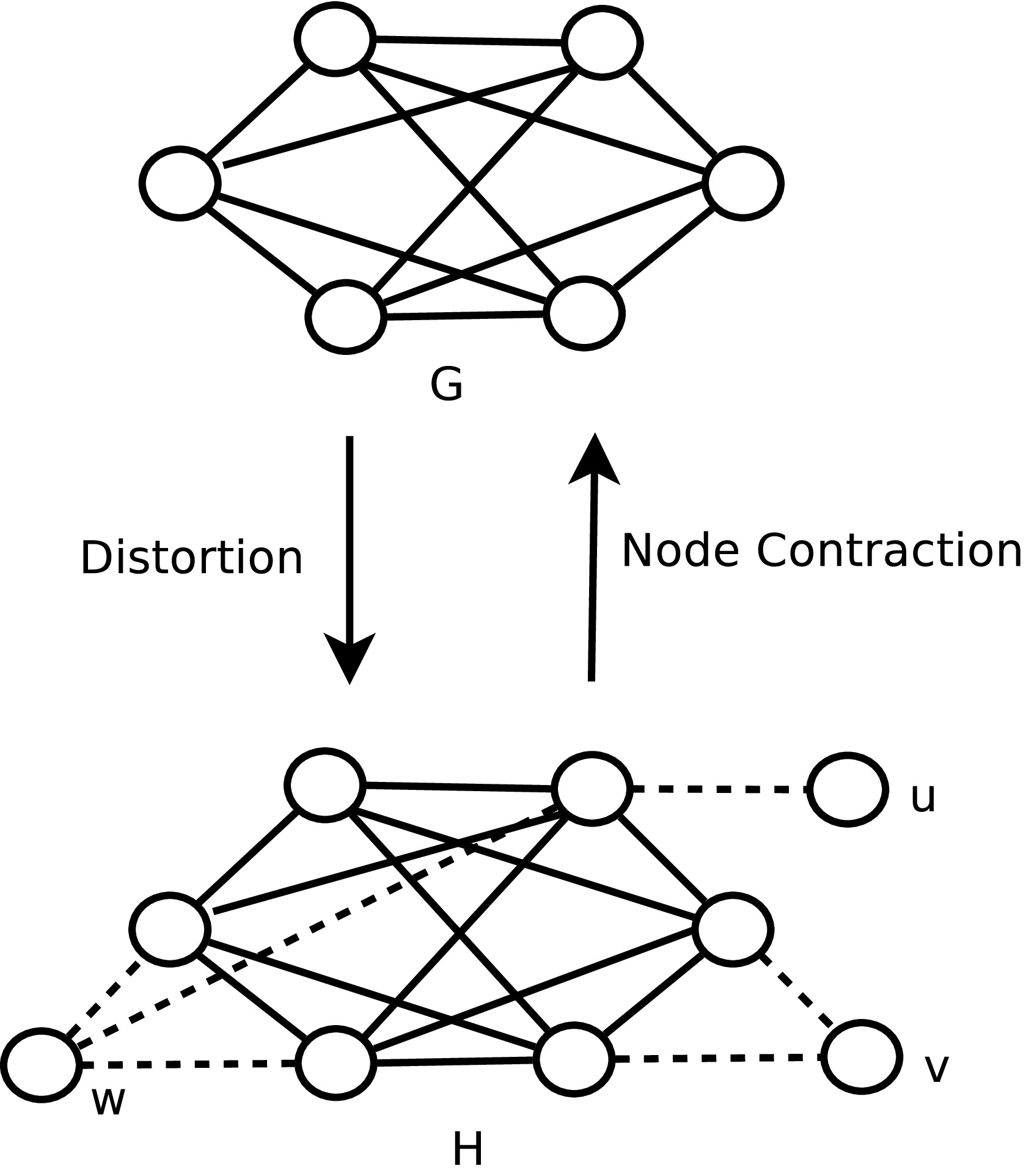}
\caption{Node contraction on the graph \textit{H}}
\label{fig:node-contraction}
\end{figure}
% From Basic concepts and motivation (end)

The node contraction's idea is to ignore the less important nodes in the graph before proceeding to perform graph matching. One of the measures of importance or centrality of nodes in a graph is degree centrality. Considering the degree centrality as a node importance indicator, we ignore or delete the nodes starting from least degree nodes. By merely removing  nodes from the graph will alter the topology of the graph drastically. To use a middle path, where the topology of the graph does not get changed abruptly, we delete the nodes and their associated edges, only when this does not disconnects the graph. In other words, we remove the smaller degree nodes provided that the total number of connected components of the graph remains the same. Keeping in mind the above point of view, we can extend the graph edit distance to ignore the smaller degree nodes. 

\begin{defn}
%\textbf{Definition 1.}
\textit{Node contraction} of a node $u$ in a graph $G$ is simply the deletion of the node $u$ and its associated links provided that the node $u$ is not a cut vertex. 
\end{defn}
Cut vertex is a vertex whose removal disconnects the graph. It is also called as cut point or articulation point.

%\textbf{Definition 2.}
\begin{defn}
\textit{$k$-degree node contraction} is the operation of performing the node contraction of all the nodes of degree $k$ of a graph $G$. We use $k$NC($G$) to denote $k$-degree node contraction on a graph $G$.
\end{defn}

%\textbf{Definition 3.}
\begin{defn}
\textit{$k^{*}$-degree node contraction} is the process of applying $k$-degree node contraction consecutively from one degree to $k$ degree on a graph $G$. 
\end{defn}

If we denote $k^{*}$-degree node contraction on a graph $G$ by $k^{*}$-$NC(G)$ then $k^{*}$-$NC(G)$=$kNC((k-1)NC((k-2)NC(...1NC(G)...)))$.

%\textbf{Definition 4.}
\begin{defn}
\textit{$k$-GED} is the the computation of graph edit distance between two graphs with $k$-degree node contraction performed on both graphs.
\end{defn}

For two graphs $G_1$ and $G_2$, we have $$k\text{-}GED(G_1, G_2)=GED(kNC(G_1), kNC(G_2))$$
% $$k\text{GED}(G_1, G_2)=\text{GED}(k\text{NC}(G_1), k\text{NC}(G_2))$$

For $k=2$, $k$-GED generalizes to HGED \citep{DwivediSingh2017}, when path contraction is performed so that all simple paths $(u_1,...,u_n)$ are replaced by $(u_1, u_n)$ where $Deg(u_i)=2$ for $i=2$ to $n-1$.

%\textbf{Definition 5.}
\begin{defn}
\textit{$k^{*}$-GED} is the computation of graph edit distance between two graphs with $k$-degree node contraction performed on both graphs starting from one-degree nodes to $k$ degree nodes.
\end{defn}

Let $G_1$ and $G_2$ be two graphs, then 
$$
 k^{*}\text{-}GED(G_1, G_2)=
 \begin{cases}
  GED(G_1, G_2), \text{if } k=0 
  \\
  GED(k^{*}\text{-}NC(G_1), k^{*}\text{-}NC(G_2)), \text{if } k \geq 1
 \end{cases}
$$

To make the effect of node contraction on graph topology more specific, we define \textit{node deletion}, as a standard removal of a node and its associated edges, regardless of whether it is a cut vertex. Similar to $k$-degree node contraction, $k^{*}$-degree node contraction ($k^{*}$-NC), $k$-GED and $k^{*}$-GED, we define  $k$-degree node deletion, $k^{*}$-degree node deletion ($k^{*}$-ND), $k$-ND-GED and $k^{*}$-ND-GED respectively, where in each definition node contraction operation is replaced by node deletion. Here, we observe that $|k^{*}$-ND$(G)| \leq |k^{*}$-NC$(G)|$, where $|G|$ denotes size or number of nodes in graph $G$. Since the removal of nodes in node deletion operation is unrestricted; therefore the number of nodes in a graph $G$ after node deletion is less than or equal to that of the number of nodes after node contraction.

\subsection{Extended Edit Cost}
To define the edit cost of $k$-GED, we can extend the edit cost of graph edit distance by adding $c(u \rightarrow \epsilon)= 0$, whenever $deg(u)=k$ and $u$ is not a cut vertex of the graph.

$k$-GED uses Euclidean distance measure and assigns the constant cost to insertion, deletion and substitution of nodes and edges. Let $G_1$ and $G_2$ be two graphs, for all nodes $u \in V_1$, $v \in V_2$ and for all edges $e \in E_1$, $f \in E_2$, we define the extended edit cost function as follows. \\
$c(u \rightarrow \epsilon)= x_{node}$ \\
$c(\epsilon \rightarrow v)= x_{node}$ \\
$c(u \rightarrow v)= y_{node}.|| \mu_1(u) - \mu_2(v)||$ \\
$c(e \rightarrow \epsilon)= x_{edge}$ \\
$c(\epsilon \rightarrow f)= x_{edge}$ \\
$c(e \rightarrow f)= y_{edge}. || \nu_1(e) - \nu_2(f)||$ \\
$c(u \rightarrow \epsilon)= 0$, if $deg(u)=k$ and $u$ is not a cut point \\
Here, $c(u \rightarrow \epsilon)$ denotes the cost of deletion of node $u$, $c(\epsilon \rightarrow v)$ stands for the cost of insertion of node $v$, $c(u \rightarrow v)$ is the cost of substitution of node $u$ by node $v$, $c(e \rightarrow \epsilon)$ denotes the cost of deletion of edge $e$, $c(\epsilon \rightarrow f)$ is the cost of insertion of edge $f$, and $x_{node}$, $y_{node}$, $x_{edge}$, $y_{edge}$ are non-negative constants.

We observe that the above cost function satisfy the non-negativity property, i.e., $c(e_i) \geq 0$, for every node and edge edit operations $e_i$ along with the triangle inequality property of insertion, deletion and substitution of nodes and edges.

\subsection{Algorithm}
In this section, we present the algorithm to perform error-tolerant graph matching using node contraction. The computation of $k^{*}$-degree node contraction of an input graph is described in Algorithm \ref{algorithm:node-contraction}. The input to the $k^{*}$-Node-Contraction algorithm is a graph $G$ and a parameter $k$. The outer \textit{for} loop of the algorithm in lines 1--16, iteratively perform $k$-degree node contraction from 1 to $k$. The \textit{for} loop of the algorithm in lines 2--8, uses a boolean flag \textit{visit}, which is set to 1 when the degree of a vertex of $G$ is $k$ else it is reset to 0. If the visit flag of a node is set to 1 and it is not a cut vertex, then the node and its associated edges are removed from $G$ and visit flag is reset to 0 in the \textit{for} loop of lines 9--15. Finally, the transformed graph $G$ is returned in line 17. This algorithm can be considered as a preprocessing phase of the proposed error-tolerant graph matching framework.

\begin{algorithm}
\caption{\bf :  $k^{*}$-Node-Contraction $(G, k)$} \label{algorithm:node-contraction}
\begin{algorithmic}[1]

\Require A graph $G = (V, E, \mu,\nu)$ and a parameter $k$
\Ensure Transformed graph after applying $k^{*}$-degree node contraction on $G$
%\Statex \textbf{Input}: A graph $G = (V, E, \mu,\nu)$ and a parameter $k$  \\             
%\Statex \textbf{Output}: Transformed graph after applying $k^{*}$-degree node contraction on $G$
    \For {$(i \leftarrow 1 \text{ to } k)$ }
     \For {each $(u \in G)$}
%   {
     \If {($(deg(u)$==$k)$)}
     \State $u.visit \leftarrow 1$ 
     \Else
     \State $u.visit \leftarrow 0$ 
     \EndIf
%   }
     \EndFor
    \For {each $(u \in G)$}
%   {
     \If {($(u.visit$==$1) \&\& (u \text{ is not cut vertex})$)} 
     \State $V \leftarrow V \setminus \{u\}$ 
     \State $E \leftarrow E \setminus \{(u,v) | (u,v) \in E, \forall v \in G \}$ 
     \State $u.visit \leftarrow 0$ 
     \EndIf
%   }
    \EndFor
   \EndFor \\
   \Return $G$
  \end{algorithmic}
\end{algorithm}

%\textbf{Proposition 1.}
\begin{prop}
$k^{*}$-Node-Contraction algorithm computes $k^{*}$-degree node contraction of $G_1$ and $G_2$.
\end{prop}
We can observe that the \textit{for} loop in lines 1--16 of the Algorithm \ref{algorithm:node-contraction}, ensures that $k$-degree node contraction starts from $i$=1 to $k$. \textit{If} loop in line 10, allow only those nodes to be removed, whose degree is $k$ and it is not a cut vertex, whereas \textit{visit} flag make sure that each vertex is considered only once for contraction.  
 
%\textbf{Proposition 2.}
\begin{prop}
$k^{*}$-Node-Contraction algorithm executes in $O(n)$ time.
\end{prop}
 We can check whether a node is cut vertex in $O(n)$, and therefore \textit{for} loop in lines 9--15 takes 
$O(n)$. \textit{For} loop in lines 2--8 also take $O(n)$ time, while the outer \textit{for} loop in lines 1--16 executes $k$ times. So the Algorithm \ref{algorithm:node-contraction} takes overall $O(k.n)$, which is $O(n)$ time.
 
The $k^{*}$-GED computation of two graphs is described in Algorithm \ref{algorithm:k-ged}. A brief description of the $k^{*}$-Graph-Edit-Distance algorithm is as follows. The input to the algorithm is two graphs $G_1$, $G_2$, and a parameter $k$, and the output is the minimum cost $k^{*}$-GED between $G_1$ and $G_2$. It calls the Algorithm \ref{algorithm:node-contraction} to perform $k^{*}$-degree node contraction on $G_1$ and $G_2$ in lines 1--2. The transformed graphs are $G_1'$ and $G_2'$ with vertex set $V_1' = \{u_1',...,u_{n'}'\}$ and $V_2' = \{v_1',...,v_{m'}'\}$ respectively. The algorithm initializes an empty set $S$ in line 3. In the \textit{for} loop in lines 4--6, $S$ is updated by substitution of vertex $u_1'$ of $G_1'$ with each vertex of $G_2'$, then deletion of $u_1'$ in line 7 is added to $S$. The \textit{while} loop in lines 8--28, is used to compute the minimum cost edit path $C_{min}$, from $S$.

\begin{algorithm}
\caption{\bf :  $k^{*}$-Graph-Edit-Distance $(G_1,G_2)$} \label{algorithm:k-ged}
\begin{algorithmic}[1]
\Require Two Graphs $G_1$, $G_2$, where $G_i = (V_i, E_i, \mu_i,\nu_i)$ for $i=1,2$ 
                       where $V_1 = \{u_1,...,u_n\}$ and $V_2 = \{v_1,...,v_m\}$ and a parameter $k$
\Ensure A min. cost $k^{*}$-GED between $G_1$ and $G_2$                       
%\Statex \textbf{Input}: Two Graphs $G_1$, $G_2$, where $G_i = (V_i, E_i, \mu_i,\nu_i)$ for $i=1,2$ 
%                       where $V_1 = \{u_1,...,u_n\}$ and $V_2 = \{v_1,...,v_m\}$ and a parameter $k$
%\Statex \textbf{Output}: A min. cost $k^{*}GED$ between $G_1$ and $G_2$
  
   \State $G_1' \leftarrow \text{$k^{*}$-Node-Contraction} (G_1, k)$  $\text{ }\text{ }\text{ }\text{ }$ // $V_1' = \{u_1',...,u_{n'}'\}$
   \State $G_2' \leftarrow \text{$k^{*}$-Node-Contraction} (G_2, k)$  $\text{ }\text{ }\text{ }\text{ }$ // $V_2' = \{v_1',...,v_{m'}'\}$
   \State $S \leftarrow \emptyset$
   \For {all $(v_i' \in V_2')$}
%   {
   \State $S \leftarrow S \cup \{ u_1' \rightarrow v_i' \}$ 
%  }
   \EndFor
   \State $S \leftarrow S \cup \{ u_1' \rightarrow \epsilon \}$ 
   \While { $(True)$ }
%   {
   \State Prune $S$ using heuristic methods 
   \State Compute $C_{min}$ the min. cost edit path from $S$ 
   \If {($C_{min}$ is a complete edit distance path)} 
       \State \textbf{return} $C_{min}$
    \Else
       \If {(every node $(u_i' \in V_1')$ is processed)}
        \For {every unprocessed $(v_j' \in V_2')$}
%         {
         \State $C_{min} \leftarrow C_{min} \cup \{ \epsilon \rightarrow v_j' \} $ 
%          }
        \EndFor
        \State $S \leftarrow S \cup \{ C_{min} \}$
        \Else
       \For {(each unprocessed node $(u_i' \in V_1')$)}
%       {
         \For {(every $(v_j' \in V_2')$)}
%         {
         \State $C_{min} \leftarrow C_{min} \cup \{ u_i' \rightarrow v_j' \} \cup \{ u_i' \rightarrow \epsilon \}$ 
%          }
         \EndFor
%          }
       \EndFor
       \State $S \leftarrow S \cup \{ C_{min} \}$
   \EndIf
   \EndIf
%   }
   \EndWhile
    
\end{algorithmic}
\end{algorithm}

Computation of minimum cost edit path is usually performed using tree-based search algorithm like $A^*$ search where the top node or root denotes the first edit operation and bottom or leaf node represents the last edit operation. Edit path from the root to leaf nodes constitutes a complete edit distance path, and it exhibits an edit path to transform a graph to another one which may or may not be optimal. If we want to return optimal edit path, then during each level of search we have to consider all the combinations of edit operations in the set $S$, which can be computationally much expensive. An alternative is to prune the set $S$ (line 9) using various heuristic techniques to consider only the partial set of edit operations which leads to a suboptimal but relatively efficient solution. For example, we can use beam search to limit the search space by considering only $w$ best possibility at each level of search, where $w$ is called as beam width. During the execution of the algorithm, if $C_{min}$ is the complete edit distance path then the algorithm returns its value in line 12; otherwise all the unprocessed nodes of $G_2'$ are added in $C_{min}$ and $S$ is updated in lines 14--18. Finally, every unprocessed node of $G_1'$ are substituted by each node of $G_2'$ and these operations are added to $C_{min}$, together with the deletion of unprocessed nodes of $G_1'$ in lines 20--24, and $S$ is updated in line 25.

%\textbf{Proposition 3.}
\begin{prop}
$k^{*}$-Graph-Edit-Distance algorithm computes error-tolerant graph matching of $G_1'$ and $G_2'$.
\end{prop}
It follows from the properties of edit operations, i.e., addition, deletion and substitution of nodes and edges along with the definition of $k^{*}$-GED. A complete edit path returned by Algorithm \ref{algorithm:k-ged} ensures that each node of $G_1'$ is uniquely mapped to $G_2'$, while Algorithm \ref{algorithm:node-contraction} makes sure that the modified vertex set $V_1'$ and $V_2'$ are subsets of $V_1$ and $V_2$ respectively.   

The worst case complexity of $k^{*}$-Graph-Edit-Distance algorithm is exponential on the number of nodes in input graphs, but a suitable parameter $k$ can be selected to reduce the overall execution time. 
Let us consider a tree-search based method implementation of $k^{*}$-Graph-Edit-Distance algorithm. At the first level of execution $u_1'$  is replaced by each of the $v_i'$ for $i=1$ to $m'$, making it $O(m')$. Suppose substitution of $u_1'$  to $v_1'$ is selected at the first level, then during the next level $u_2'$ is again replaced by each of the $v_i'$ for $i=2$ to $m'$, which is $O(m')$, making total time $O(m'^2)$ up to the second level. Similarly, at $n'$th level, the worst-case execution time would become $O(m'^{n'})$. As discussed in Algorithm \ref{algorithm:k-ged}, a heuristic method like beam search can be used to reduce the average case execution time of algorithm at the cost of getting a suboptimal solution. 

To consider the effect of $k$ on the size of the transformed graphs $n'$  and $m'$. We observe that value of $n'$ and $m'$ is highly dependent on the topology and size of the input graphs. Let $G(n,p)$ be a random graph, where $n$ is the number of nodes and $p$ is the probability of edges between each pair of vertices. Then the vertex degree distribution, which is the number of vertices of degree $k$ is given by $p_k = {n-1 \choose k} p^k (1-p)^{n-k-1}$. Now, we have the following loose relationship between $n$ and $n'$. 

%\textbf{Proposition 4.} 
\begin{prop}
For a random graph $G(n,p)$, the value of $n$ and $n'$ as used in Algorithm \ref{algorithm:k-ged} satisfy the following inequality
$$ n-\sum_{n=1}^k {n-1 \choose k} p^k (1-p)^{n-k-1} \leq n' \leq n-(n-1)p(1-p)^{n-2}.$$
\end{prop}
\begin{proof}
Here, $n$ is the number of nodes in $G_1$ and $n'$ is the number of nodes in $G_1'$ obtained after $k^{*}$-NC$(G_1)$. So, the number of contracted nodes are $n-n'$. We have, number of nodes removed during $k^{*}$-NC$(G_1) \leq n-n' \leq$ number of nodes removed during $k^{*}$-ND$(G_1)$, which implies 
$ (n-1)p(1-p)^{n-2} \leq n-n' \leq  \sum_{n=1}^k {n-1 \choose k} p^k (1-p)^{n-k-1}$, where left most term is the number of nodes removed during $1^{*}$-NC or $1^{*}$-ND, since both are equivalent operations. Now multiplying the above expression by -1 and rearranging the terms  $ -\sum_{n=1}^k {n-1 \choose k} p^k (1-p)^{n-k-1} \leq n'-n \leq -(n-1)p(1-p)^{n-2}$, by adding $n$ in above expression we get \\
$ n-\sum_{n=1}^k {n-1 \choose k} p^k (1-p)^{n-k-1} \leq n' \leq n-(n-1)p(1-p)^{n-2}$.
\end{proof}

\section{Results and Discussion}
% Results of ETGM using Homeomorphism
\subsection{Homeomorphic Graph Edit Distance}
We executed Homeomorphic-Graph-Edit-Distance algorithm on random graphs with a given number of nodes. The comparison of execution times of homeomorphic graph edit distance computation versus graph edit distance computation without any optimization or pruning technique is shown in Figure \ref{fig:ged-vs-hged}. Here, each execution time is the mean value of several execution times taken by graph edit distance computation and homeomorphic graph edit distance computation for different random graphs with fixed nodes. Simple graph edit distance computation usually did not terminate beyond 10 number of nodes, even though homeomorphic graph edit distance computation was able to complete the execution well beyond this size.  

\begin{figure}[!t]
\centering
 \includegraphics[scale=.6]{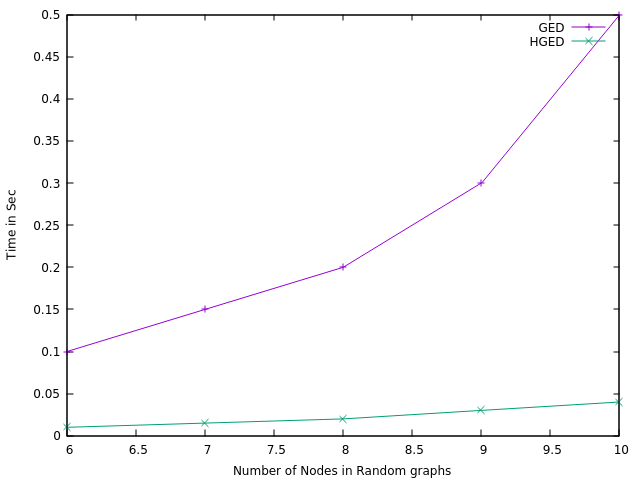}
 \caption{GED vs HGED computation} \label{fig:ged-vs-hged}
\end{figure}

The homeomorphism is basically a topological concept, which encapsulates the topological similarity of different objects. In fact, two homeomorphic objects or spaces are called topologically equivalent. This notion can be extended to measure the structural similarity of graphs. A set of homeomorphic graphs, exhibit some kind of similarity, which is denoted by homeomorphic graph edit distance. All graphs which are homeomorphic with respect to each other, will have same homeomorphic graph edit distance. 
For example, the homeomorphic graphs shown in Figure \ref{fig:homeomorphic-graphs}, have same homeomorphic graph edit distance. 

\begin{figure}[!t]
\centering
 \includegraphics[scale=.5]{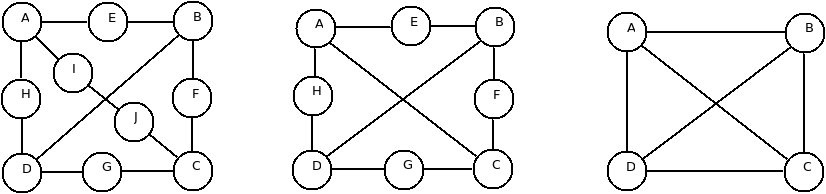} 
 \caption{Homeomorphic Graphs} \label{fig:homeomorphic-graphs}
\end{figure}

This idea can be particularly helpful in real world applications, where data can be modified by the presence of noise and distortion. We can use Homeomorphic-Graph-Edit-Distance algorithm for error-tolerant graph matching, which has a wide range of applications ranging from pattern recognition, biometric identification, image processing to biological can chemical applications.

The proposed homeomorphic graph edit distance can be seen as 2-degree path contraction, since it is obtained by removing all the vertices along a path having degree 2. It can further be generalized to $n$-degree vertices path contraction, where a given graph will be transformed to another graph by removing all the vertices having up to degree $n$ along any simple path of the input graph. The parameter $n$ can be selected in an application specific manner.

% Results of ETGM using node contraction
\subsection{Extended Graph Edit Distance}

%\section{Experimental Validation}
In this section, we compare the proposed graph matching framework with other important graph matching methods. To evaluate the execution time and accuracy of $k^{*}$-Graph-Edit-Distance, we use the IAM graph database \citep{RiesenBunke2008} and graph matching toolkit \citep{Riesenetal2013}. We use the letter database and AIDS database of IAM graph repository for the results.
\subsubsection{Dataset Description}
 Letter graphs consist of capital letters of alphabets, drawn using straight lines only. It includes 15 classes of capital letters namely A, E, F, H, J, K, L, M, N, T, V, W, X, Y, and Z. For each prototype graph, distortions of three distinct levels, i.e., high, medium and low are applied to generate various graph datasets. The graphs of letter dataset are uniformly distributed across the 15 letters. All nodes are labeled with an $(x, y)$ coordinate representing its position in a reference plane. The average number of nodes and edges per graph in letter graph data of high distortion level is 4.7 nodes and 4.5 edges respectively. The maximum number of nodes and edges per graph for this graph dataset are both 9. AIDS dataset contains graph characterizing chemical compounds. It consists of two different class of molecular compounds, i.e., confirmed active and confirmed inactive. Active class molecules show activity against HIV, whereas inactive class represents inactivity against HIV. Labels on node denote chemical symbol whereas labels on edges exhibit valence. In AIDS dataset average number of nodes per graph is 15.7 nodes while the average number of edges per graph is 16.2 edges. The maximum number of nodes and edges per graph is 95 nodes and 103 edges respectively. Table \ref{table:letter-aids-datasets-summary} contains a summary of the descriptions of the letter and AIDS datasets. . In Figure \ref{fig:fraction-of-graph-size-aids} we can observe the fraction of graphs concerning their size for active and inactive molecules of the training set of AIDS dataset. We note that the number of nodes in active molecules is usually more than that of inactive molecules.

 \begin{table}[!t]
\renewcommand{\arraystretch}{1.0}
\caption{Summary of letter and AIDS datasets} \label{table:letter-aids-datasets-summary}
\label{table }
\begin{center}
\begin{tabular}{|c| c| c|} 
\hline
Description & Letter dataset & AIDS dataset  \\ [1ex] \hline\hline 
Patterns & Letter line drawings & Chemical compounds \\ \hline
Node labels & $(x,y)$ coordinates & Chemical symbol \\ \hline
Edge labels & none & none \\ \hline
Average per graph & 4.7 nodes, 4.5 edges & 15.7 nodes, 16.2 edges \\ \hline
Maximum per graph & 9 nodes, 9 edges & 95 nodes, 103 edges \\ [1ex] \hline
\end{tabular} 
\end{center}
\end{table}
 
\begin{figure}[!t]
\centering
\includegraphics[scale=.6]{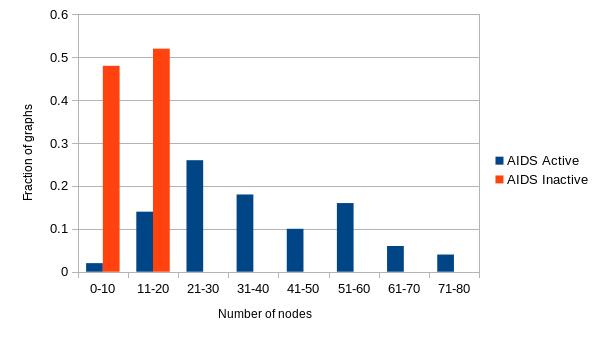}
\caption{Fraction of graphs with their size for AIDS dataset} \label{fig:fraction-of-graph-size-aids}
\end{figure}

\subsubsection{Execution Time Comparison}
All results in this section are computed using the system having 9.8 GB of memory and running the processor at 3.40 GHz. Comparison of average execution time of graph matching in milliseconds for GED and $k^{*}$-GED (where $k$=1,2 and 3) for letter graphs, using $A^*$ and with beam search optimization having beam width $w=10$ is shown in Figure \ref{fig:ged-vs-kged-letter-time}. There is a sharp change in running time from GED to $1^{*}$-GED using $A^*$. This decrease in execution time accounts for reduction in 1-degree nodes in the input graphs. 

\begin{figure}[!t]
\centering
\includegraphics[scale=.6]{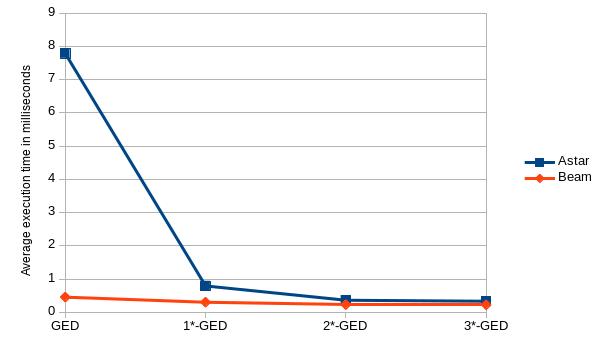}
\caption{GED vs. $k^{*}$-GED computation on letter graphs} \label{fig:ged-vs-kged-letter-time}
\end{figure}

Comparison of running time for GED and $k^{*}$-GED for AIDS graphs using $A^*$ is given in Figure \ref{fig:kged-aids-time}, whereas Figure \ref{fig:kged-aids-beam-time} shows the time in $ms$ with beam search heuristic. Here also, there is a sudden decrease in execution time from GED to $1^{*}$-GED using $A^*$, which becomes lesser from $1^{*}$-GED to $2^{*}$-GED and further from $2^{*}$-GED to $3^{*}$-GED. This is primarily due to reason that, we can not remove easily the 2-degree and 3-degree nodes from the graph as compared to 1-degree nodes without making the graph disconnected. 
%To reduce this shortcoming we can use a variant of $k^{*}$-GED, which allow the deletion of an articulation point in the definition of node contraction.

\begin{figure}[!t]
\centering
\includegraphics[scale=.6]{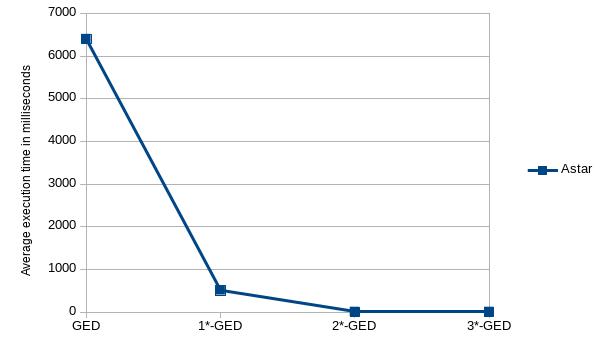}
\caption{\textit{GED} vs. \textit{k*GED} computation on AIDS graphs} \label{fig:kged-aids-time}
\end{figure}

\begin{figure}[!t]
\centering
\includegraphics[scale=.15]{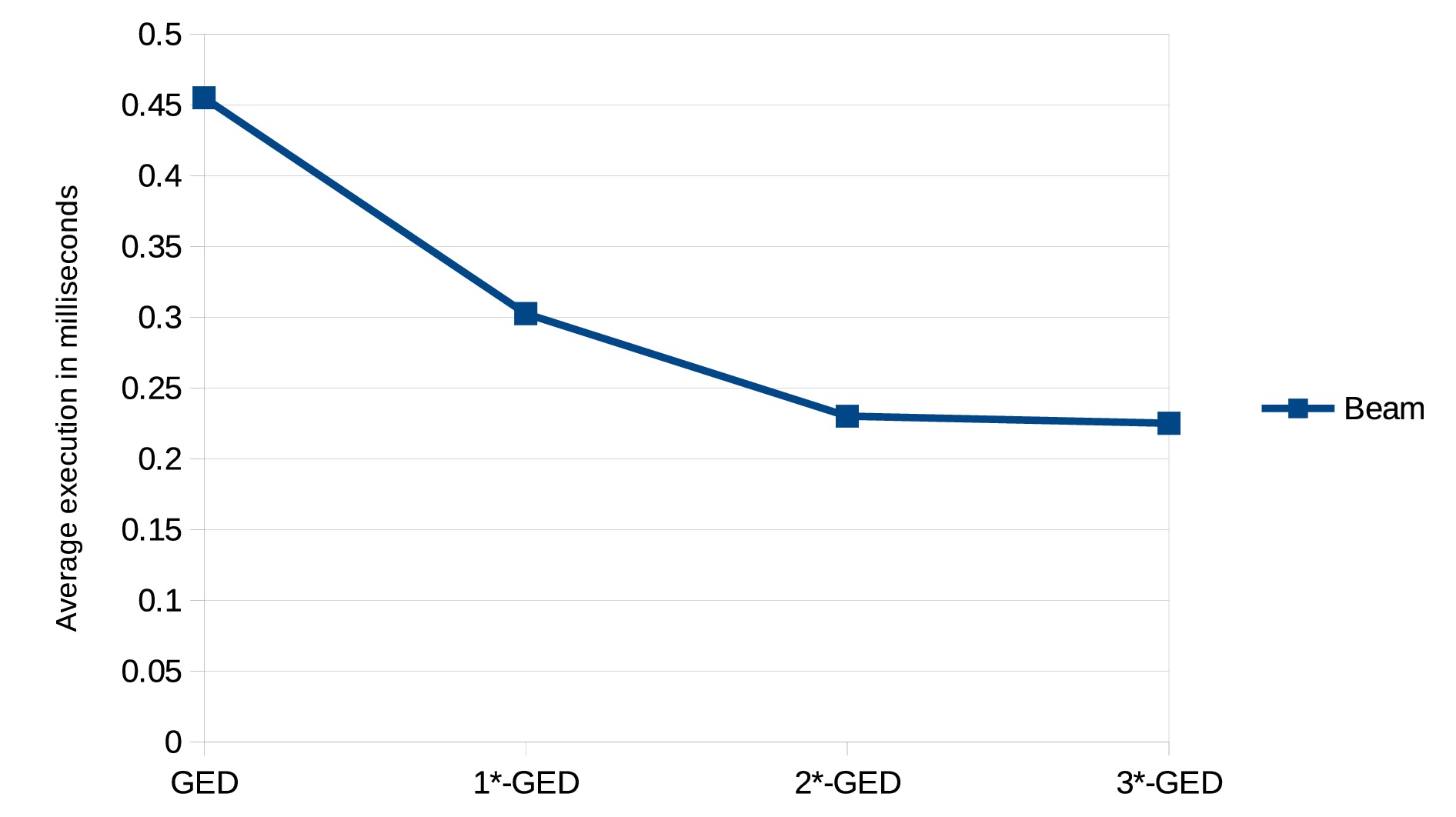}
\caption{\textit{GED} vs. \textit{k*GED} computation on AIDS graphs using beam search} \label{fig:kged-aids-beam-time}
\end{figure}

\subsubsection{Accuracy Comparison}
For accuracy evaluation, we consider the problem of classification of letter graphs using nearest neighbor classifier. We use 750 letter graphs of high distortion of the test dataset, which consists of 50 graphs for each of the 15 classes of letter dataset. We also use 750 graphs of training dataset, which again includes 50 graphs for each of the 15 letters. Comparison of classification accuracy for each of the 15 letters using $A^*$ based GED and $k^{*}$-GED for $k=1$ to $3$ is shown in Table \ref{table:accuracy-letter-ged-kged}, whereas classification accuracy of GED and $k^{*}$-GED based on beam search is shown in Table \ref{table:accuracy-letter-ged-kged-beam}. Here, we can observe that letters having less number of straight lines have less accuracy while letter having more straight lines have usually higher accuracy using $k^{*}$-GED. For example letters I,L,T, and V have less accuracy whereas letters M,W,Y, and Z have higher accuracy. We also note that the accuracy of $k^{*}$-GED remains almost same with or without beam search, as the number of nodes in the letter graphs obtained after $k^{*}$-GED becomes so small such that the edit path considered by beam search becomes identical to that of $A^*$.  

% \begin{table}[!t]
% \renewcommand{\arraystretch}{1.0}
% \caption{Accuracy on letter dataset}
% \label{table }
% \begin{center}
% \begin{tabular}{|c| c| c| c| c| c| c| c| c|} 
% \hline
% % Class & $GED$ & $1^{*}GED$ & $2^{*}GED$ & $3^{*}GED$ & BG & B1*G & B2*G & B3*G  \\ [1ex] \hline\hline 
%   & G  & 1G & 2G & 3G & BG & B1G & B2G & B3G  \\ [1ex] \hline\hline 
% A & 94 & 70 & 64 & 64 & 90 & 70  & 64  & 64 \\ \hline
% E & 80 & 54 & 54 & 54 & 86 & 54  & 54  & 54 \\ \hline
% F & 80 & 64 & 44 & 42 & 84 & 64  & 44  & 42 \\ \hline
% H & 70 & 42 & 30 & 30 & 60 & 42  & 30  & 30 \\ \hline
% I & 94 & 06 & 06 & 06 & 94 & 06  & 06  & 06 \\ \hline
% K & 86 & 44 & 38 & 38 & 80 & 44  & 38  & 38 \\ \hline
% L & 86 & 56 & 36 & 36 & 86 & 56  & 36  & 36 \\ \hline
% M & 96 & 90 & 60 & 60 & 94 & 90  & 60  & 60 \\ \hline
% N & 86 & 64 & 52 & 52 & 84 & 64  & 52  & 52 \\ \hline
% T & 90 & 50 & 46 & 44 & 88 & 50  & 46  & 44 \\ \hline
% V & 90 & 48 & 46 & 46 & 88 & 48  & 46  & 46 \\ \hline
% W & 94 & 78 & 56 & 54 & 92 & 78  & 56  & 54 \\ \hline
% X & 78 & 52 & 50 & 50 & 76 & 52  & 50  & 50 \\ \hline
% Y & 92 & 76 & 52 & 52 & 92 & 76  & 52  & 52 \\ \hline
% Z & 90 & 80 & 64 & 64 & 90 & 80  & 64  & 64 \\ [1ex] \hline
% 
% \end{tabular} 
% \end{center}
% \end{table} 

\begin{table}[!t]
\renewcommand{\arraystretch}{1.0}
\caption{Accuracy on letter dataset using GED and $k^{*}$-GED} \label{table:accuracy-letter-ged-kged}
\label{table }
\begin{center}
\begin{tabular}{|c| c| c| c| c|} 
\hline
 Class & GED & $1^{*}$-GED & $2^{*}$-GED & $3^{*}$-GED \\ [1ex] \hline\hline 
A & 94 & 70 & 64 & 64 \\ \hline
E & 80 & 54 & 54 & 54 \\ \hline
F & 80 & 64 & 44 & 42 \\ \hline
H & 70 & 42 & 30 & 30 \\ \hline
I & 94 & 06 & 06 & 06 \\ \hline
K & 86 & 44 & 38 & 38 \\ \hline
L & 86 & 56 & 36 & 36 \\ \hline
M & 96 & 90 & 60 & 60 \\ \hline
N & 86 & 64 & 52 & 52 \\ \hline
T & 90 & 50 & 46 & 44 \\ \hline
V & 90 & 48 & 46 & 46 \\ \hline
W & 94 & 78 & 56 & 54 \\ \hline
X & 78 & 52 & 50 & 50 \\ \hline
Y & 92 & 76 & 52 & 52 \\ \hline
Z & 90 & 80 & 64 & 64 \\ [1ex] \hline

\end{tabular} 
\end{center}
\end{table}

\begin{table}[!t]
\renewcommand{\arraystretch}{1.0}
\caption{Accuracy on letter dataset using GED and $k^{*}$-GED with beam} \label{table:accuracy-letter-ged-kged-beam}
\label{table }
\begin{center}
\begin{tabular}{|c| c| c| c| c|} 
\hline
Class & GED & $1^{*}$-GED & $2^{*}$-GED & $3^{*}$-GED  \\ [1ex] \hline\hline 
A & 90 & 70  & 64  & 64 \\ \hline
E & 86 & 54  & 54  & 54 \\ \hline
F & 84 & 64  & 44  & 42 \\ \hline
H & 60 & 42  & 30  & 30 \\ \hline
I & 94 & 06  & 06  & 06 \\ \hline
K & 80 & 44  & 38  & 38 \\ \hline
L & 86 & 56  & 36  & 36 \\ \hline
M & 94 & 90  & 60  & 60 \\ \hline
N & 84 & 64  & 52  & 52 \\ \hline
T & 88 & 50  & 46  & 44 \\ \hline
V & 88 & 48  & 46  & 46 \\ \hline
W & 92 & 78  & 56  & 54 \\ \hline
X & 76 & 52  & 50  & 50 \\ \hline
Y & 92 & 76  & 52  & 52 \\ \hline
Z & 90 & 80  & 64  & 64 \\ [1ex] \hline

\end{tabular} 
\end{center}
\end{table}

To evaluate accuracy on AIDS dataset, We use 1500 graphs of test dataset including 300 graphs from the active class and 1200 graphs from the inactive class of AIDS molecules. We also use 250 graphs of training dataset consisting of 50 graphs of active class and 200 graphs of inactive class. Comparison of classification accuracy of AIDS dataset for GED using beam search heuristic with $k^{*}$-GED  for $k=1$ to $4$ using beam search heuristic having beam width $w=10$ is shown in Table \ref{table:accuracy-aids-ged-kged}. Also, the comparison of classification accuracy of AIDS dataset using bipartite GED with bipartite $k^{*}$-GED  for $k=1$ to $4$ is shown in Table \ref{table:aids-bipartite-ged-kged}. We can observe that classification accuracy for AIDS dataset is more than that of letter dataset. This increased accuracy shows that this graph matching model is more relevant for large graphs, as the contraction of nodes in small graphs tends to modify the graph more as compared to graphs with a large number of nodes.  

\begin{table}[!t]
\renewcommand{\arraystretch}{1.0}
\caption{Accuracy on AIDS dataset using GED and $k^{*}$-GED with beam} \label{table:accuracy-aids-ged-kged}
\label{table }
\begin{center}
\begin{tabular}{|c| c| c|} 
\hline
Class & Active & Inactive  \\ [1ex] \hline\hline 
GED with beam & 98.6 & 99.9 \\ \hline
$1^{*}$-GED with beam & 98.6 & 99.3 \\ \hline
$2^{*}$-GED with beam & 97.3 & 99.2 \\ \hline
$3^{*}$-GED with beam & 96.6 & 99.1 \\ \hline
$4^{*}$-GED with beam & 96.6 & 99.1 \\ [1ex] \hline

\end{tabular} 
\end{center}
\end{table}

\begin{table}[!t]
\renewcommand{\arraystretch}{1.0}
\caption{Accuracy on AIDS dataset using Bipartite GED and $k^{*}$-GED} \label{table:aids-bipartite-ged-kged}
\label{table }
\begin{center}
\begin{tabular}{|c| c| c|} 
\hline
Class & Active & Inactive  \\ [1ex] \hline\hline 
Bipartite GED & 98.6 & 99.9 \\ \hline
Bipartite $1^{*}$-GED  & 98.6 & 99.3 \\ \hline
Bipartite $2^{*}$-GED  & 97.3 & 99.2 \\ \hline
Bipartite $3^{*}$-GED  & 97.0 & 99.2 \\ \hline
Bipartite $4^{*}$-GED & 96.6 & 99.1 \\ [1ex] \hline

\end{tabular} 
\end{center}
\end{table} 

\subsubsection{Topological Considerations and Discussion}
Node contraction operation changes the topology of the input graphs. We can observe the effect of $k^{*}$-NC on the graph topology of the training set of letter A in Figure \ref{fig:topology-a}. For comparison purpose, we have included the effect of $k^{*}$-ND on the structure of the same graphs. We can also examine the effect of $k^{*}$-NC and $k^{*}$-ND on the topology of graphs of the training set of active molecules of AIDS dataset in the Figure \ref{fig:topology-aids}. Here, we can notice the abrupt change in the average number of nodes in $k^{*}$-ND from $k=1$ to $3$, as it deletes almost the entire structure of graphs.  Therefore $k^{*}$-NC can be used as a compromise between change in topology and efficiency because it changes the structure of the graphs up to a certain extent only.

\begin{figure}[!t]
\centering
\includegraphics[scale=0.15]{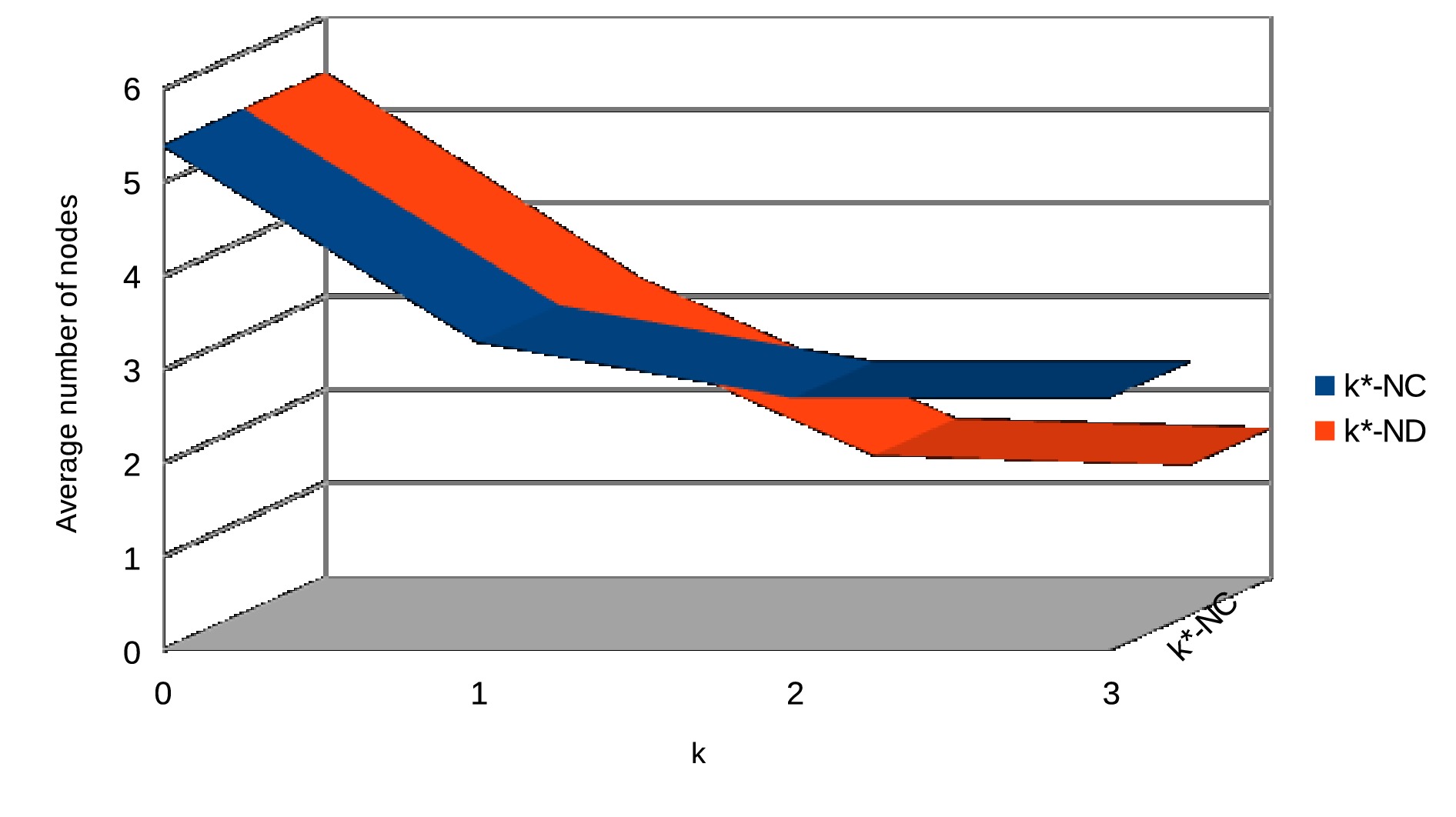}
\caption{Effect of $k^{*}$-NC and $k^{*}$-ND on the topology of Letter A} \label{fig:topology-a}
\end{figure}

\begin{figure}[!t]
\centering
\includegraphics[scale=0.15]{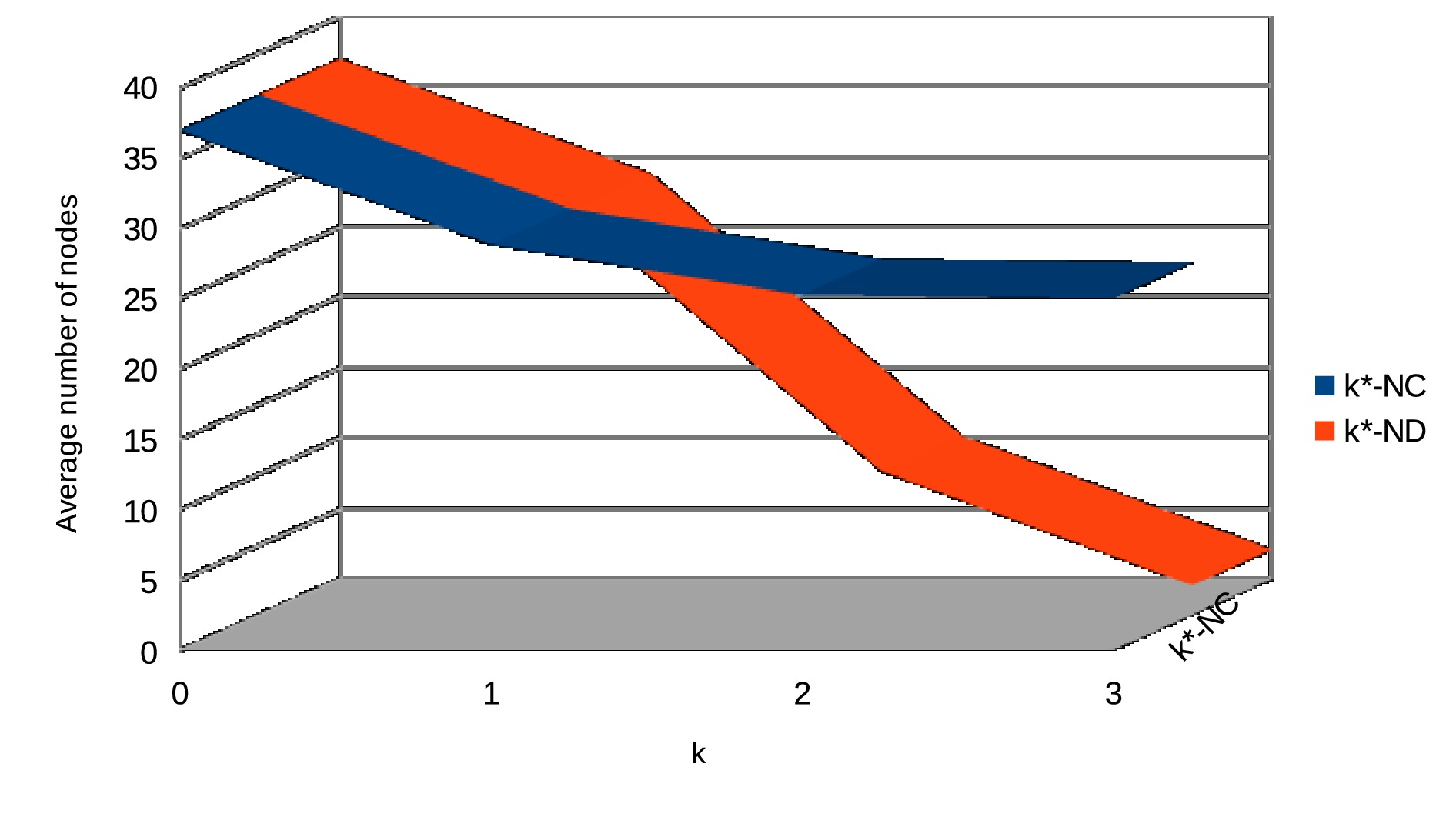}
\caption{Effect of $k^{*}$-NC and $k^{*}$-ND on the topology of active molecules of AIDS dataset} \label{fig:topology-aids}
\end{figure}

The proposed method can be useful in applications, where graph representation is relatively dense and have a large number of nodes. Depending on the requirement of time and exactness of similarity, we can proceed from $k=0$ to $k=n$. If exact matching is must regardless of time spent, we have to use $k=0$, which is nothing but standard graph edit distance. If the exact similarity is more important, then we can proceed from $k=1$, otherwise if time is more crucial, then we can proceed from $k=n$. Where $n$ is the minimum of the second highest degree of nodes between $G_1$ and $G_2$. It can also be used as an online or anytime graph matching scheme, where we need to find a matching under some time constraint. As mentioned above, $k^{*}$-ND-GED which allows the deletion of nodes and its associated edges, even if it is a cut vertex can provide a further efficiency but may be at the cost of much less accuracy.

\section{Summary}
%Conclusion 1
In this chapter, we described an approach to error-tolerant graph matching using the concept of graph homeomorphism. We presented Homeomorphic-Graph-Edit-Distance algorithm, which uses path contraction to decrease the number of vertices in the input graphs leading to the reduction in the computation of graph edit distance.   

%Conclusion 2
We also presented an approach to error-tolerant graph matching using node contraction. It works by removing the nodes based on their degree centrality as a measure of importance. We deleted the least degree nodes provided this does not increase the connected components. $k^{*}$-Node-Contraction can be used as a preprocessing step on the top of any graph matching algorithms. We implemented the proposed method on the letter and AIDS dataset, and the result shows efficiency in execution time at the cost of a decrease in accuracy. Results on AIDS dataset are more promising, which shows that the approach is more suitable for large graphs. We have also described the effect of node contraction on the topology of graphs. In general, the algorithm achieves efficiency without disturbing the topology of graphs too much.

% Chapter Template

\chapter{Graph Matching utilizing Node Centrality Information} % Main chapter title

\label{Chapter4} % Change X to a consecutive number; for referencing this chapter elsewhere, use \ref{ChapterX}

\lhead{Chapter 4. \emph{Graph Matching utilizing Node Centrality Information}} % Change X to a consecutive number; this is for the header on each page - perhaps a shortened title

%----------------------------------------------------------------------------------------
%	SECTION 1
%----------------------------------------------------------------------------------------

\section{Introduction}

The applications of exact graph matching to real-world applications is rather limited due to the presence of noise or error during the processing of the graphs. Error-tolerant graph matching offers an alternative to performing approximate graph matching. Due to exponential complexity associated with graph edit distance, other methods have been introduced to perform efficient graph matching at the cost of a slight decrease in accuracy. The node contraction technique described in chapter 3, is based on removing the nodes based on their degree centrality to decrease the size of the matching graphs. However, the degree centrality may not always be the best criteria to ignore the nodes. Depending on the structure and properties of the different dataset, we can select the appropriate centrality measure to delete the nodes for reducing the size of the graphs. In this chapter, we use eigenvector, betweenness and PageRank centrality in addition to degree centrality to reduce the size of the graphs for estimating an early approximate graph matching between two graphs \citep{DwivediSingh2020}.

Now we briefly explain the above centrality measures \citep{Newman2010}. The centrality of a node in the graph signifies its relative importance in the graph. The centrality measures aim to find the most important or central nodes of a graph or network. Simplest centrality measure is \textit{degree centrality}, which simply refers to the degree of the given node. A node with more adjacent nodes will have higher degree centrality as compared to nodes with a fewer connection. \textit{Betweenness} centrality of a node is based on the extent by which this node lies on the paths between other nodes. \textit{Eigenvector} centrality is a generalization of degree centrality, which assigns each node a value proportionate to the sum of the values of its neighbors. For a node $u_i$ its eigenvector centrality is given by $x_i=\kappa_1^{-1}\sum_j A_{ij} x_j$, where $\kappa_1$ is the largest eigenvalue of adjacency matrix $A$ and $A_{ij}$ is an element of $A$. In PageRank centrality, the centrality of a node is proportionate to the centrality of its neighbors divided by their outgoing degree. The \textit{PageRank} centrality is defined by $x_i = \alpha \sum_j A_{ij} \frac{x_j}{k_j} + \gamma$, where $\alpha$ is a free parameter, $k_j$ is the outgoing degree and $\gamma$ is a constant.

\begin{figure}[!t]
\centering
 \includegraphics[scale=.5]{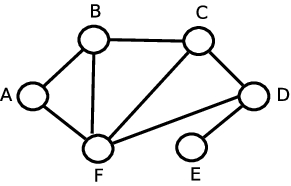}
 \caption{Graph $G$} \label{fig:graph-example}
\end{figure}

For example, consider the graph $G$ shown in Figure \ref{fig:graph-example}, consisting of 6 nodes. The degree centrality of nodes $A,B,C,D,E$ and $F$ are $2,3,3,3,1$ and $4$ respectively. The betweenness centrality of nodes $A$ and $E$ of graph $G$ is $0$ since these nodes do not lie on the shortest path between any two nodes of $G$. Since node $B$ occurs in one of two shortest path between $A$ and $C$, its centrality is $0.5$. Similarly, betweenness centrality of vertices $C,D$ and $F$ are $1,4$ and $3.5$ respectively. Eigenvector centrality of vertices $A,B,C,D,E$ and $F$ of graph $G$ obtained by multiplying its adjacency matrix with degree vector is $7,9,10,8,3$ and $11$ respectively. The PageRank centrality of nodes in the graph $G$, when we consider each node of $G$ to be a web page and each edge to be a bidirectional edge is $0.09, 0.18, 0.15, 0.26, 0.05$ and $0.25$ for vertices $A,B,C,D,E$ and $F$ respectively.

This chapter is organized as follows. Section 4.2, presents error-tolerant graph matching using centrality measures and introduces $r$-centrality graph edit distance. Section 4.3, describes experimental evaluation. Lastly, section 4.4 provides a summary.

\section{Error-Tolerant Graph Matching using Centrality Measures}
To reduce the computation time of error-tolerant graph matching, we ignore the nodes from the graphs with less centrality value before computing a similarity score using GED between two graphs.

\begin{defn}
 $r$-centrality node contraction is the process of contracting $r$ fractions of nodes from a graph $G$ with least centrality values of a given centrality measure.
\end{defn}

In the above definition, $r$ stands for ratio, which is equal to the decimal equivalent of the percentage of nodes from a graph $G$. If $r=0.1$, then 10\% of nodes will be contracted from $G$. 

The above definition implies that starting from the node with the lowest centrality value in a graph $G$, up to $r.|G|$ nodes are deleted provided they are not a cut vertex. Where $|G|$ is the number of nodes in the graph $G$. Depending on the centrality measure used, $r$-centrality node contraction ($r$-NC) can be $r$-degree centrality node contraction ($r$DC-NC), $r$-betweenness centrality node contraction ($r$-BC-NC), $r$-eigenvector node contraction ($r$-EV-NC) and $r$-PageRank node contraction ($r$-PR-NC).   

\begin{defn}
 $r$-degree centrality node contraction is the operation of contracting $r.|G|$ nodes of the smallest degree from graph $G$.
\end{defn}

When $r.|G|$ is equal to the number of nodes of degree $k$ in a graph, then $r$-degree centrality node contraction corresponds to $k$-degree node contraction.

\begin{defn}
 $r$-betweenness centrality node contraction is the operation of contracting $r.|G|$ nodes with the lowest betweenness score from graph $G$.
\end{defn}

\begin{defn}
 $r$-eigenvector centrality node contraction is the process of contracting $r.|G|$ nodes having the lowest eigenvector centrality from graph $G$.
\end{defn}

\begin{defn}
 $r$-PageRank centrality node contraction is the process of contracting $r.|G|$ nodes with the lowest PageRank score from graph $G$.
\end{defn}

\begin{defn}
 $r$-centrality GED computation between two graphs $G_1$ and $G_2$ is defined as GED between these graphs, when $r.|G_1|$ nodes of  $G_1$ and $r.|G_2|$ of $G_2$ of least centrality value have been contracted.
\end{defn}

In the above definition depending on the actual centrality criteria used $r$-centrality GED computation ($r$-GED) corresponds to $r$-degree centrality GED computation ($r$-DC-GED), $r$-betweenness centrality GED computation ($r$-BC-GED), $r$-eigenvector GED computation ($r$-EV-GED) and $r$- PageRank GED computation ($r$-PR-GED). When $r=0$, $r$-GED corresponds to standard GED computation.

\subsection{Edit Cost}
We can define the edit cost of $r$-GED by using an additional operation $c(u \rightarrow \epsilon)= 0$, for $r.|G|$ vertices of the graph $G$ having the lowest score of the given centrality measure.   

$r$-GED utilizes the Euclidean distance and allocates the constant cost to insertion, deletion and substitution of vertices and links. For two graphs $G_1$ and $G_2$, having vertices $u \in V_1$, $v \in V_2$ and links $e \in E_1$, $f \in E_2$, we specify the extended edit cost function as given below. \\
$c(u \rightarrow \epsilon)= x_{node}$ \\
$c(\epsilon \rightarrow v)= x_{node}$ \\
$c(u \rightarrow v)= y_{node}.|| \mu_1(u) - \mu_2(v)||$ \\
$c(e \rightarrow \epsilon)= x_{edge}$ \\
$c(\epsilon \rightarrow f)= x_{edge}$ \\
$c(e \rightarrow f)= y_{edge}. || \nu_1(e) - \nu_2(f)||$ \\
$c(u \rightarrow \epsilon)= 0$, if $u$ is one of the $r.|G|$ nodes of the lowest centrality value and is not a cut vertex. \\
Here $x_{node}$, $y_{node}$, $x_{edge}$, $y_{edge}$ are positive constants.

\subsection{Algorithm}

The computation of error-tolerant graph matching using $r$-centrality node contraction is outlined in Algorithm \ref{algorithm:r-centrality-ged}. The input to the $r$-Centrality-Graph-Edit-Distance algorithm is two graphs $G_1=(V_1, E_1, \mu_1,\nu_1)$, $G_2=(V_2, E_2, \mu_2,\nu_2)$ and a parameter $r$. The output to the algorithm is the minimum cost $r$-GED between $G_1$ and $G_2$. It calls Algorithm \ref{algorithm:r-centrality-node-contraction} in line 1 to remove $\lceil r.n \rceil$ nodes in $G_1$ having the lowest centrality value provided they are not cut vertex. Here, $\lceil  \rceil$ is a \textit{ceiling} function, which returns the least integer greater than equal to $rn$. Similarly line 2 to remove $\lceil r.m \rceil$ in $G_2$ having the lowest centrality value. $G_1'$ and $G_2'$ are the resultant graphs obtained after performing $r$-Centrality-Node-Contraction on $G_1$ and $G_2$ respectively, such that $V_1' = \{u_1',...,u_{n'}'\}$ and $V_2' = \{v_1',...,v_{m'}'\}$. Line 3 initializes an empty set $A$. The vertex $u_1'$ of $G_1'$ is substituted by each vertex $v_j'$ of $G_2'$ in the \textit{for} loop of lines 3--6, and deletion of $u_1'$ is performed in line 7. The computation of the minimum cost edit path is performed in the \textit{while} loop of lines 8--27. \textit{If} loop in line 10 checks, whether $C_{min}$ is a complete edit path so that it completely transform $G_1'$ to $G_2'$.  If all nodes $V_1'$ are processed (line 13), then remaining nodes of $V_2'$ are simply inserted in $C_{min}$ in \textit{for} loop of lines 14--16. Similarly, all unprocessed vertices of $V_1'$ are substituted by all vertices of $V_2'$ along with the deletion of vertices of $V_1'$ in the \textit{for} loop of lines 19--23, and $A$ is updated in line 24. 

\begin{algorithm}
\caption{\bf :  $r$-Centrality-Graph-Edit-Distance $(G_1,G_2)$} \label{algorithm:r-centrality-ged}
\begin{algorithmic}[1]
\Require  Two Graphs $G_1$, $G_2$, %where $G_i = (V_i, E_i, \mu_i,\nu_i)$ for $i=1,2$ 
                       where $V_1 = \{u_1,...,u_n\}$ and $V_2 = \{v_1,...,v_m\}$ and a parameter $r$
\Ensure  A minimum cost $r$-GED between $G_1$ and $G_2$
  
   \State $G_1' \leftarrow r$-Centrality-Node-Contraction $(G_1, \lceil r.n \rceil)$
   \State $G_2' \leftarrow r$-Centrality-Node-Contraction $(G_2, \lceil r.m \rceil)$
   \State $A \leftarrow \emptyset$
   \For {each $(v_j' \in V_2')$}
%   {
   \State $A \leftarrow A \cup \{ u_1' \rightarrow v_j' \}$ 
%   }
   \EndFor
   \State $A \leftarrow A \cup \{ u_1' \rightarrow \epsilon \}$ 
   \While { (True) }
%   {
%   \State Prune $A$ using optimizing techniques 
   \State Compute minimum cost edit path $C_{min}$ from $A$ 
   \If {($C_{min}$ is a complete edit path)} 
    \State \textbf{return} $C_{min}$
   %\Return $C_{min}$
   \Else
       \If {(all vertices $(u_i' \in V_1')$ are visited)}
        \For {all unvisited $(v_j' \in V_2')$}
%         {
         \State $C_{min} \leftarrow C_{min} \cup \{ \epsilon \rightarrow v_j' \} $ 
%          }
        \EndFor
        \State $A \leftarrow A \cup \{ C_{min} \}$
        \Else
       \For {(all unvisited vertices $(u_i' \in V_1')$)}
%       {
         \For {(each $(v_j' \in V_2')$)}
%         {
         \State $C_{min} \leftarrow C_{min} \cup \{ u_i' \rightarrow v_j' \} \cup \{ u_i' \rightarrow \epsilon \}$ 
%          }
         \EndFor
%          }
       \EndFor
       \State $A \leftarrow A \cup \{ C_{min} \}$
   \EndIf
   \EndIf
%   }
   \EndWhile
%  \Procedure{$r$-\textbf{Centrality-Node-Contraction}}{$G, \lceil r.|G| \rceil$}
%   \For {$(i \leftarrow 1$ to $\lceil r.|G| \rceil)$ }
%   \State Select node $u$ with minimum centrality
%    \If {($u$  is not cut vertex)} 
%      \State $V \leftarrow V \setminus \{u\}$ 
%      \State $E \leftarrow E \setminus \{(u,v) | (u,v) \in E, \forall v \in G \}$ 
%     \EndIf
%   \EndFor 
%   \State \textbf{return} $G$
%  %\Return $G$ 
%  \EndProcedure
\end{algorithmic}
\end{algorithm}

Algorithm \ref{algorithm:r-centrality-node-contraction} describes the steps to perform $r$-centrality node contraction. The \textit{for} loop in lines 1--7 iterates up to $\lceil r.|G| \rceil)$ times to check for nodes in $G$ to be a cut vertex. In case the node is a cut vertex it deletes the node and its connected edges and updates the graph $G$. 

\begin{prop}
 $r$-Centrality-Graph-Edit-Distance algorithm performs error-tolerant graph matching of $G_1'$ and $G_2'$.
\end{prop}
Using the properties of the edit costs of $r$-GED, the Algorithm \ref{algorithm:r-centrality-ged} returns the minimum cost of complete edit path which transform input graph $G_1'$ to output graph $G_2'$, so that every vertex of $G_1'$ uniquely corresponds to a vertex of $G_2'$. Also the algorithm $r$-Centrality-Node-Contraction ensures that $V_1' \subset V_1$ and $V_2' \subset V_2$.

\begin{prop}
 $r$-Centrality-Node-Contraction algorithm executes in $O(n)$ time.
\end{prop}
We can check whether a node $u$ is a cut vertex in $O(n)$ time. Therefore the \textit{for} loop of the algorithm takes $O(r.|G|.n)$ time, that is $O(n)$.

The worst case computational complexity of the $r$-Centrality-Graph-Edit-Distance algorithm is exponential in the number of vertices in input graphs. We can use an appropriate variable $r$ to minimize the overall computation time.

\begin{algorithm}
\caption{\bf :  $r$-Centrality-Node-Contraction $(G, r)$} \label{algorithm:r-centrality-node-contraction}
\begin{algorithmic}[1]
\Require  A Graph $G$ and a parameter $r$
\Ensure  Transformed graph after applying $r$-Centrality-Node-Contraction on $G$
% \Procedure{$r$-\textbf{Centrality-Node-Contraction}}{$G, \lceil r.|G| \rceil$}
  \For {$(i \leftarrow 1$ to $\lceil r.|G| \rceil)$ }
  \State Select node $u$ with minimum centrality
   \If {($u$  is not cut vertex)} 
     \State $V \leftarrow V \setminus \{u\}$ 
     \State $E \leftarrow E \setminus \{(u,v) | (u,v) \in E, \forall v \in G \}$ 
    \EndIf
  \EndFor 
  \State \textbf{return} $G$
 %\Return $G$ 
% \EndProcedure
 
\end{algorithmic}
\end{algorithm}

\subsection{$t$-Centrality Graph Edit Distance}
%Introduction

\begin{defn}
 $t$-centrality node contraction is the process of contracting $t$ nodes from a graph $G$ with least centrality values of a given centrality measure.
\end{defn}

The above definition implies that starting from the node with the lowest centrality value in a graph $G$, up to $t$ nodes are deleted provided they are not a cut vertex. Depending on the centrality measure used $t$-centrality node contraction ($t$-NC) can be $t$-degree centrality node contraction ($t$DC-NC), $t$-betweenness centrality node contraction ($t$-BC-NC), $t$-eigenvector node contraction ($t$-EV-NC) and $t$-PageRank node contraction ($t$-PR-NC).   

\begin{defn}
 $t$-degree centrality node contraction is the operation of contracting $t$ nodes of the smallest degree from a graph $G$.
\end{defn}

When $t$ is equal to the number of nodes of degree $k$ in a graph, then $t$-degree centrality node contraction corresponds to $k$-degree node contraction.

\begin{defn}
 $t$-betweenness centrality node contraction is the operation of contracting $t$ nodes with the lowest betweenness score from a graph $G$.
\end{defn}

\begin{defn}
 $t$-eigenvector centrality node contraction is the process of contracting $t$ nodes with the lowest eigenvector centrality from a graph $G$.
\end{defn}

\begin{defn}
 $t$-PageRank centrality node contraction is the process of contracting $t$ nodes with the lowest PageRank score from a graph $G$.
\end{defn}

\begin{defn}
 $t$-centrality GED computation between two graphs $G_1$ and $G_2$ is defined as GED between these graphs, when $t$ nodes of least centrality of both graphs $G_1$ and $G_2$ have been contracted.
\end{defn}

In the above definition depending on the actual centrality criteria used $t$-centrality GED computation ($t$-GED) corresponds to $t$-degree centrality GED computation ($t$-DC-GED), $t$-betweenness centrality GED computation ($t$-BC-GED), $t$-eigenvector GED computation ($t$-EV-GED) and $t$- PageRank GED computation ($t$-PR-GED).

The computation of error-tolerant GM using $t$-centrality node contraction is outlined in Algorithm \ref{algorithm:t-centrality-ged}. The input to the $t$-Centrality-Graph-Edit-Distance algorithm is two graphs $G_1=(V_1, E_1, \mu_1,\nu_1)$, $G_2=(V_2, E_2, \mu_2,\nu_2)$ and a parameter $t$. The output to the algorithm is the minimum cost $t$-GED between $G_1$ and $G_2$. The algorithm calls the procedure $t$-Centrality-Node-Contraction in lines 1--2 for graphs $G_1$ and $G_2$ respectively to remove $t$ nodes having the lowest centrality value provided they are not cut vertex. $G_1'$ and $G_2'$ are the resultant graphs obtained after performing $t$-Centrality-Node-Contraction on $G_1$ and $G_2$ respectively, such that $V_1' = \{u_1',...,u_{n'}'\}$ and $V_2' = \{v_1',...,v_{m'}'\}$. Line 3 initializes an empty set $A$.	 The vertex $u_1'$ of $G_1'$ is substituted by each vertex $v_j'$ of $G_2'$ in the \textit{for} loop of lines 3--6, and deletion of $u_1'$ is performed in line 7. The computation of the minimum cost edit path is performed in the \textit{while} loop of lines 8--27. \textit{If} loop in line 10 check, whether $C_{min}$ is a complete edit path, so that it completely transform $G_1'$ to $G_2'$.  If all nodes $V_1'$ are processed (line 13), then remaining nodes of $V_2'$ are simply inserted in $C_{min}$ in \textit{for} loop of lines 14--16. Similarly, all unprocessed vertices of $V_1'$ is substituted by all vertices of $V_2'$ along with the deletion of vertices of $V_1'$ in the \textit{for} loop of lines 19--23, and $A$ is updated in line 24.

\begin{algorithm}
\caption{\bf :  $t$-Centrality-Graph-Edit-Distance $(G_1,G_2)$} \label{algorithm:t-centrality-ged}
\begin{algorithmic}[1]
\Require  Two Graphs $G_1$, $G_2$, %where $G_i = (V_i, E_i, \mu_i,\nu_i)$ for $i=1,2$ 
                       where $V_1 = \{u_1,...,u_n\}$ and $V_2 = \{v_1,...,v_m\}$ and a parameter $t$
\Ensure  A minimum cost $t$-GED between $G_1$ and $G_2$
  
   \State $G_1' \leftarrow t$-CENTRALITY-NODE-CONTRACTION $(G_1, t)$
   \State $G_2' \leftarrow t$-CENTRALITY-NODE-CONTRACTION $(G_2, t)$
   \State $A \leftarrow \emptyset$
   \For {each $(v_j' \in V_2')$}
%   {
   \State $A \leftarrow A \cup \{ u_1' \rightarrow v_j' \}$ 
%   }
   \EndFor
   \State $A \leftarrow A \cup \{ u_1' \rightarrow \epsilon \}$ 
   \While { (True) }
%   {
%   \State Prune $A$ using optimizing techniques 
   \State Compute minimum cost edit path $C_{min}$ from $A$ 
   \If {($C_{min}$ is a complete edit path)} 
    \State \textbf{return} $C_{min}$
   %\Return $C_{min}$
   \Else
       \If {(all vertices $(u_i' \in V_1')$ are visited)}
        \For {all unvisited $(v_j' \in V_2')$}
%         {
         \State $C_{min} \leftarrow C_{min} \cup \{ \epsilon \rightarrow v_j' \} $ 
%          }
        \EndFor
        \State $A \leftarrow A \cup \{ C_{min} \}$
        \Else
       \For {(all unvisited vertices $(u_i' \in V_1')$)}
%       {
         \For {(each $(v_j' \in V_2')$)}
%         {
         \State $C_{min} \leftarrow C_{min} \cup \{ u_i' \rightarrow v_j' \} \cup \{ u_i' \rightarrow \epsilon \}$ 
%          }
         \EndFor
%          }
       \EndFor
       \State $A \leftarrow A \cup \{ C_{min} \}$
   \EndIf
   \EndIf
%   }
   \EndWhile
    
 \Procedure{$t$-\textbf{Centrality-Node-Contraction}}{$G, t$}
  \For {$(i \leftarrow 1$ to $t)$ }
  \State Select node $u$ with minimum centrality
   \If {($u$  is not cut vertex)} 
     \State $V \leftarrow V \setminus \{u\}$ 
     \State $E \leftarrow E \setminus \{(u,v) | (u,v) \in E, \forall v \in G \}$ 
    \EndIf
  \EndFor 
  \State \textbf{return} $G$
 %\Return $G$ 
 \EndProcedure
 
\end{algorithmic}
\end{algorithm}

\section{Experimental Evaluation}
In this section, we apply $r$-Centrality-Graph-Edit-Distance algorithm for error-tolerant graph matching using the degree, betweenness, eigenvector and PageRank centrality. We use IAM graph database \citep{RiesenBunke2008} for the comparison of execution time and accuracy obtained by these centrality techniques. We use letter and AIDS dataset for the evaluation of the proposed error-tolerant graph matching scheme.

\subsection{Execution Time Comparison}
%All execution time in this section is computed using a system possessing 4 GB of memory and running at 2.20 GHz$\times$8 processor. 
For the comparison purpose, we have used three distinct values of $r$ in $r$-GED, which are $r=0.1, r=0.3$ and $r=0.5$. We have used these three values of $r$ to compute $0.1$-GED,$0.2$-GED and $0.3$-GED. Comparison of the average execution time of graph matching in milliseconds using $r$-Centrality-Graph-Edit-Distance algorithm as applied to letter A and E of high distortion letter dataset using different centrality measures is shown in Figure \ref{fig:time-letter-a} and Figure \ref{fig:time-letter-e} respectively. 

\begin{figure}[!t]
\centering
%  \vspace{2.5cm}
 \includegraphics[scale=.15]{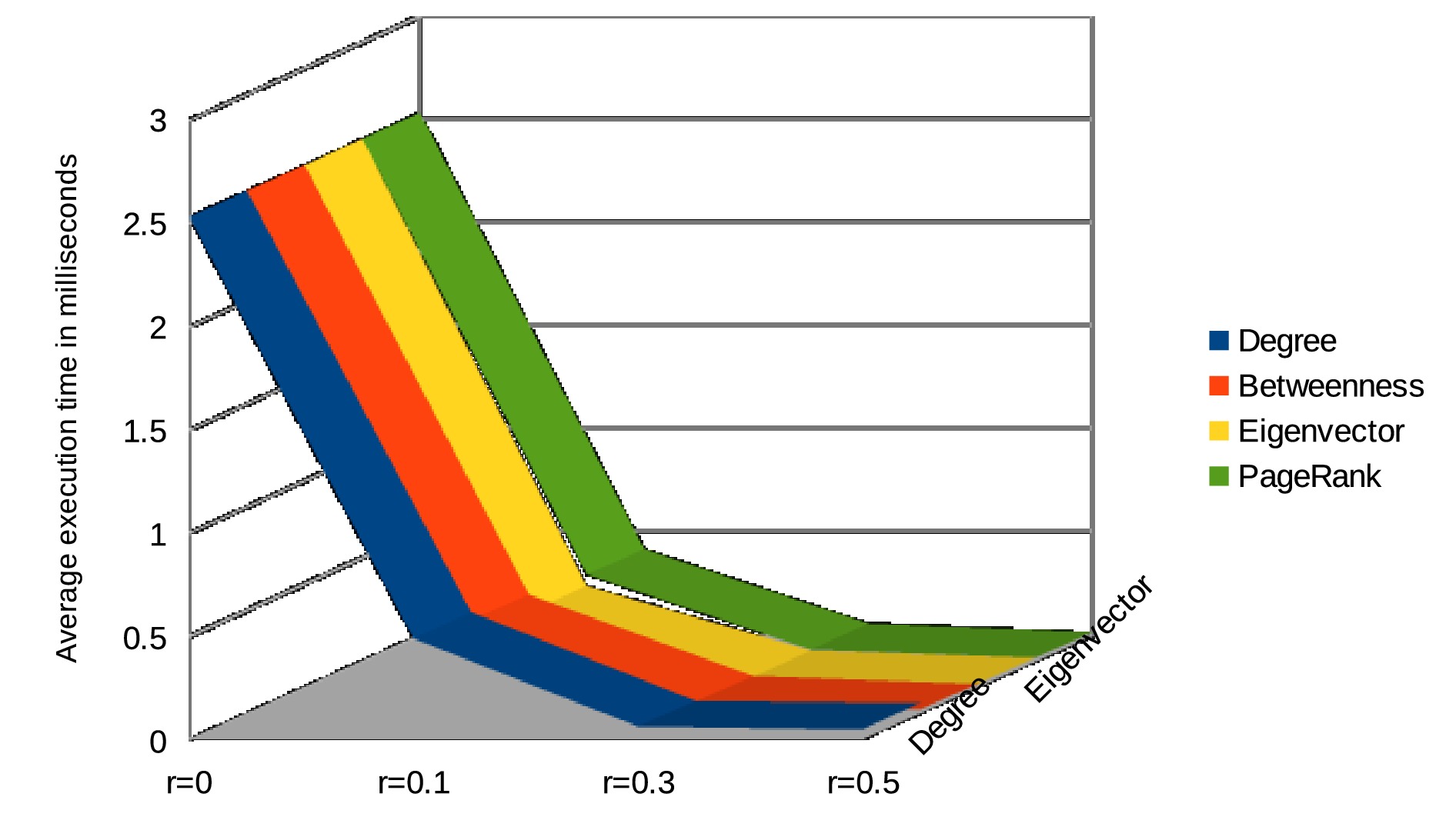}
 \caption{Comparison of execution time for letter A dataset}
 \label{fig:time-letter-a}
\end{figure}

We can observe that graph matching time using eigenvector criteria is least, whereas time using degree centrality is large. Computation time for letter E is higher as it contains more nodes than letter A. 
\begin{figure}[!t]
\centering
%  \vspace{2.5cm}
 \includegraphics[scale=.15]{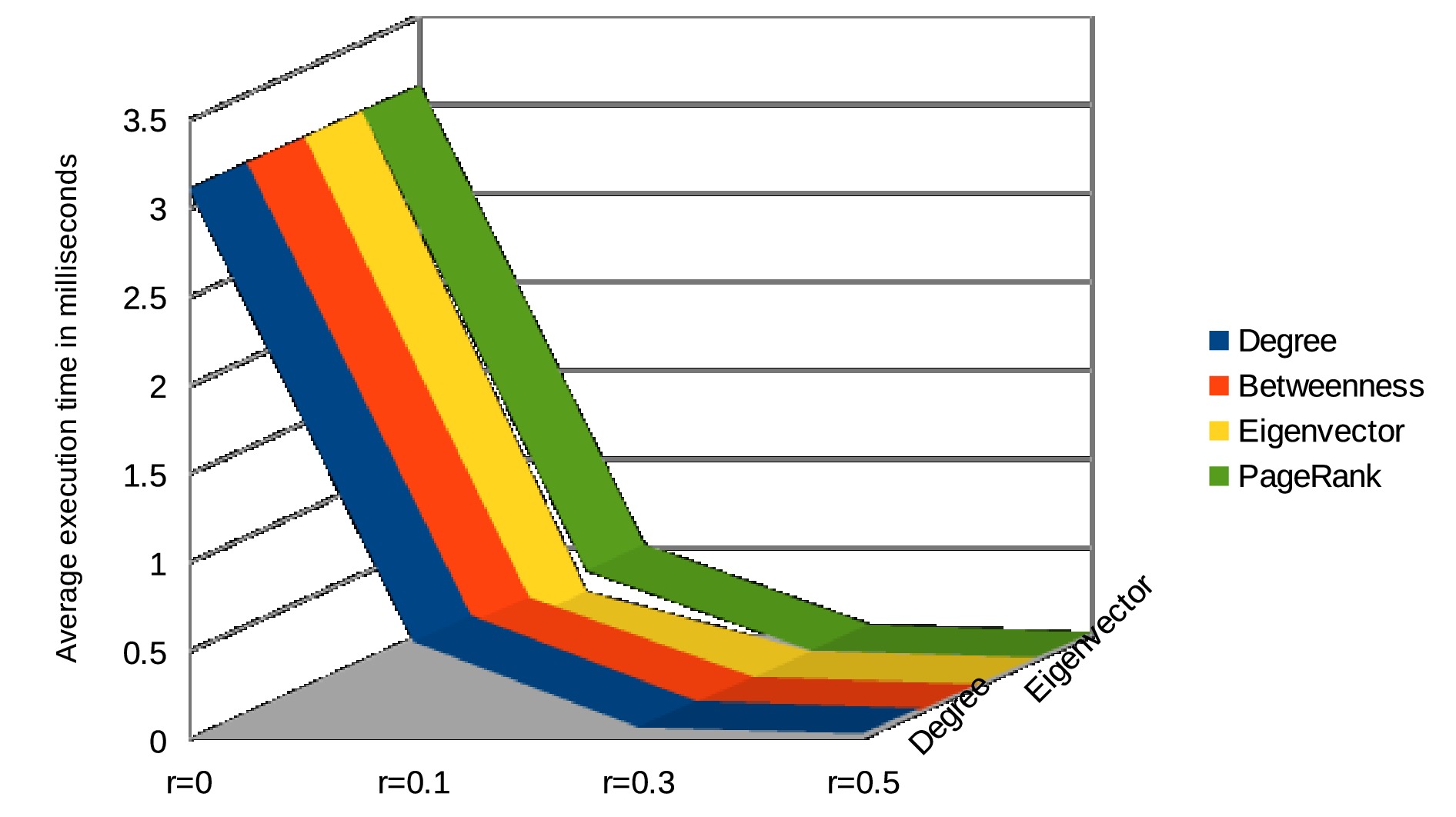}
 \caption{Comparison of execution time for letter E dataset}
 \label{fig:time-letter-e}
\end{figure}

Comparison of the average running time of graph matching in milliseconds using beam search heuristic (beam width $w=10$) for the four different centrality measures for the active class of AIDS dataset are shown in Figure \ref{fig:time-active}. From this figure, we observe that Algorithm \ref{algorithm:r-centrality-ged} usually takes less time using eigenvector and betweenness centrality as compared to the degree and PageRank centrality.

\begin{figure}[!t]
\centering
%  \vspace{2.5cm}
 \includegraphics[scale=.15]{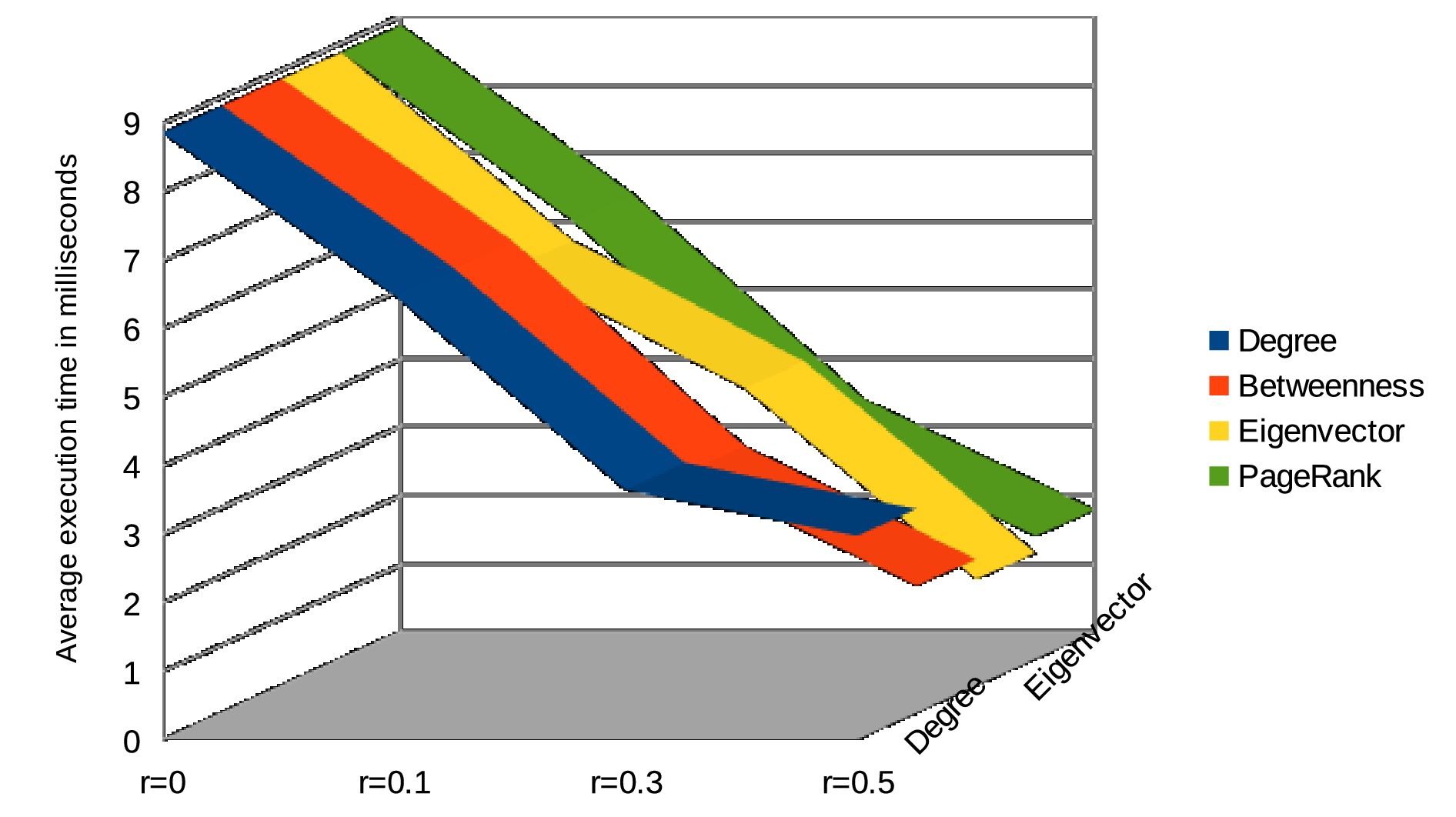}
 \caption{Comparison of execution time for active class of AIDS dataset}
 \label{fig:time-active}
\end{figure}

Figure \ref{fig:time-inactive} shows the corresponding average execution time of graphs for inactive AIDS dataset using the four centrality measures. Here again, the computation time using eigenvector takes less time than the degree, betweenness, and PageRank centrality measures. Running time using betweenness centrality is usually less than that of degree and PageRank centrality measures.  

\begin{figure}[!t]
\centering
%  \vspace{2.5cm}
 \includegraphics[scale=.15]{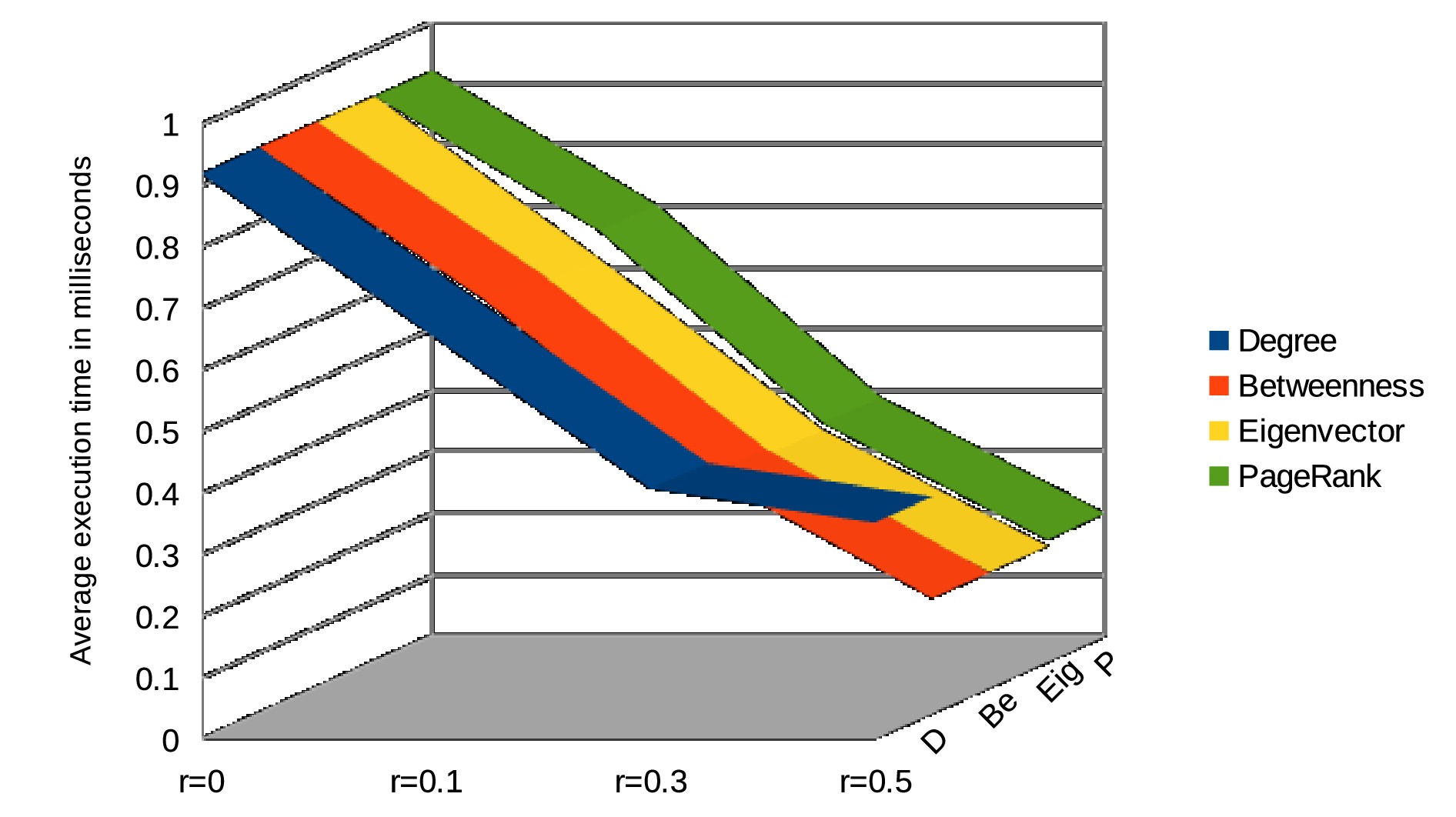}
 \caption{Comparison of execution time for inactive class of AIDS dataset}
 \label{fig:time-inactive}
\end{figure}

\subsection{Accuracy Comparison}
For accuracy assessment, we consider the problem of classification of graphs by the nearest neighbor classifier. Letter dataset of high distortion level consists of 750 graphs for each of training as well as test sets.  Each of these training, as well as test dataset, contains 50 graphs for every 15 letters. Classification accuracy of the proposed graph matching for letter A of high distortion using the four centrality indicators is given in Figure \ref{fig:accuracy-a}, while the accuracy of graph matching for letter E for the same measures is shown in Figure \ref{fig:accuracy-e}. 

\begin{figure}[!t]
\centering
%  \vspace{2.5cm}
 \includegraphics[scale=.15]{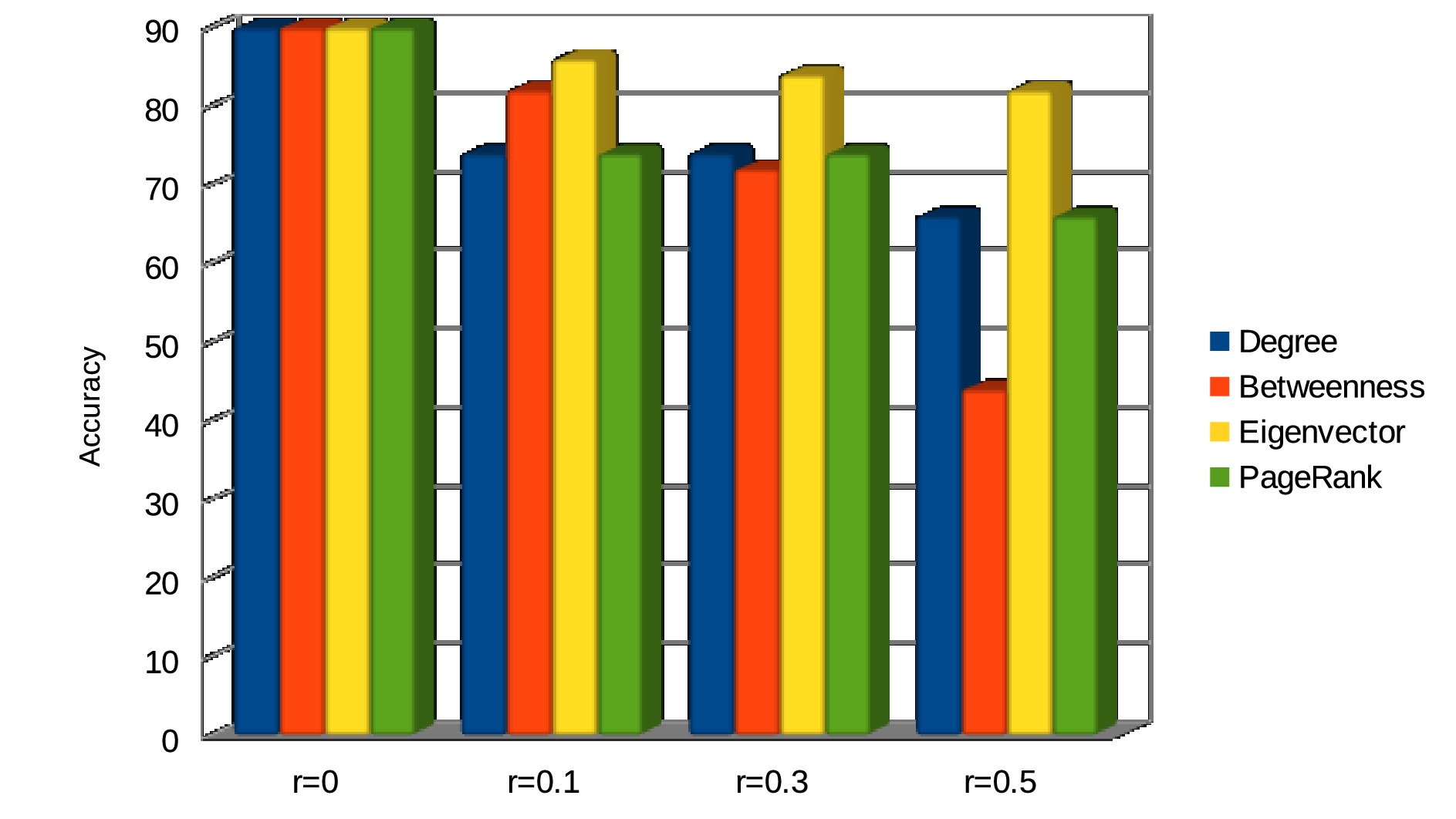}
 \caption{Comparison of accuracy ratio of letter A dataset}
 \label{fig:accuracy-a}
\end{figure}

% Here we note that accuracy of letter A for degree centrality is lower than the other three measures by contracting $t$ nodes, where t is equal to nodes with degree 1 in the input graphs ($1^*$-GED). We can also observe that for letter E, the accuracy ratio using betweenness and PageRank is usually higher than that of degree centrality even though they take less computation time.

Here we note that the accuracy of both letters A and E using eigenvector centrality is higher than that of the other three measures. It shows that eigenvector centrality better captures the relative importance of nodes for letter dataset. Table \ref{table:accuracy-letter-degree} provides the accuracy results of all fifteen letters using degree centrality for $r=0.1, 0.3$ and $0.5$. Similarly, Tables \ref{table:accuracy-letter-betw}, \ref{table:accuracy-letter-eigen} and \ref{table:accuracy-letter-pagerank} show accuracy results for all letters using betweenness, eigenvector and PageRank centrality respectively. 

\begin{figure}[!t]
\centering
%  \vspace{2.5cm}
 \includegraphics[scale=.15]{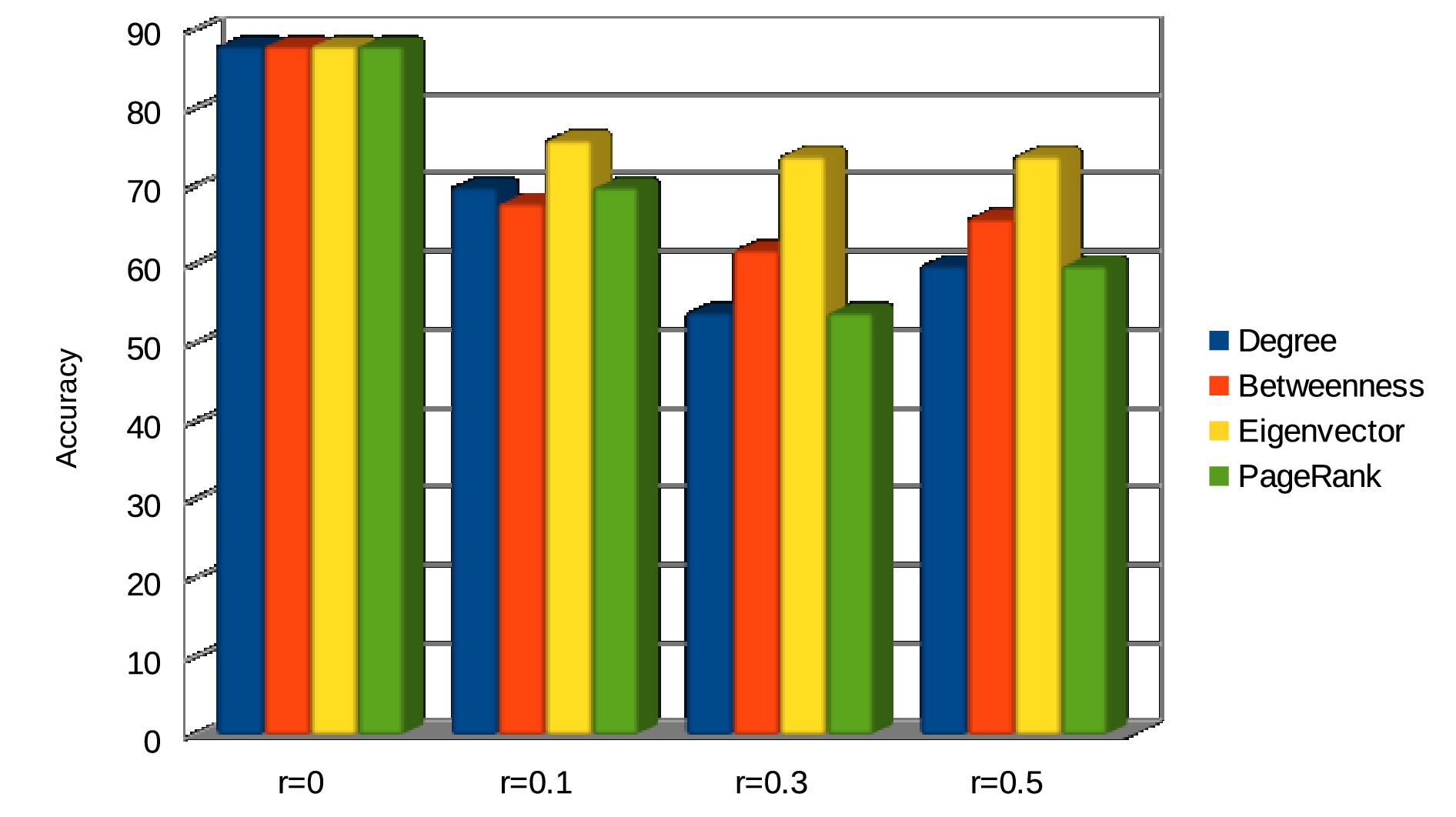} 
 \caption{Comparison of accuracy ratio of letter E dataset}
 \label{fig:accuracy-e}
\end{figure}

%Table 1
\begin{table}[!t]
\renewcommand{\arraystretch}{1.0}
\caption{Accuracy on letter dataset using degree centrality} \label{table:accuracy-letter-degree}
\label{table }
\begin{center}
\begin{tabular}{|c| c| c| c|} 
\hline
 Class & $r=0.1$ & $r=0.2$ & $r=0.3$ \\ [1ex] \hline\hline 
A & 74 & 74 & 66  \\ \hline
E & 70 & 54 & 60  \\ \hline
F & 72 & 44 & 42  \\ \hline
H & 58 & 42 & 26  \\ \hline
I & 90 & 86 & 84  \\ \hline
K & 68 & 58 & 44  \\ \hline
L & 78 & 64 & 54  \\ \hline
M & 82 & 50 & 66  \\ \hline
N & 58 & 50 & 50  \\ \hline
T & 70 & 46 & 52  \\ \hline
V & 88 & 54 & 60  \\ \hline
W & 88 & 74 & 60  \\ \hline
X & 76 & 52 & 40  \\ \hline
Y & 80 & 62 & 52  \\ \hline
Z & 78 & 62 & 64  \\ [1ex] \hline
\end{tabular} 
\end{center}
\end{table} 

%Table 2
\begin{table}[!t]
\renewcommand{\arraystretch}{1.0}
\caption{Accuracy on letter dataset using betweenness centrality} \label{table:accuracy-letter-betw}
\label{table }
\begin{center}
\begin{tabular}{|c| c| c| c|} 
\hline
 Class & $r=0.1$ & $r=0.2$ & $r=0.3$ \\ [1ex] \hline\hline 
A & 82 & 72 & 44  \\ \hline
E & 68 & 62 & 66  \\ \hline
F & 62 & 54 & 40  \\ \hline
H & 54 & 38 & 40  \\ \hline
I & 94 & 92 & 74  \\ \hline
K & 76 & 36 & 38  \\ \hline
L & 82 & 68 & 52  \\ \hline
M & 86 & 70 & 32  \\ \hline
N & 62 & 50 & 30  \\ \hline
T & 68 & 56 & 62  \\ \hline
V & 78 & 52 & 80  \\ \hline
W & 84 & 60 & 46  \\ \hline
X & 58 & 48 & 34  \\ \hline
Y & 82 & 76 & 74  \\ \hline
Z & 62 & 50 & 48  \\ [1ex] \hline
\end{tabular} 
\end{center}
\end{table} 

%Table 3
\begin{table}[!t]
\renewcommand{\arraystretch}{1.0}
\caption{Accuracy on letter dataset using eigenvector centrality} \label{table:accuracy-letter-eigen}
\label{table }
\begin{center}
\begin{tabular}{|c| c| c| c|} 
\hline
 Class & $r=0.1$ & $r=0.2$ & $r=0.3$ \\ [1ex] \hline\hline 
A & 86 & 84 & 82  \\ \hline
E & 99 & 96 & 92  \\ \hline
F & 58 & 42 & 22  \\ \hline
H & 40 & 32 & 42  \\ \hline
I & 96 & 92 & 58  \\ \hline
K & 68 & 38 & 42  \\ \hline
L & 76 & 60 & 56  \\ \hline
M & 84 & 54 & 30  \\ \hline
N & 64 & 38 & 30  \\ \hline
T & 54 & 34 & 36  \\ \hline
V & 72 & 36 & 66  \\ \hline
W & 90 & 62 & 34  \\ \hline
X & 62 & 34 & 34  \\ \hline
Y & 74 & 80 & 72  \\ \hline
Z & 62 & 56 & 36  \\ [1ex] \hline
\end{tabular} 
\end{center}
\end{table} 

%Table 4
\begin{table}[!t]
\renewcommand{\arraystretch}{1.0}
\caption{Accuracy on letter dataset using PageRank centrality} \label{table:accuracy-letter-pagerank}
\label{table }
\begin{center}
\begin{tabular}{|c| c| c| c|} 
\hline
 Class & $r=0.1$ & $r=0.2$ & $r=0.3$ \\ [1ex] \hline\hline 
A & 74 & 74 & 66  \\ \hline
E & 70 & 54 & 60  \\ \hline
F & 72 & 44 & 42  \\ \hline
H & 58 & 42 & 26  \\ \hline
I & 90 & 86 & 84  \\ \hline
K & 68 & 58 & 44  \\ \hline
L & 78 & 64 & 54  \\ \hline
M & 82 & 50 & 66  \\ \hline
N & 58 & 50 & 50  \\ \hline
T & 70 & 46 & 52  \\ \hline
V & 88 & 54 & 60  \\ \hline
W & 88 & 74 & 60  \\ \hline
X & 76 & 52 & 40  \\ \hline
Y & 80 & 62 & 52  \\ \hline
Z & 78 & 62 & 64  \\ [1ex] \hline
\end{tabular} 
\end{center}
\end{table}

To find the accuracy on AIDS dataset, we utilize test dataset consisting of 300 graphs from active class and 1200 graphs from inactive class, whereas training dataset consists of 50 graphs from active class and 200 graphs from the inactive class of AIDS dataset. We can observe the accuracy ratio of the proposed error-tolerant scheme using the four different centrality measure for the active class of AIDS dataset in Figure \ref{fig:active-accuracy}. In this figure, we observe that the accuracy ratio for the active class of AIDS dataset using betweenness centrality is usually higher than other centrality measures.

%In this figure, we observe that the accuracy obtained using degree and PageRank centrality are generally higher than that of eigenvector and betweenness centrality. Here we notice the time versus accuracy trade-off, the centrality criteria which takes less time leads to less accuracy, whereas the centrality techniques which are more accurate take more computation time.

\begin{figure}[!t]
\centering
%  \vspace{2.5cm}
 \includegraphics[scale=.15]{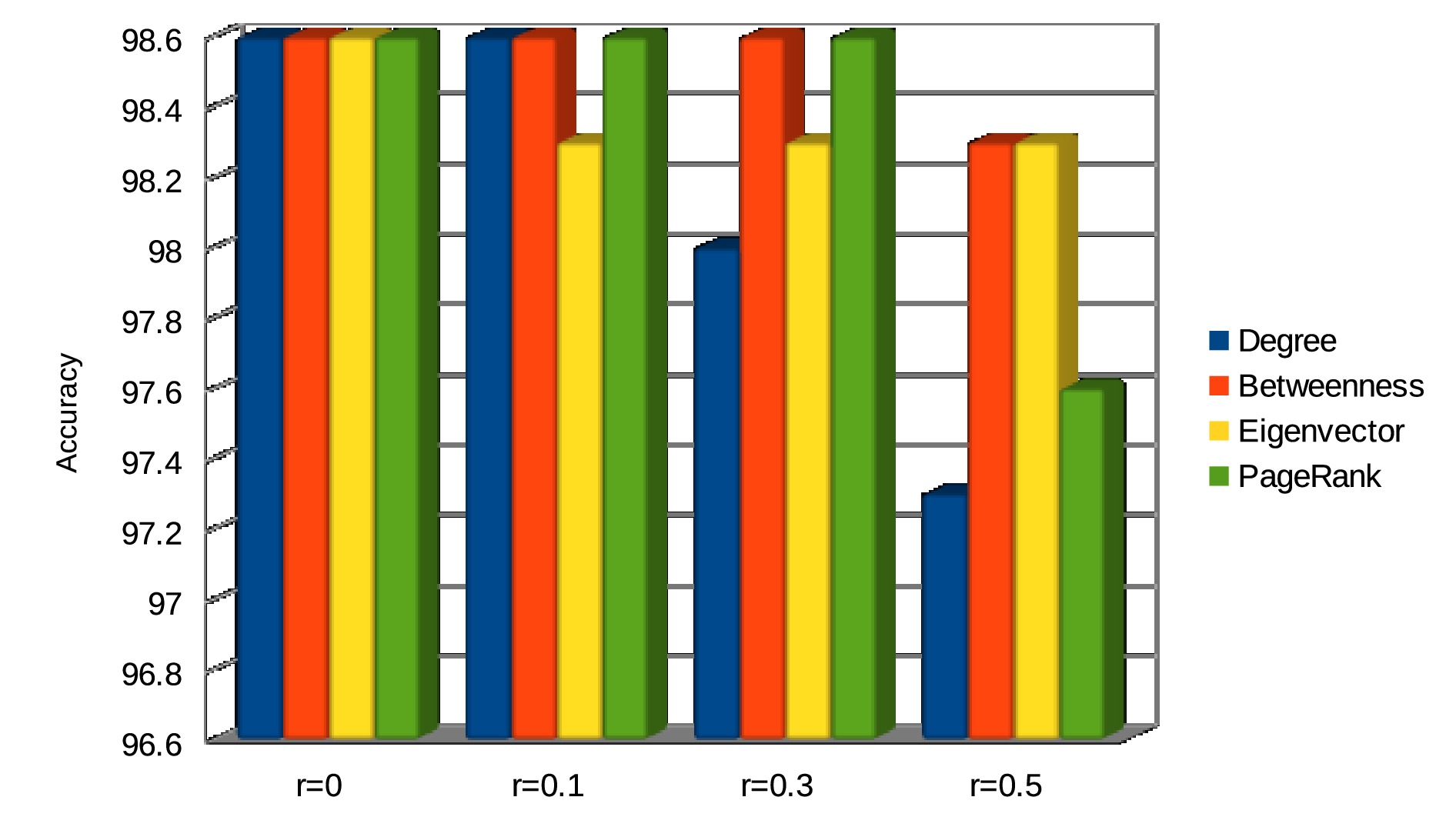}
 \caption{Comparison of accuracy for active class of AIDS dataset}
 \label{fig:active-accuracy}
\end{figure}

Figure \ref{fig:inactive-accuracy} shows the comparison of accuracy for the inactive class of AIDS dataset using the four centrality measures. In this figure, we observe that the accuracy ratio of inactive class AIDS dataset using eigenvector centrality is highest followed by betweenness centrality.

%In this figure also degree and PageRank criteria lead to higher accuracy for the classification of graphs of AIDS dataset.

\begin{figure}[!t]
\centering
%  \vspace{2.5cm}
 \includegraphics[scale=.15]{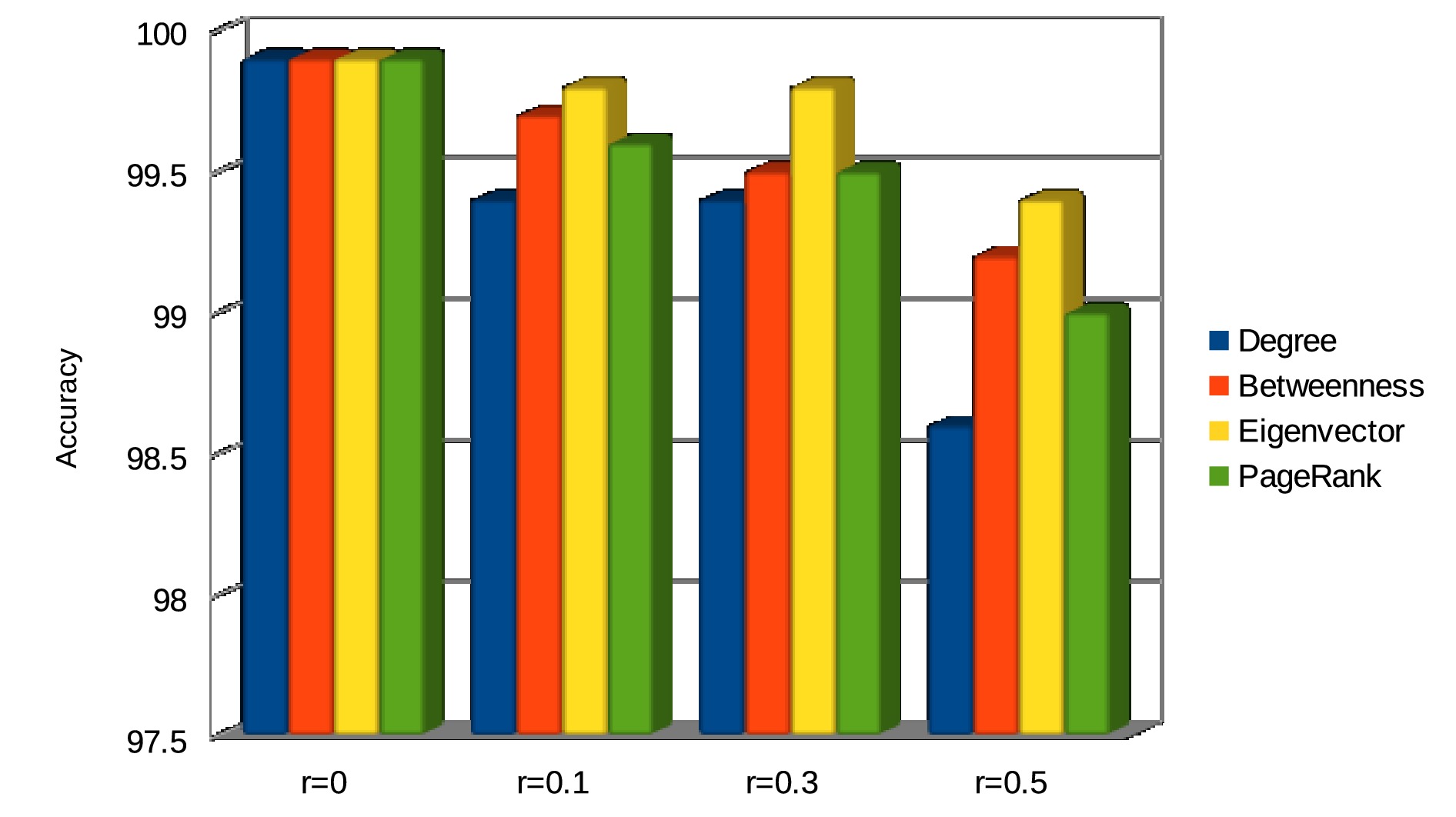}
 \caption{Comparison of accuracy for inactive class of AIDS dataset}
 \label{fig:inactive-accuracy}
\end{figure}

Tables \ref{table:aids-degree}--\ref{table:aids-pagerank} show the accuracy ratio of AIDS dataset using degree, betweenness, eigenvector and PageRank centrality for three different values of $r=0.1, 0.3$ and $0.5$.

%Table 5
\begin{table}[!t]
\renewcommand{\arraystretch}{1.0}
\caption{Accuracy on AIDS dataset using degree centrality} \label{table:aids-degree}
\label{table }
\begin{center}
\begin{tabular}{|c| c| c| c|} 
\hline
 Class & $r=0.1$ & $r=0.2$ & $r=0.3$ \\ [1ex] \hline\hline 
Active & 98.6 & 98 & 97.3  \\ \hline
Inactive & 99.4 & 99.4 & 98.6  \\ [1ex] \hline
\end{tabular} 
\end{center}
\end{table} 

%Table 6
\begin{table}[!t]
\renewcommand{\arraystretch}{1.0}
\caption{Accuracy on AIDS dataset using betweenness centrality}
\label{table }
\begin{center}
\begin{tabular}{|c| c| c| c|} 
\hline
 Class & $r=0.1$ & $r=0.2$ & $r=0.3$ \\ [1ex] \hline\hline 
Active & 98.6 & 98.6 & 98.3  \\ \hline
Inactive & 99.7 & 99.5 & 99.2  \\ [1ex] \hline
\end{tabular} 
\end{center}
\end{table} 

%Table 7
\begin{table}[!t]
\renewcommand{\arraystretch}{1.0}
\caption{Accuracy on AIDS dataset using eigenvector centrality}
\label{table }
\begin{center}
\begin{tabular}{|c| c| c| c|} 
\hline
 Class & $r=0.1$ & $r=0.2$ & $r=0.3$ \\ [1ex] \hline\hline 
Active & 98.3 & 98.3 & 98.3  \\ \hline
Inactive & 99.8 & 99.8 & 99.4  \\ [1ex] \hline
\end{tabular} 
\end{center}
\end{table} 

%Table 8
\begin{table}[!t]
\renewcommand{\arraystretch}{1.0}
\caption{Accuracy on AIDS dataset using PageRank centrality} \label{table:aids-pagerank}
\label{table }
\begin{center}
\begin{tabular}{|c| c| c| c|} 
\hline
 Class & $r=0.1$ & $r=0.2$ & $r=0.3$ \\ [1ex] \hline\hline 
Active & 98.6 & 98.6 & 97.6  \\ \hline
Inactive & 99.6 & 99.5 & 99  \\ [1ex] \hline
\end{tabular} 
\end{center}
\end{table}

\subsection{$t$-Centrality Graph Edit Distance}

For the comparison purpose, we have used the value of $t^*$ in $t^*$-GED to be equal to the number of nodes which would be considered for contraction in $k^*$-degree node contraction. Therefore the value of $1^*$ in $1^*$-GED is the number of nodes of degree 1, value of $2^*$ in $2^*$-GED is the number of nodes of degree 1 followed by degree 2, similarly the value of $3^*$ in $3^*$-GED is the number of nodes of degree 1 followed by degree 2 and degree 3. Comparison of the average execution time of GM in milliseconds using $t$-Centrality-Graph-Edit-Distance algorithm as applied to letter A and E of high distortion letter dataset using different centrality measures in shown in Figure \ref{fig:letter-a-time2} and Figure \ref{fig:letter-e-time2} respectively.

\begin{figure}[!t]
\centering
%  \vspace{2.5cm}
 \includegraphics[scale=.15]{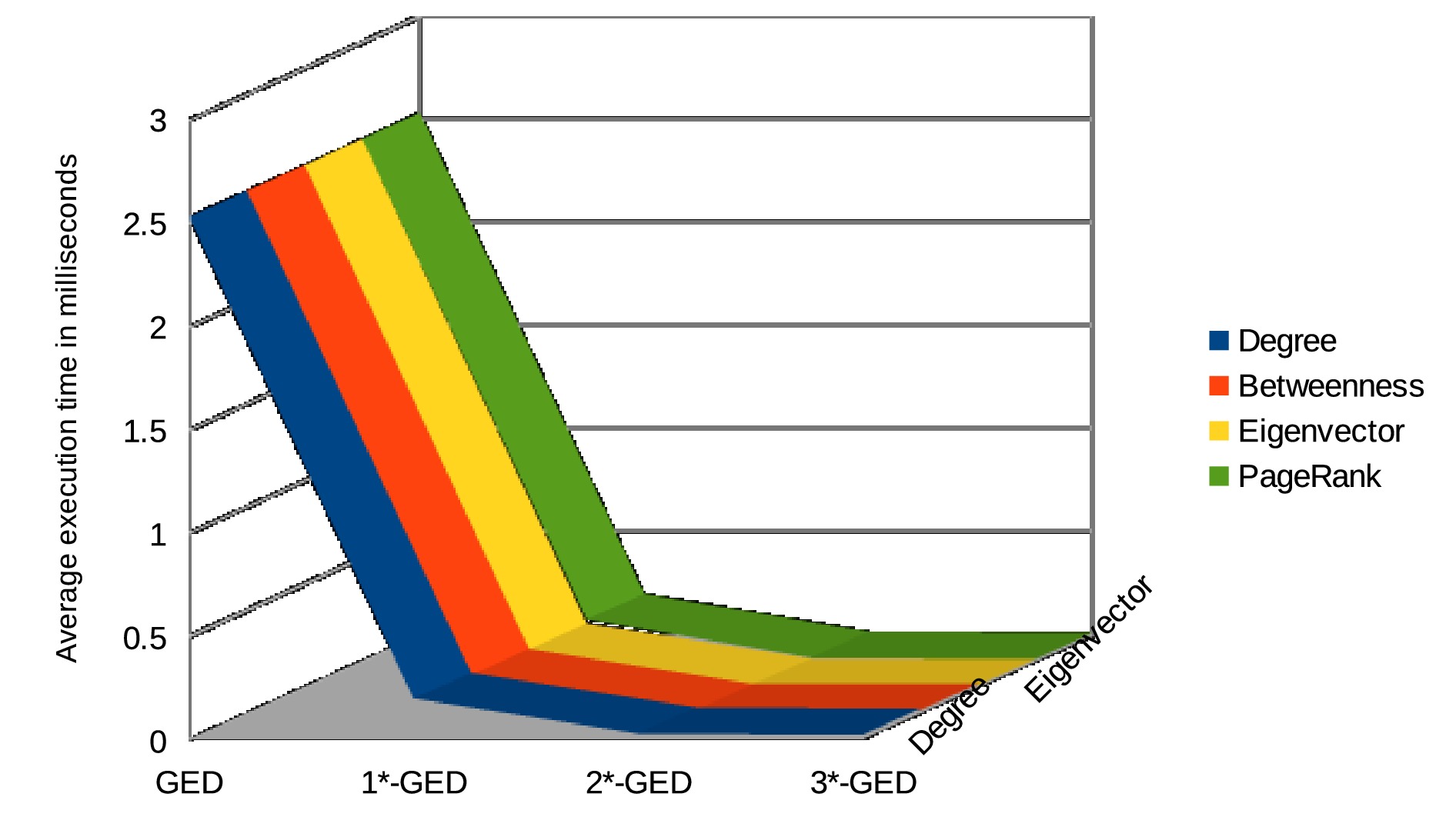}
 \caption{Comparison of execution time for letter A dataset}
 \label{fig:letter-a-time2}
\end{figure}

We can observe that GM time using eigenvector criteria is least, whereas time using degree centrality is higher. Computation time for letter E is higher as it contains more nodes than letter A. 

\begin{figure}[!t]
\centering
%  \vspace{2.5cm}
 \includegraphics[scale=.15]{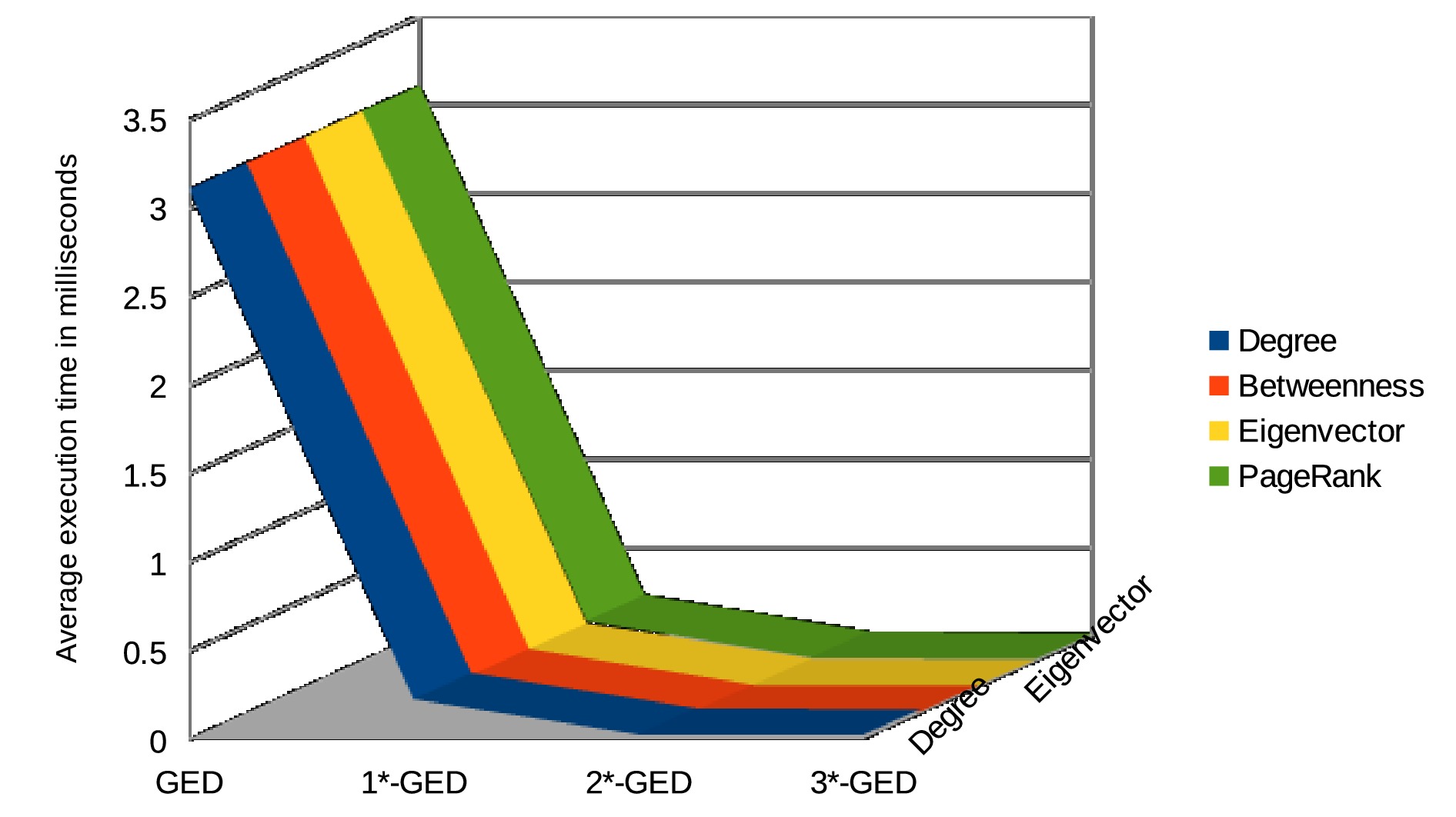}
 \caption{Comparison of execution time for letter E dataset}
 \label{fig:letter-e-time2}
\end{figure}

Comparison of the average running time of GM in milliseconds using beam search heuristic (beam width $w=10$) for the four different centrality measures for the active class of AIDS dataset are shown in Figure \ref{fig:aids-active-time2}. From this figure, we observe that Algorithm 1 usually takes less time using eigenvector and betweenness centrality as compared to the degree and PageRank centrality.

\begin{figure}[!t]
\centering
%  \vspace{2.5cm}
 \includegraphics[scale=.6]{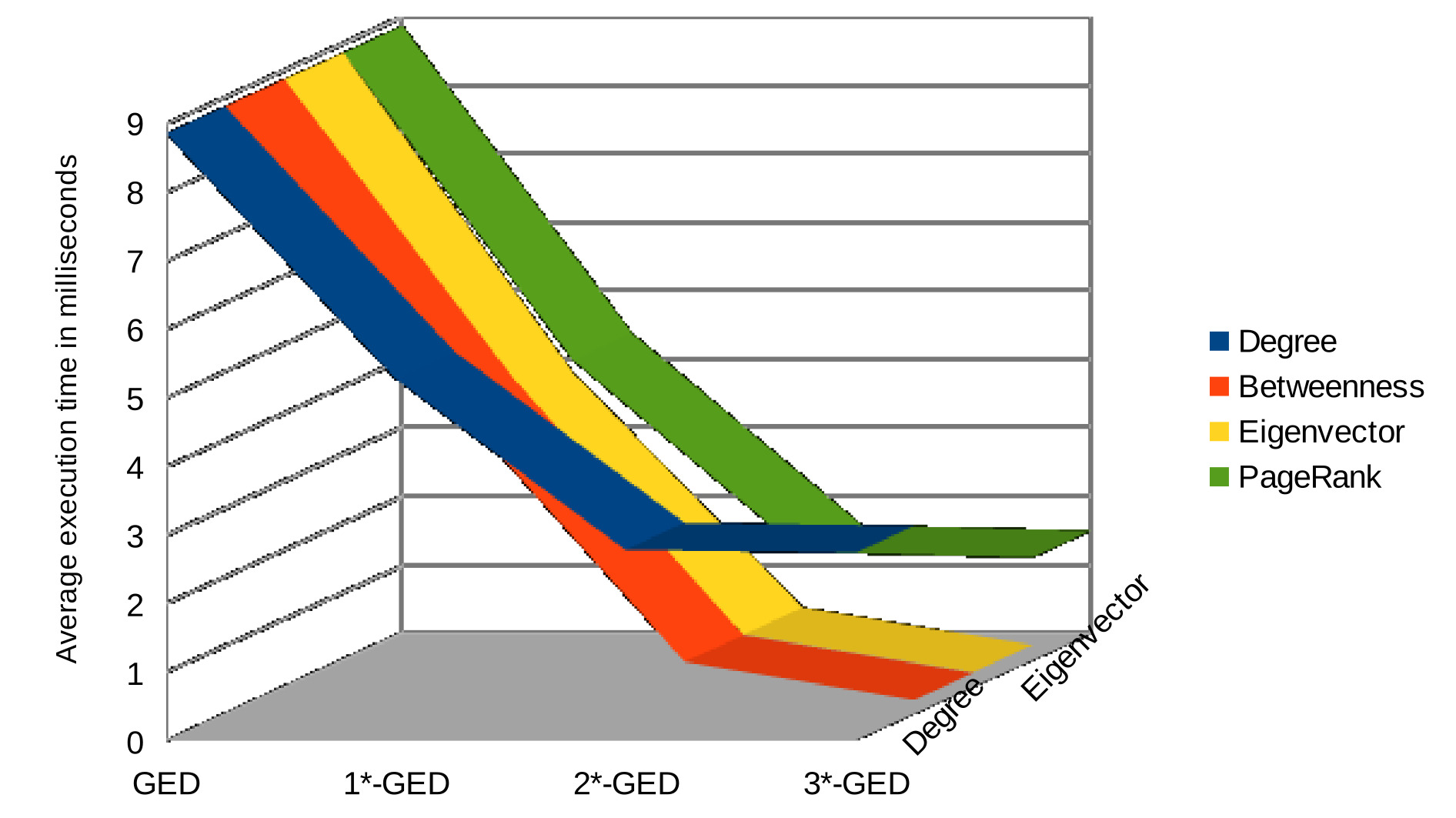}
 \caption{Comparison of execution time for active class of AIDS dataset}
 \label{fig:aids-active-time2}
\end{figure}

Figure \ref{fig:aids-inactive-time2} shows the corresponding average execution time of graphs for inactive AIDS dataset using the four centrality measures. Here again, the computation time using eigenvector and betweenness criteria take less time than the degree and PageRank, and between these two the average time using PageRank is less than degree centrality. 

\begin{figure}[!t]
\centering
%  \vspace{2.5cm}
 \includegraphics[scale=.15]{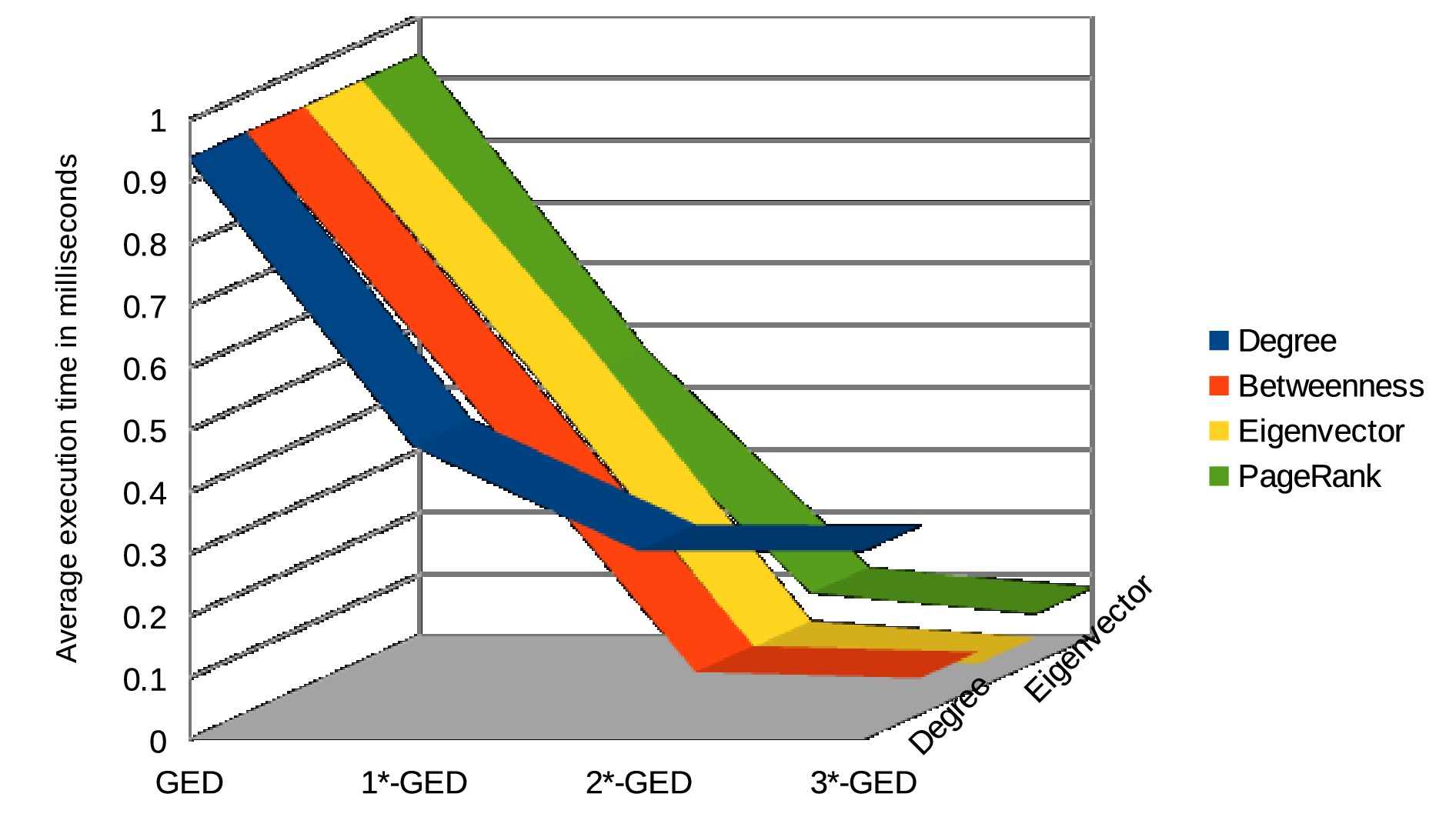}
 \caption{Comparison of execution time for inactive class of AIDS dataset}
 \label{fig:aids-inactive-time2}
\end{figure}

For accuracy assessment, we consider the problem of classification of graphs by the nearest neighbor classifier. Letter dataset of high distortion level consists of 750 graphs for both training as well as test sets.  Each of these training, as well as test dataset, contains 50 graphs for every 15 letters. Classification accuracy of proposed GM for letter A of high distortion using the four centrality indicators is given in Figure \ref{fig:letter-a-accuracy2}, while the accuracy of GM for letter E for the same measures are shown in Figure \ref{fig:letter-e-accuracy2}. 

\begin{figure}[!t]
\centering
%  \vspace{2.5cm}
 \includegraphics[scale=.15]{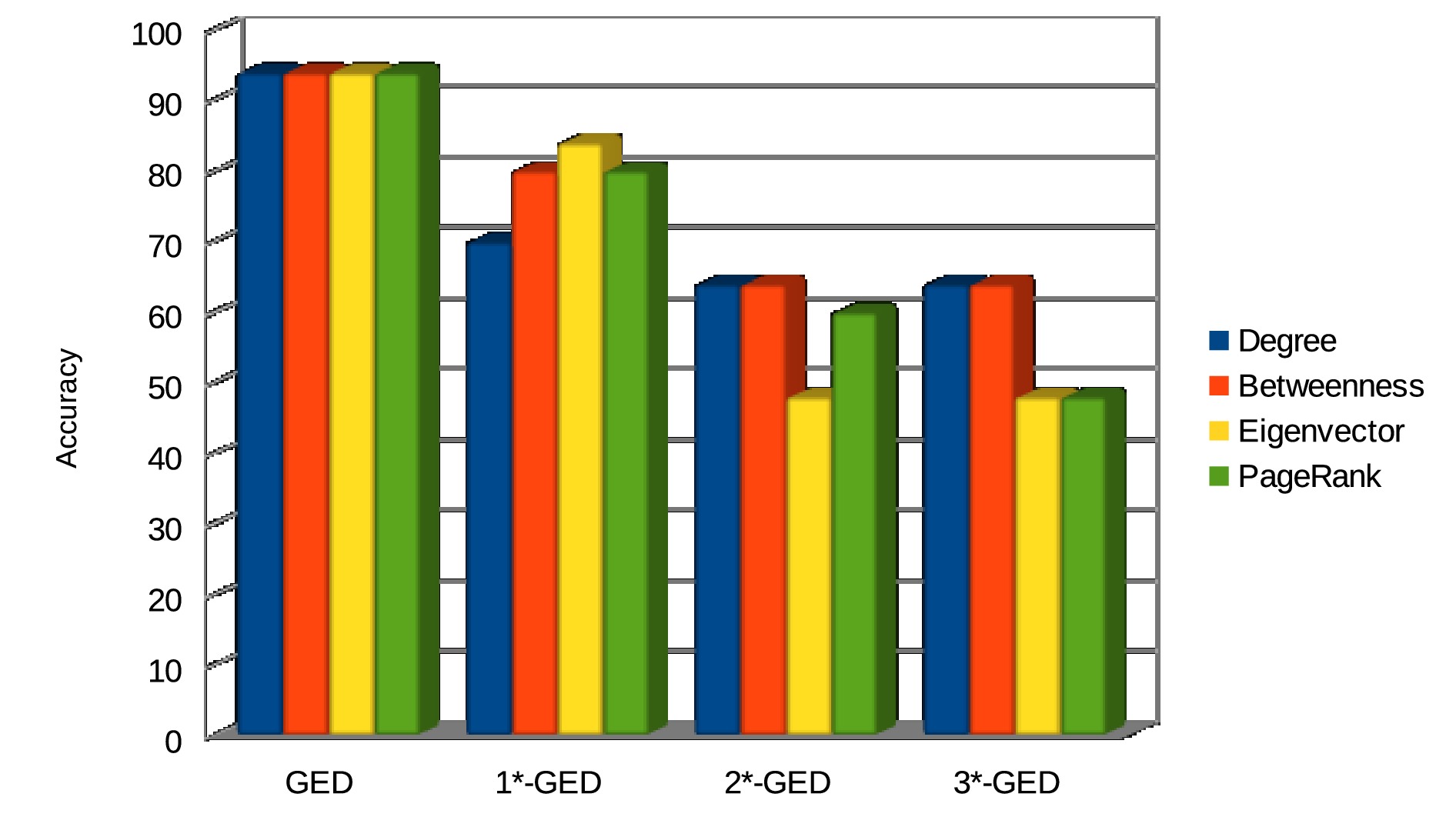}
 \caption{Comparison of accuracy ratio of letter A dataset}
 \label{fig:letter-a-accuracy2}
\end{figure}

Here we note that accuracy of letter A for degree centrality is lower than the other three measures by contracting $t$ nodes, where t is equal to nodes with degree 1 in the input graphs ($1^*$-GED). We can also observe that for letter E, the accuracy ratio using betweenness and PageRank is usually higher than that of degree centrality even though they take less computation time.

\begin{figure}[!t]
\centering
%  \vspace{2.5cm}
 \includegraphics[scale=.15]{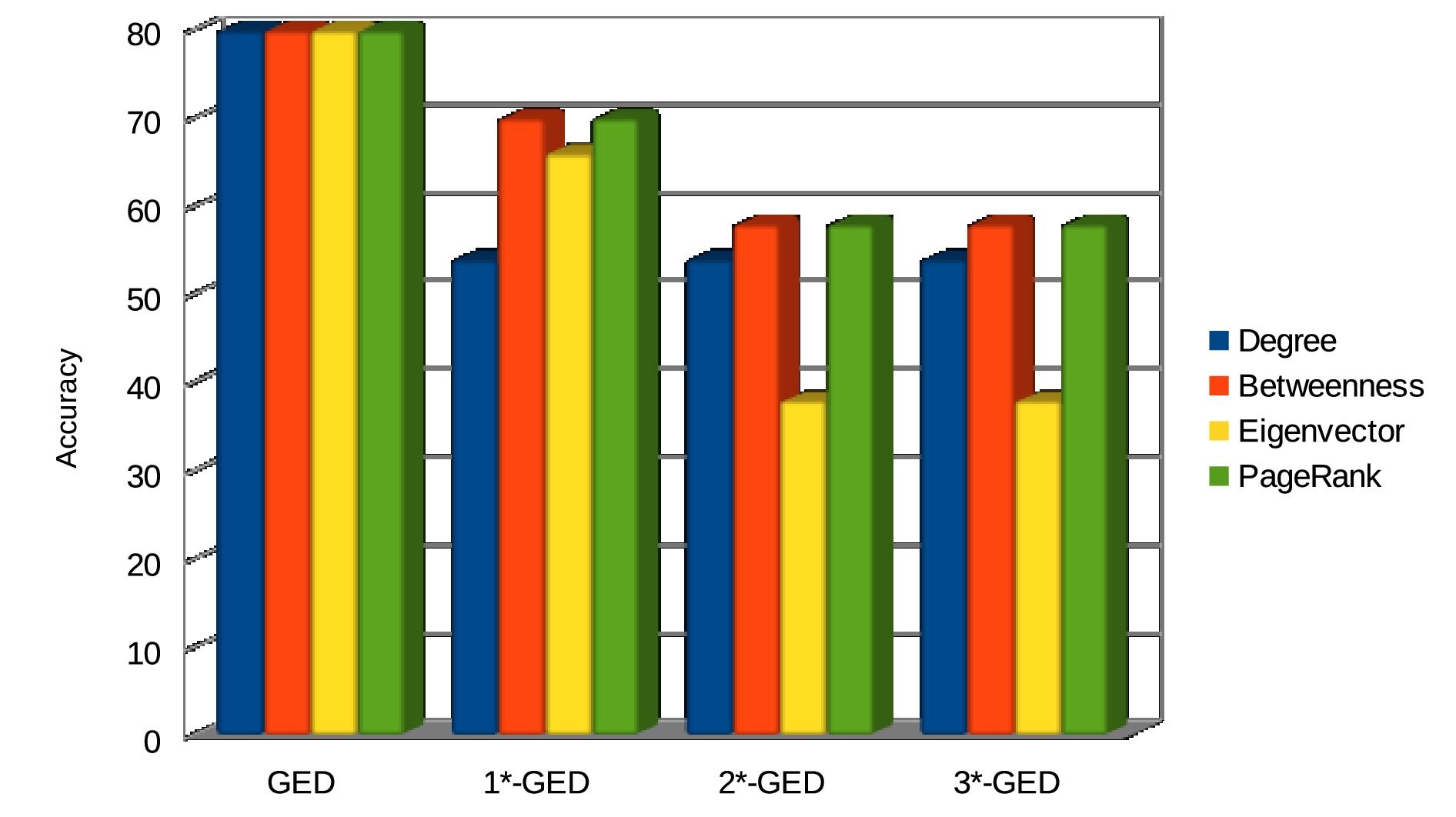}
 \caption{Comparison of accuracy ratio of letter E dataset}
 \label{fig:letter-e-accuracy2}
\end{figure}

To find the accuracy on AIDS dataset, we utilize test dataset consisting of 300 graphs from active class and 1200 graphs from inactive class, whereas training dataset consists of 50 graphs from active class and 200 graphs from the inactive class of AIDS dataset. We can observe the accuracy ratio of the proposed error-tolerant scheme using the four different centrality measure in Figure \ref{fig:aids-active-accuracy2}. In this figure, we observe that the accuracy obtained using degree and PageRank centrality are generally higher than that of eigenvector and betweenness centrality. Here we notice the time versus accuracy trade-off, the centrality criteria which takes less time leads to less accuracy, whereas the centrality techniques which are more accurate take more computation time.

\begin{figure}[!t]
\centering
%  \vspace{2.5cm}
 \includegraphics[scale=.6]{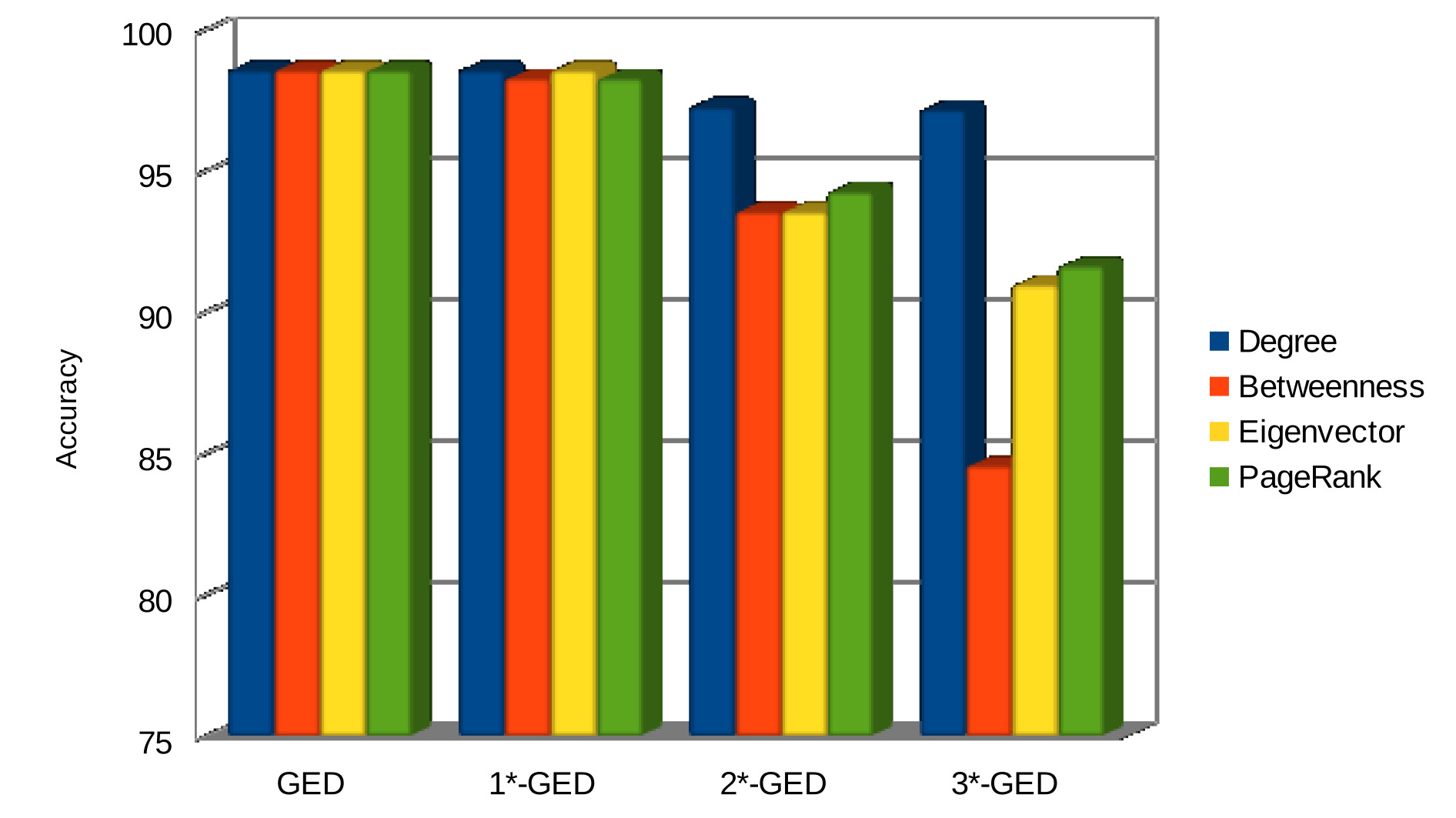}
 \caption{Comparison of accuracy for active class of AIDS dataset}
 \label{fig:aids-active-accuracy2}
\end{figure}

Figure \ref{fig:aids-inactive-accuracy2} shows the comparison of accuracy for the inactive class of AIDS dataset using the four centrality measures. In this figure also degree and PageRank criteria lead to higher accuracy for the classification of graphs of AIDS dataset.

\begin{figure}[!t]
\centering
%  \vspace{2.5cm}
 \includegraphics[scale=.15]{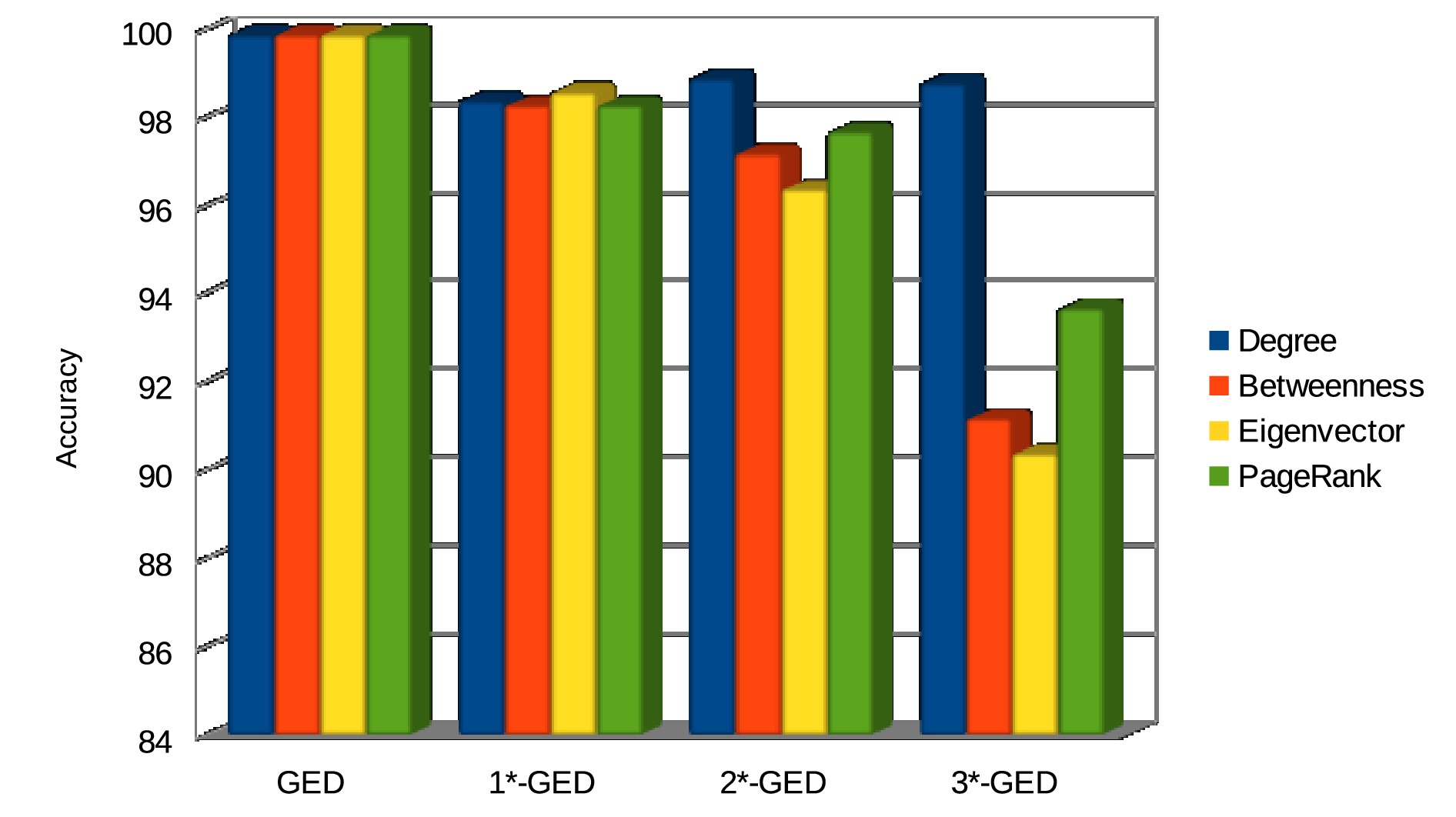}
 \caption{Comparison of accuracy for inactive class of AIDS dataset}
 \label{fig:aids-inactive-accuracy2}
\end{figure}

From the above experiments, we observe that node contraction using various centrality measure can have a different effect on the execution time and accuracy ratio for a given graph dataset. Therefore we can select the best suitable centrality measure for a given dataset depending on the requirement of execution time and accuracy. 

\section{Summary}
In this chapter, we presented techniques to approximate graph matching utilizing the concept of centrality measure to reduce the size of the graphs by ignoring the nodes with a lower value of given centrality criteria. In particular, we have used eigenvector, betweenness and PageRank centrality apart from degree centrality to perform the node contraction for the computation for error-tolerant graph matching. Experimental results show that these centrality criteria can be used as computation time versus accuracy trade-off for different graph dataset. 
 
% Chapter Template

\chapter{Geometric Graph Matching} % Main chapter title

\label{Chapter5} % Change X to a consecutive number; for referencing this chapter elsewhere, use \ref{ChapterX}

\lhead{Chapter 5. \emph{Geometric Graph Matching}} % Change X to a consecutive number; this is for the header on each page - perhaps a shortened title

%----------------------------------------------------------------------------------------
%	SECTION 1
%----------------------------------------------------------------------------------------

\section{Introduction}

%----------------Abstract-----------------------
Geometric graph matching is the process of evaluating the similarity between geometric graphs, where every vertex has an associated coordinate point in a plane. Exact matching computes a strict correspondence between two geometric graphs, whereas in error-tolerant matching an approximate similarity between two geometric graphs is calculated. In this chapter, we present an approach to error-tolerant graph matching using geometric graphs. We introduce the vertex distance or dissimilarity and edge distance or dissimilarity between two geometric graphs and use it to compute graph distance. Finally, we use graph distance to perform exact and error-tolerant geometric graph matching.

%--------------Introduction----------------------
A graph is said to be a geometric graph where each of the vertices is assigned a coordinate point in a two-dimensional or three-dimensional plane. Geometric graph matching is the process of computing the similarity between the two geometric graphs. Geometric graph matching is a particular type of graph matching in which each input graph is modeled as a geometric graph. Depending on the nature of the matching, the graph matching process is broadly divided into exact and error-tolerant graph matching. Exact graph matching requires a strict correspondence between nodes and edges of the two geometric graphs. It is like a geometric graph isomorphism, which in addition to preserving the structure of the two geometric graphs, also preserves the coordinates of the corresponding vertices. 

Exact geometric graph matching although theoretically appealing, may not be useful in many real-world applications, as due to the presence of noise or distortion during the processing, the input graph data may be altered. In such situations, we use error-tolerant geometric graph matching which is also known as inexact graph matching due to its flexibility to accommodate errors during the process of matching \citep{Conteetal2004}.

As each vertex of a geometric graph has a unique coordinate representation, we can use this information along with other geometric transformation properties to perform geometric graph matching. Kuramochi and Karypis \citep{KuramochiKarypis2007} in 2007 proposed an algorithm to compute geometric graph isomorphism in polynomial time. The 2009 paper by Cheong \textit{et al.} \citep{Cheongetal2009} have shown geometric graph matching using edit distance method to be $NP$-hard. Approximate solution for geometric graph matching using vertex edit distance is described in \citep{Armiti2014}. In \citep{Pinheiroetal2017} authors have described geometric graph matching by applying Monte Carlo tree search. 

% End para of introduction
The graph matching problem seems to be inherently linked to geometry and topology of graphs. In this chapter, we extend the work given in \citep{DwivediSingh2018a} and describe a simple yet promising framework for geometric graph similarity and matching. We describe a graph matching algorithm using geometric graphs, which is both error-tolerant as well as computable in polynomial time \citep{DwivediSingh2019}. We define the vertex distance between two geometric graphs (not between two vertices) using the position of their vertices and show it to be metric over the set of all graphs with vertices only. We define edge distance between two graphs based on the angular orientation and length of the edges. We also define a modified version of edge distance, which uses the position of edges as an additional attribute, and demonstrate it to be a metric. Then we combine the notion of vertex distance and edge distance to define the graph distance between two geometric graphs and show some of its properties. Finally, we use this graph distance to perform error-tolerant graph matching on letter and AIDS dataset, in which all the nodes have a coordinate position in the two-dimensional plane \citep{DwivediSingh2019}.

This chapter is organized as follows. Section 5.2, provides preliminaries and motivation. Section 5.3, presents the novel geometric graph similarity framework. Section 5.4, introduce geometric graph matching based on the proposed geometric graph similarity framework. It describes both exact as well as error-tolerant graph matching. Section 5.5, contains an experimental evaluation of the proposed geometric graph matching scheme. Finally, section 5.6, provides a summary.

%----------------------------------------------------------------------------------------
%	SECTION 2
%----------------------------------------------------------------------------------------

\section{Preliminaries and Motivation}
In this section, we review the basic definitions and notations used in this chapter. We also explain the motivation for the proposed work.

A \textit{geometric graph} $G$ is defined as $G= (V,E,l,c)$, where $V$ is the set of nodes, $E$ is the set of edges, $l$ is a labeling function $l: \{V \cup E \} \rightarrow \Sigma$ which assigns a label from $\Sigma$ to each vertex and edge, $c$ is a function $c: V \rightarrow \mathbb{R}^2$ which assigns a coordinate point to each vertex of $G$. If $\Sigma = \emptyset$ then $G$ is called the unlabeled geometric graph.

Let $X$ be a nonempty set. A mapping $d: X \times X \rightarrow \mathbb{R}$ is defined to be a \textit{metric} on $X$ if $d$ satisfy the following conditions:
(i) $d(u, v) \geq 0$ for all $u, v \in X$ and $d(u,v)=0$ if and only if $u=v$ \\
(ii) $d(u,v) = d(v,u)$ for all $u, v \in X$ (Symmetry) \\
(iii) $d(u,w) \leq d(u, v) + d(v, w)$ for all $u, v, w \in X$ (Triangle inequality). 
The pair $(X,d)$ is called a \textit{metric space}.

In this chapter, we consider the problem of finding a similarity score between two geometric graphs. Given two geometric graphs $G_1$,$G_2$ in graph domain $\mathcal{G}$ of the two-dimensional space, we would like to find a distance $d: \mathcal{G} \times \mathcal{G} \rightarrow \mathbb{R}$ which computes the similarity between $G_1$ and $G_2$. For general graphs, it is also known as graph comparison problem. In this chapter, we consider graph comparison problem synonymous to graph matching problem.

Edit-distance-based approach to graph matching, as well as geometric graph matching, have no efficient exact solution. For example, graph edit distance \citep{Zeng2009} and geometric graph matching based on edit distances \citep{Cheongetal2009} are $NP$-hard and therefore no efficient algorithms are available. Since our goal is not to reconstruct $G_2$ from $G_1$, therefore we can avoid the intractability of edit distance based methods for the problem of geometric graph similarity. Instead, we use the Linear Sum Assignment Problem (LSAP) formulation-based approach to geometric graph distance using the various attributes of vertices and edges of the geometric graphs. 

Geometric graph isomorphism, as opposed to graph isomorphism, can be solved efficiently in polynomial time \citep{KuramochiKarypis2007} so we can take advantage of this to find the best geometric representation of a geometric graph $G_2$ with respect to an input geometric graph $G_1$ which make them similar. We define geometric graph similarity which intuitively captures the minimum error required to make the two geometric graphs similar.

\section{Geometric Graph Similarity}
In this section, we introduce vertex distance, edge distance between the geometric graphs $G_1$ and $G_2$. We use these distance measures to compute the dissimilarity between two geometric graphs.

 Initially, we consider two geometric graphs $G_1$ and $G_2$ with vertices only without connections. To find similarity between these graphs, one approach is to assign each vertex of $G_1$ to one of the vertices of $G_2$, so that the total distance is minimum. Suppose vertex coordinate points of $G_1$ be $\{(a_1, b_1), (a_2, b_2), \dots, (a_n, b_n)\}$ and coordinate points of $G_2$ be $\{(x_1, y_1), (x_2, y_2), \dots, (x_n, y_n)\}$, as shown in Figure \ref{fig:geometric-graphs}. If $l_i = \{\sqrt{(a_i-x_j)^2 + (b_i-y_j)^2} | 1 \leq j \leq n \}, \forall 1 \leq i \leq n$. Then vertex distance between $G_1$ and $G_2$ is $\sum_{i=1}^n l_i = \sum_{i=1}^n  \min_{1 \leq j \leq n} \{ \sqrt{(a_i-x_j)^2 + (b_i-y_j)^2} \}$. One limitation of this expression is that more than one vertices of $G_1$ can be assigned to the same vertex of $G_2$.
 
 For a more realistic optimization problem formulation, let $C=(c_{ij})$ be a $n \times n$ cost matrix, where $c_{ij}$ is the Euclidean distance from $i$-th vertex of $G_1$ to $j$-th vertex of $G_2$. We have, $c_{ij}= \sqrt{(a_i-x_j)^2 + (b_i-y_j)^2}$. Now, we can minimize the objective function $\sum_{i=1}^n c_{i \varphi(i)}$ to get the minimum distance between $G_1$ and $G_2$ as $\min_{\varphi \in S_n} \sum_{i=1}^n c_{i \varphi(i)}$. Where $S_n$ is the set of all assignments $\varphi$ from $V_1$ of $G_1$ to $V_2$ of $G_2$. This is the standard LSAP formulation of vertex assignments from $V_1$ to $V_2$. The LSAP can be solved using Hungarian algorithm in $O(n^3)$ time \citep{Burkard2009}.

 In the following definition of vertex distance, we assume that the number of vertices in two graphs are equal. For arbitrary graph in which the number of vertices is not same, we can insert additional vertices in the smaller graph, with coordinates equal to mean of all its existing coordinate values, to make the vertex set of both graphs equal.  

\begin{defn}
%\textbf{Definition 1.}
 Let $G_1= (V_1,E_1,c_1)$ and $G_2= (V_2,E_2,c_2)$ be two geometric graphs with $|V_1|=|V_2|=n$. Let coordinate points of $V_1$ be $\{(a_1, b_1), (a_2, b_2), \dots, (a_n, b_n)\}$ and coordinate points of $V_2$ be $\{(x_1, y_1), (x_2, y_2), \dots, (x_n, y_n)\}$ then the vertex distance between the two geometric graphs $G_1$ and $G_2$ is defined as 
 \begin{equation}
  VD(G_1, G_2)= \min_{\varphi \in S_n} \sum_{i=1}^n \sqrt{(a_i-x_{\varphi(i)})^2 + (b_i-y_{\varphi(i)})^2}
 \end{equation}
 Where $S_n$ and $\varphi$ are as defined above.
\end{defn}

\begin{figure}[!t]
\centering
\includegraphics[scale=0.1]{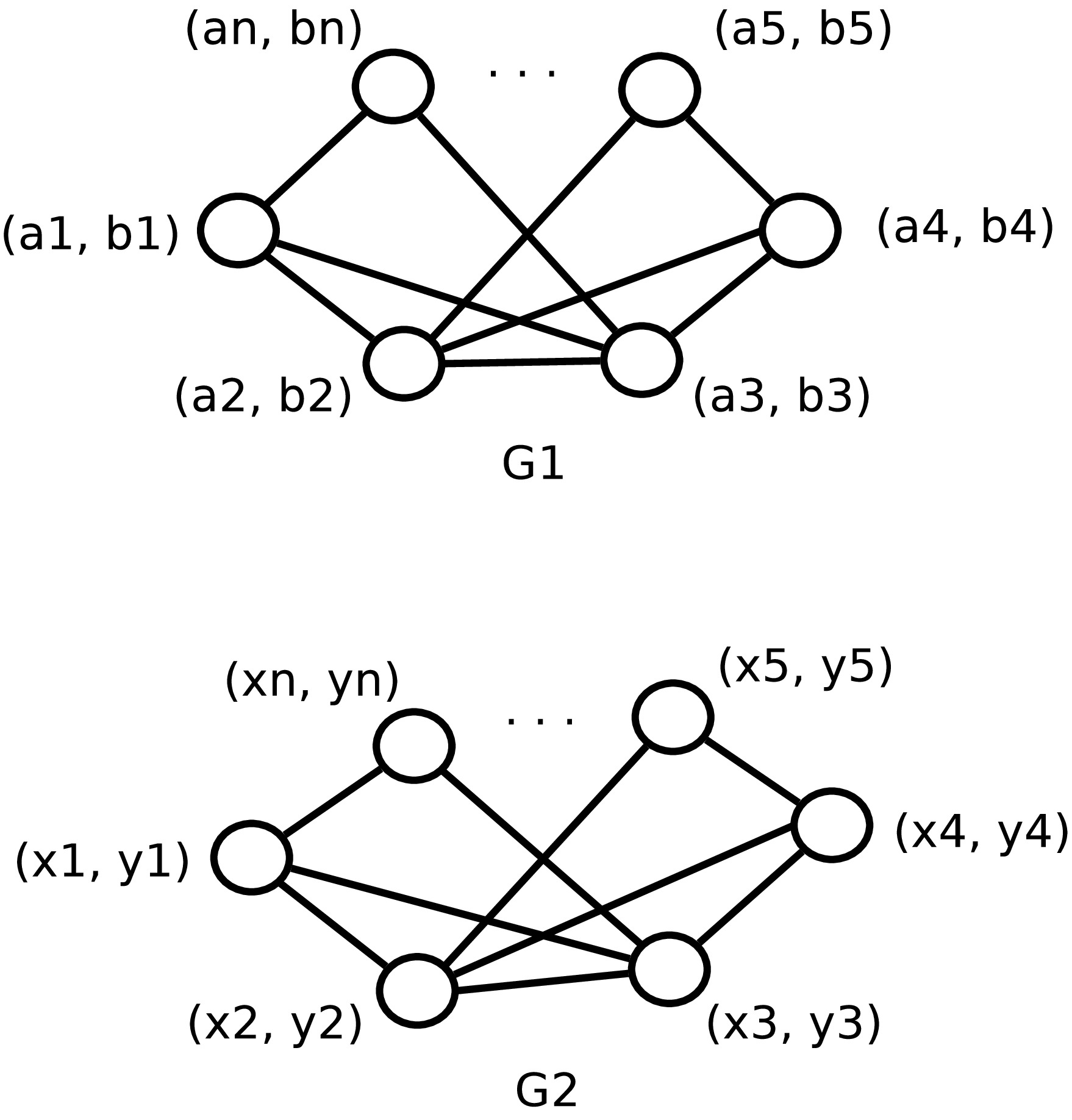}
\caption{Geometric Graphs $G_1$ and $G_2$}
\label{fig:geometric-graphs}
\end{figure}

Here, \textit{VD} represents an LSAP formulation for the assignment of the vertex set $V_1$ of $G_1$ to the vertex set $V_2$ of $G_2$. The geometric graphs having a similar position of vertices will have less \textit{VD}, whereas the graphs having a very different position of vertices have large \textit{VD}. In the above definition, we can also represent vertex set $V_1$ and $V_2$ using their offsets from the mean position of all vertex coordinates of $G_1$ and $G_2$, respectively. 
%Here, $VD$ represents the sum of the distance of each pair of assigned vertices from $V_1$ to $V_2$. More the deviation of corresponding coordinates between $G_1$ and $G_2$ more will be $VD$ value.

\begin{figure}[!t]
\centering
\includegraphics[scale=0.4]{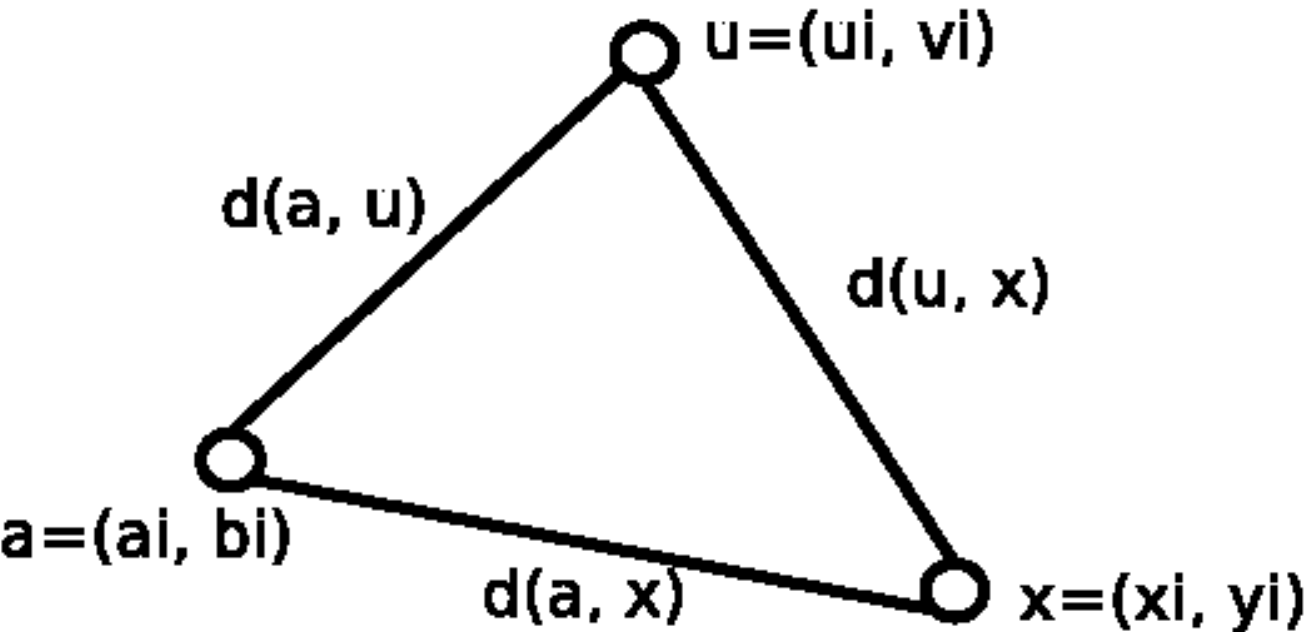}
\caption{Triangle inequality: $d(a,x) \leq d(a,u) + d(u,x)$}
\label{fig:triangle-inequality}
\end{figure}

% Lemma 1
\begin{lem} \label{lem:lemma1}
$VD(G_1, G_2)$ is a metric on the set of all geometric graphs $\mathcal{G}$ having edge set $E=\emptyset$.
\end{lem}

\begin{proof}
Property (i) of metric:  Here by definition $VD(G_1, G_2) \geq 0$, If $G_1= G_2$ then clearly $VD(G_1, G_2)= \min_{\varphi \in S_n} \sum_{i=1}^n [(a_i-x_{\varphi(i)})^2 + (b_i-y_{\varphi(i)})^2]^{1/2} = 0$, since in each component of summation we have $a_i=x_{\varphi(i)}$ and $b_i=y_{\varphi(i)}$. \\ Conversely, if $VD(G_1, G_2)=0$ then $\min_{\varphi \in S_n} \sum_{i=1}^n [(a_i-x_{\varphi(i)})^2 + (b_i-y_{\varphi(i)})^2]^{1/2} = 0$ implies that for some combination of $i,{\varphi(i)}$ each term of summation \\
$\sum_{i=1}^n [(a_i-x_{\varphi(i)})^2 + (b_i-y_{\varphi(i)})^2]^{1/2}$ is $0$, \\ $\Rightarrow (a_i-x_{\varphi(i)})^2=0$ and $(b_i-y_{\varphi(i)})^2=0$, \\ $\Rightarrow a_i=x_{\varphi(i)}$ and $b_i=y_{\varphi(i)}$. Hence $G_1=G_2$ \\
Property (ii): \\
$VD(G_1, G_2)= \min_{\varphi \in S_n} \sum_{i=1}^n [(a_i-x_{\varphi(i)})^2 + (b_i-y_{\varphi(i)})^2]^{1/2}$ \\
$= \min_{\varphi \in S_n} \sum_{i=1}^n [(x_{\varphi(i)}-a_i)^2 + (y_{\varphi(i)}-b_i)^2]^{1/2} = VD(G_2, G_1) $  \\
Reversibility of $\varphi$ can also be demonstrated. Let $V_1 = {u_1, u_2, \cdots, u_n}$ and $V_1 = {v_1, v_2, \cdots, v_n}$. Let  $\phi$ from $V_1 \rightarrow V_2$ using LSAP assign $u_1 \rightarrow v_1,  u_2 \rightarrow v_2, \cdots,  u_n \rightarrow v_n$. Now suppose $\phi^{-1}$ from $V_2 \rightarrow V_1$ assigns $v_1 \rightarrow x_1,  v_2 \rightarrow x_2, \cdots, v_n \rightarrow x_n$. There are three cases to consider. \\
Case 1: Cost of $\phi$ from $V_1 \rightarrow V_2$ = cost of $\phi^{-1}$ from $V_2 \rightarrow V_1$ then $VD(G_1, G_2) = VD(G_2, G_1)$. \\ Case 2: Cost of $\phi$ from $V_1 \rightarrow V_2 >$  cost of $\phi^{-1}$ from $V_2 \rightarrow V_1$
              Then $x_1 \neq u_1, x_2 \neq u_2, \cdots,  x_n \neq u_n$.
              It is a contradiction since cost using $\phi$ was supposed to be minimum using LSAP.
              Here the cost is a distance.  So $\phi$ could have selected $x_1$ instead of $u_1$, $x_2$ instead of $u_2$      
              and so on, to return a lower value. \\
Case 3: Cost of $\phi$ from $V_1 \rightarrow V_2 <$  cost of $\phi^{-1}$ from $V_2 \rightarrow V_1$. 
              Arguments similar to Case 2 applies here.              
\\
Property (iii): \\ 
Let $G_3= (V_3,E_3,c_3)$ be a geometric graph with $|V_3|=n$, $E_3=\emptyset$ and coordinate points of $V_3$ be $\{(u_1, v_1), (u_2, v_2), \dots, (u_n, v_n)\}$. Figure \ref{fig:triangle-inequality} illustrate geometrically the triangle inequality among the points $(a_i,b_i), (x_i, y_i)$ and $(u_i, v_i)$ of the graphs $G_1, G_2$ and $G_3$ respectively. We can observe that \\
$VD(G_1, G_3) + VD(G_3, G_2)$ \\
$=\min_{\varphi \in S_n} \sum_{i=1}^n [(a_i-u_{\varphi(i)})^2 + (b_i-v_{\varphi(i)})^2]^{1/2} $ \\
$\text{ }\text{ } + \min_{\varphi' \in S_n} \sum_{i=1}^n [(u_i-x_{\varphi'(i)})^2 + (v_i-y_{\varphi'(i)})^2]^{1/2}$ \\
let the minimum function over all ${\varphi(i)}$ and ${\varphi'(i)}$ returns the following\\
$=\sum_{i=1}^n [(a_i-u_{\varphi(i)})^2 + (b_i-v_{\varphi(i)})^2]^{1/2} + \sum_{i=1}^n [(u_i-x_{\varphi'(i)})^2 + (v_i-y_{\varphi'(i)})^2]^{1/2}$ \\
In the first term of the above expression, for $i=1$, let $\varphi(1)=k$. In the second term suppose $u_{\varphi(1)}=u_k$ in $G_3$ is assigned to $x_{\varphi'(k)}$ for some $x \in G_2$. So, for $i=1$ we have the above expression \\
$ [(a_1-u_{\varphi(1)})^2 + (b_1-v_{\varphi(1)})^2]^{1/2} + [(u_k-x_{\varphi'(k)})^2 + (v_k-y_{\varphi'(k)})^2]^{1/2}$ \\ $
= [(a_1-u_{k})^2 + (b_1-v_{k})^2]^{1/2} + [(u_k-x_{\varphi'(k)})^2 + (v_k-y_{\varphi'(k)})^2]^{1/2}$ \\
let $p_1 = a_1-u_k, p_2=b_1-v_k,$ \\ 
$q_1=u_k-x_{\varphi'(k)}, q_2=v_k-y_{\varphi'(k)} $, then we have \\
$[(a_1-u_k)^2 + (b_1-v_k)^2]^{1/2} + [(u_k-x_{\varphi'(k)})^2 + (v_k-y_{\varphi'(k)})^2]^{1/2}$ \\
$=[(p_1)^2 + (p_2)^2]^{1/2} + [(q_1)^2 + (q_2)^2]^{1/2}$ \\
using Minkowski inequality \\
$\sum_{i=1}^n [(x_i + y_i)^2]^{1/2} \leq (\sum_{i=1}^n x_i^2)^{1/2} +(\sum_{i=1}^n y_i^2)^{1/2}$ \\
and putting the values of $p_1,q_1,p_2,q_2$ we get \\
$[(p_1)^2 + (p_2)^2]^{1/2} + [(q_1)^2 + (q_2)^2]^{1/2}$ \\
$\geq = [(p_1 + q_1)^2 + (p_2 + q_2)^2]^{1/2} $ \\
$= [(a_1 - u_k + u_k - x_{\varphi'(k)})^2 + (b_1 - v_k + v_k - y_{\varphi'(k)})^2]^{1/2} $ \\
$= [(a_1 - x_{\varphi'(k)})^2 + (b_1 - y_{\varphi'(k)})^2]^{1/2} $ \\
taking sum over all $i$-th term and a $\varphi'' \in S_n$ \\
$\sum_{i=1}^n [(a_i-x_{\varphi''(i)})^2 + (b_i-y_{\varphi''(i)})^2]^{1/2}$ \\
and taking minimum of summation over all $i$ and a ${\varphi''(i)}$ we get\\
$\min_{\varphi'' \in S_n} \sum_{i=1}^n [(a_i-x_{\varphi''(i)})^2 + (b_i-y_{\varphi''(i)})^2]^{1/2} = VD(G_1, G_2)$ \\
hence, $VD(G_1, G_3) + VD(G_3, G_2) \geq VD(G_1, G_2)$ \\
Note that if either $G_3=G_1$ or $G_3=G_2$ then  \\
$VD(G_1, G_3) + VD(G_3, G_2) = VD(G_1, G_2)$ \\
and when $G_3 \neq G_1$ and $G_3 \neq G_2$ then \\
$VD(G_1, G_3) + VD(G_3, G_2) > VD(G_1, G_2)$ . \\% $\Box$ 
\end{proof}

\textit{VD} considers only the position of vertices of geometric graphs in the plane. Now, we use the features of edges of geometric graphs to define the edge distance. The main features of an edge of a geometric graph include its orientation and length along with the coordinates of its endpoints. 

 Let $ \theta_{\{(a,b),(c,d)\}}$ denote the angle subtended between the extended line joining the coordinate points $(a,b),(c,d)$ and positive $x$-axis. In the following definition of edge distance, we assume that the number of edges in the two graphs is equal. For graphs with unequal edge set, we can append additional empty edges to the smaller graph, to make the total number of edges in both graphs same.

 \begin{defn}
 %\textbf{Definition 2.}
 Let $G_1= (V_1,E_1,c_1)$ and $G_2= (V_2,E_2,c_2)$ be two geometric graphs with $|E_1|=|E_2|=m$. Here $c_1$ and $c_2$ are the set of coordinate points of vertex set $V_1$ and $V_2$ respectively. Then the edge distance between the two geometric graphs $G_1$ and $G_2$ is defined as 
 \begin{equation}
  ED(G_1, G_2)= \min_{\varphi \in S_m} \sum_{i=1}^m c_{i\varphi(i)}  
 \end{equation}
  such that, $$ c_{ij}=\sqrt{((\varTheta_{i}-\varTheta^{}_{j})\frac{\pi}{180^{\circ}})^2} + \sqrt{(d_{i}-D_{j})^2} $$
  Where, $ \varTheta_{i} = \theta_{\{(a_i,b_i),(a_{p},b_{p}) | ((a_i,b_i),(a_{p},b_{p})) \in E_1\}}$, \\
  $ \varTheta^{}_{j} = \theta_{\{(x_j,y_j),(x_{q},y_{q}) | ((x_j,y_j),(x_{q},y_{q})) \in E_2\}}$, \\
 $d_{i} = \{ \sqrt{(a_i-a_{p})^2 + (b_i-b_{p})^2} | ((a_i,b_i),(a_{p},b_{p})) \in E_1 \}$, \\ 
 $D_{j} = \{ \sqrt{(x_j-x_{q})^2 + (y_j-y_{q})^2} | ((x_j,y_j),(x_{q},y_{q})) \in E_2 \}$\\
 and $S_m$ is the set of all assignments $\varphi$ from $E_1$ of $G_1$ to $E_2$ of $G_2$.
\end{defn}

The above definition of $ED$ finds the best assignment of edges from $E_1$ to $E_2$. The cost function $C=(c_{ij})$ is itself made up of two components. The first term $\sqrt{((\varTheta_{i}-\varTheta^{}_{j})\frac{\pi}{180^{\circ}})^2}$ calculates difference in orientation of $E_{i}$ in $E_1$ and $E'_{j}$ in $E_2$. We denote this term of $ED$ by $E_{ij}^{A}$, where $A$ stands for an angular difference. It is translation and scaling invariant, as we can observe in Figure \ref{fig:edge-distance-zero} that $G_1$ and $G_4$ have identical value of this term. The second term $\sqrt{(d_{i}-D_{j})^2}$ of $ED$ denotes the difference of length between $E_{i}$ and $E'_{j}$. We call this component as the length term and represent it by $E_{ij}^{L}$, where $L$ is used for the length of edges. $E_{ij}^{L}$ is translation and rotation invariant, as we can see in Figure \ref{fig:edge-distance-zero} that all four graphs $G_1$ to $G_4$ have the equivalent value of the second term of $ED$. Following properties of $ED$ are easy to establish.   

%The first term in the above definition of $ED$ accounts for the angular difference in radian between each pair of corresponding edges selected from $E_1$ and $E_2$, whereas the second term of $ED$ represents the difference of edge length between each pair of assigned edges. % Similar to $VD$, we can show that both first and second term of $ED$ is a metric and therefore $ED$ is also a metric. 
\begin{figure}[!t]
\centering
\includegraphics[scale=0.1]{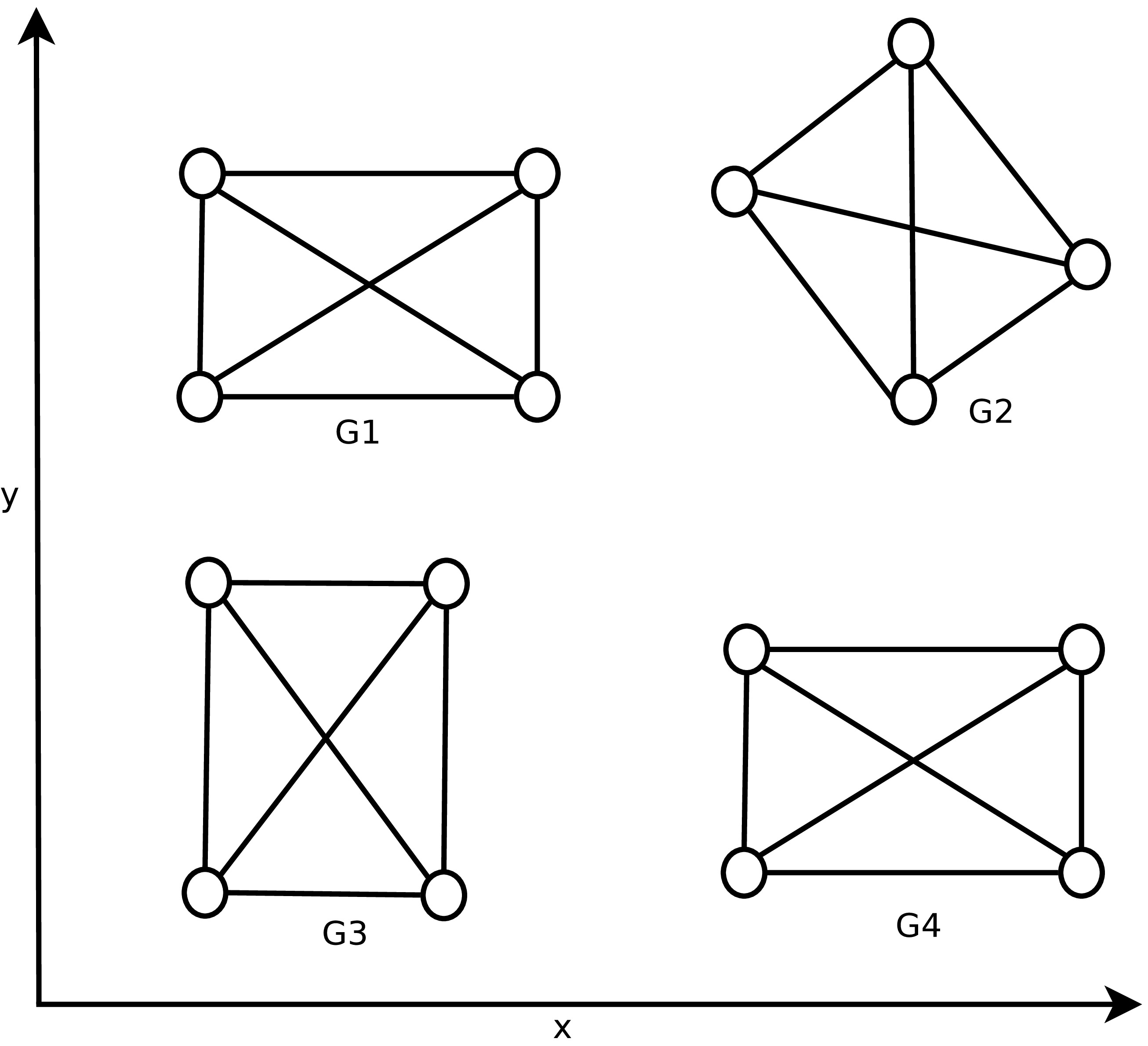}
\caption{Edge Distance from $G_1$ to $G_4$ is 0}
\label{fig:edge-distance-zero}
\end{figure}

% Lemma 2
\begin{lem} \label{lem:lemma2}
 $ED(G_1, G_2)$ satisfy the following properties: \\
 (i) $ED(G_1, G_2) \geq 0$, if $G_1 = G_2$ then $ED(G_1, G_2) = 0$ \\
 (ii) $ED(G_1, G_2) = ED(G_2, G_1)$ \\
 (iii) $ED(G_1, G_2) \leq ED(G_1, G_3) + ED(G_3, G_2)$
\end{lem}
\begin{proof}
(i) $ED(G_1, G_2) \geq 0$, since both terms in the definition of $ED$ is always non-negative. Also when $G_1 = G_2$, we have $\varTheta_{i} = \varTheta^{}_{j}$ and $d_{i} = D_{j}$ for each term of summation. Therefore $ED(G_1, G_2) = 0$. Note that converse may not be true i.e., $ED(G_1, G_2) = 0$ not necessarily imply $G_1=G_2$. We can observe in Figure \ref{fig:edge-distance-zero} that graphs $G_2$ and $G_3$ are rotation followed by translation of $G_1$, while $G_4$ is a translation of $G1$. Here $ED(G_1, G_4)=0$. Thus $ED$ between two translated version of a graph turns out to be 0. \\
(ii) $ED(G_1, G_2)= \min_{\varphi \in S_m} \sum_{i=1}^m ([((\varTheta_{i}-\varTheta^{}_{\varphi(i)})\frac{\pi}{180^{\circ}})^2]^{1/2} + [(d_{i}-D_{\varphi(i)})^2]^{1/2}) = \min_{\varphi \in S_m} \sum_{i=1}^m ([((\varTheta^{}_{\varphi(i)}-\varTheta_{i})\frac{\pi}{180^{\circ}})^2]^{1/2} + [(D_{\varphi(i)}-d_{i})^2]^{1/2}) = ED(G_2, G_1) $ \\
(iii) Let $G_3= (V_3,E_3,c_3)$ be a geometric graph with $|V_3|=n$ and coordinate points of $V_3$ be $\{(u_1, v_1), (u_2, v_2), \dots, (u_n, v_n)\}$. Suppose $ \varTheta^{}_{k} = \theta_{\{(u_k,v_k),(u_{r},v_{r})| (u_k,v_k),(u_{r},v_{r}) \in E_3\}}$ and $D^{}_{k} =\{ \sqrt{(u_k-u_{r})^2 + (v_k-v_{r})^2} | (u_k,v_k),(u_{r},v_{r}) \in E_3\}$, then 
 $ED(G_1, G_3) + ED(G_3, G_2)$ \\
 $= \min_{\varphi \in S_m} \sum_{i=1}^m ([((\varTheta_{i}-\varTheta^{}_{k})\frac{\pi}{180^{\circ}})^2]^{1/2} + [(d_{i}-D^{}_{k})^2]^{1/2})$ \\
 $\text{ }\text{ } + \min_{\varphi \in S_m} \sum_{k=1}^m ([((\varTheta^{}_{k}-\varTheta^{}_{j})\frac{\pi}{180^{\circ}})^2]^{1/2} + [(D^{}_{k}-D_{j})^2]^{1/2})$ \\
 $=\min_{\varphi \in S_m} \sum_{i=1}^m (|(\varTheta_{i}-\varTheta^{}_{k})\frac{\pi}{180^{\circ}}| + |d_{i}-D^{}_{k}|) $ \\
 $\text{ }\text{ } + \min_{\varphi \in S_m} \sum_{k=1}^m (|(\varTheta^{}_{k}-\varTheta^{}_{j})\frac{\pi}{180^{\circ}}| + |D^{}_{k}-D_{j}|) $ \\
 Let the minimum function over all $i$ and $\varphi(i)$ returns the following \\
 $\sum_{i=1}^m (|(\varTheta_{i}-\varTheta^{}_{k})\frac{\pi}{180^{\circ}}| + |d_{i}-D^{}_{k}|) + \sum_{k=1}^m (|(\varTheta^{}_{k}-\varTheta^{}_{j})\frac{\pi}{180^{\circ}}| + |D^{}_{k}-D_{j}|) $ \\
 now, using triangle inequality $|x|+|y| \geq |x+y|$ \\
 and inequality of summation $\sum_{x \in X} |f(x)| \geq |\sum_{x \in X}f(x)| $ \\
 $\sum_{i=1}^m (|(\varTheta_{i}-\varTheta^{}_{k})\frac{\pi}{180^{\circ}}| + |d_{i}-D^{}_{k}|) + \sum_{k=1}^m (|(\varTheta^{}_{k}-\varTheta^{}_{j})\frac{\pi}{180^{\circ}}| + |D^{}_{k}-D_{j}|) $ \\
 $= \sum_{i=1}^m |(\varTheta_{i}-\varTheta^{}_{k})\frac{\pi}{180^{\circ}}| + \sum_{i=1}^m |d_{i}-D^{}_{k}| + \sum_{k=1}^m |(\varTheta^{}_{k}-\varTheta^{}_{j})\frac{\pi}{180^{\circ}}| + \sum_{k=1}^m |D^{}_{k}-D_{j}|$ \\
 $\geq \sum_{i,k=1}^m (|(\varTheta_{i}-\varTheta^{}_{k}+\varTheta^{}_{k}-\varTheta^{}_{j})\frac{\pi}{180^{\circ}}| + |d_{i}-D^{}_{k}+D^{}_{k}-D_{j}|)$ \\
 $= \sum_{i=1}^m (|(\varTheta_{i}-\varTheta^{}_{j})\frac{\pi}{180^{\circ}}| + |d_{i}-D_{j}|)$ \\
 and therefore $ ED(G_1, G_3) + ED(G_3, G_2) \geq ED(G_1, G_2)$. \\
\end{proof}

We observe that $VD$ express the relative position of the vertices between two geometric graphs, whereas $ED$ captures the orientation and length of edges between two geometric graphs. Nevertheless, $ED$ does not encapsulate the relative position of edges of graphs in the two-dimensional plane. For example in Figure \ref{fig:graph-distance-zero}, two geometric graphs $G_1$ and $G_2$ are shown such that $VD(G_1, G_2)=ED(G_1, G_2)=0$ even though geometrically $G_1 \neq G_2$, because $ED$ does not capture the relative position of edges. To accommodate the relative position of edges, we introduce a third term in the definition of $ED$ called position component denoted by $E_{ij}^{P}$.  

Let $i$ and $j$ denote the $i$-th and $j$-th edge of $G_1$ and $G_2$ respectively, such that $i$-th edge of $G_1$ is $\{(a_i, b_i), (a_{p}, b_{p})\}$ and $j$-th edge of $G_2$ is $\{(x_j, y_j), (x_{q}, y_{q})\}$. We define the position term of $ED$ by 
$$ E_{ij}^{P}=(\sqrt{(a_i-x_{j})^2 + (b_i-y_{j})}^2 + \sqrt{(a_{p}-x_{q})^2 + (b_{p}-y_{q})}^2)/2 $$
The term $E_{ij}^{P}$ is similar to $VD$, which associate the position of one vertex of $G_1$ to another vertex of $G_2$. $E_{ij}^{P}$ compares position of one edge of $G_1$ to another edge of $G_2$. Now, we define Edge Distance Modified or Edge Distance Metric ($EDM$), which use an objective function to combine all three features of edges.

\begin{figure}[!t]
\centering
\includegraphics[scale=0.15]{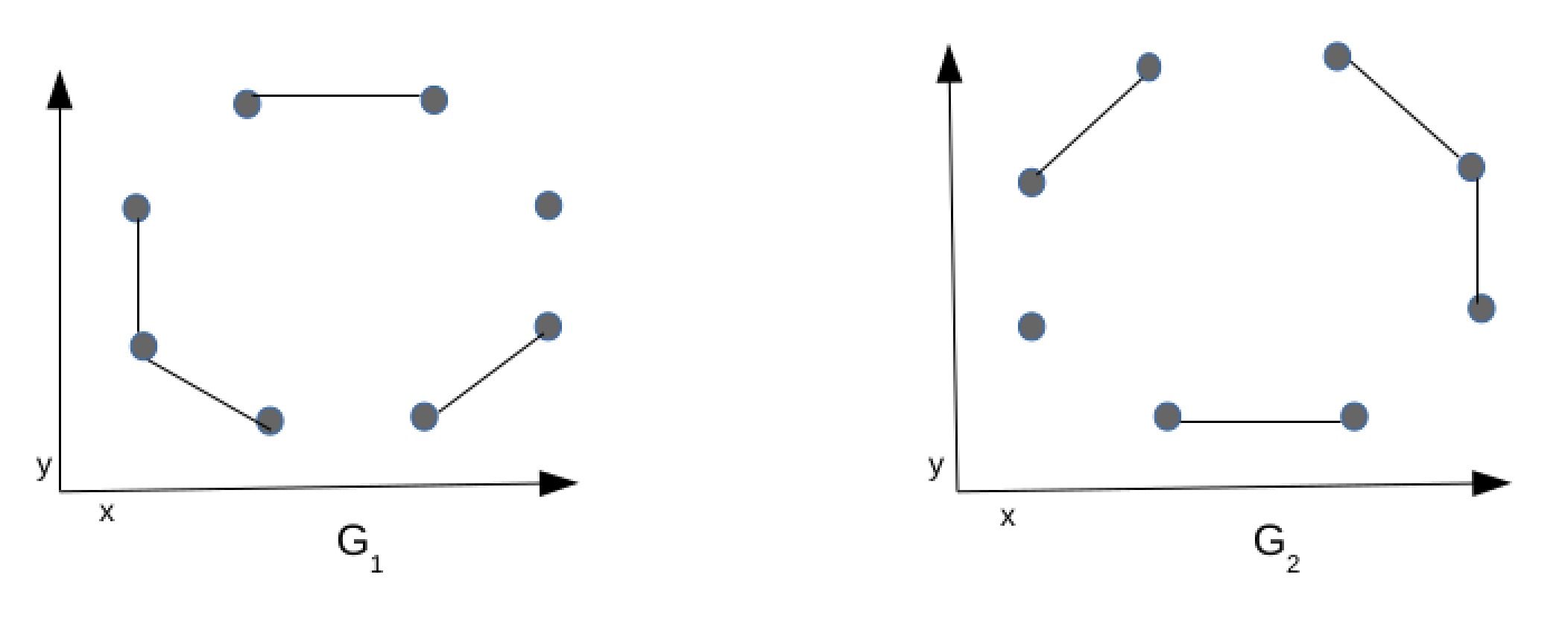}
\caption{Geometric graphs with $VD$ and $ED$ both 0}
\label{fig:graph-distance-zero}
\end{figure}

 \begin{defn}
 %\textbf{Definition 3.}
 Let $G_1= (V_1,E_1,c_1)$ and $G_2= (V_2,E_2,c_2)$ be two geometric graphs with $|E_1|=|E_2|=m$. Then the edge distance metric between the two geometric graphs $G_1$ and $G_2$ is defined as 
 \begin{equation}
  EDM(G_1, G_2)= \min_{\varphi \in S_m} \sum_{i=1}^m c_{i\varphi(i)}  
 \end{equation}
  such that, $$ c_{ij}= E_{ij}^{A} + E_{ij}^{L} + E_{ij}^{P} $$ 
  Where, $E_{ij}^{A}=\sqrt{((\varTheta_{i}-\varTheta^{}_{j})\frac{\pi}{180^{\circ}})^2}$, 
  $E_{ij}^{L}=\sqrt{(d_{i}-D_{j})^2} $, \\ 
  $E_{ij}^{P}=(\sqrt{(a_i-x_{j})^2 + (b_i-y_{j})}^2 + \sqrt{(a_{p}-x_{q})^2 + (b_{p}-y_{q})}^2)/2$ \\
  and $S_m$ is the set of all assignments $\varphi$ from $E_1$ of $G_1$ to $E_2$ of $G_2$.
\end{defn}
% lemma 3
\begin{lem} \label{lem:lemma3}
 $EDM(G_1, G_2)$ is metric over the set of all geometric graphs $\mathcal{G}$ without isolated vertices. 
\end{lem}
\begin{proof}
Property (i) of metric: $EDM(G_1, G_2)\geq 0$, since $E_{ij}^{A}$, $E_{ij}^{L}$, and $E_{ij}^{P}$ are all non-negative. Now, when $G_1 = G_2$, we have $ c_{ij}= E_{ij}^{A} + E_{ij}^{L} + E_{ij}^{P}=0$, hence $EDM(G_1, G_2)=0$. Conversely, suppose $EDM(G_1, G_2)=0$. Then for some ${\varphi \in S_m}$, we get $\sum_{i=1}^m c_{i\varphi(i)}=0$. So, $E_{ij}^{A} = E_{ij}^{L} = E_{ij}^{P}=0$. We notice that $E_{ij}^{A}=0$ implies $\varTheta_{i}=\varTheta^{'}_{j}$ and since $E_{ij}^{A}$ is translation and scaling invariant therefore $E_{ij}^{A}=0$ indicates that rotation is preserved between $G_1$ and $G_2$. Also $E_{ij}^{L}$ is translation and rotation invariant so $E_{ij}^{L}=0$ denotes that scaling is preserved. Similarly, $E_{ij}^{P}=0$ implies $a_i=x_{j}, b_i=y_{j},a_{p}=x_{q},b_{p}=y_{q}$ and therefore translation is also preserved between $G_1$ and $G_2$. Hence $EDM(G_1, G_2)=0$ implies that $G_1=G_2$ as long as $G_1$ and $G_2$ does not have 0-degree vertices. For example, if the position of the isolated vertex of graph $G_1$ in Figure \ref{fig:graph-distance-zero} is changed to construct another graph $G_1'$ then also we get $EDM(G_1, G_1')=0$. \\
Property (ii): In the definition of $EDM$, we have $ c_{ij}= E_{ij}^{A} + E_{ij}^{L} + E_{ij}^{P}= E_{ji}^{A} + E_{ji}^{L} + E_{ji}^{P}=c_{ji}$. Hence $EDM(G_1, G_2)=EDM(G_1, G_2)$. \\
Property (iii): Let $G_3= (V_3,E_3,c_3)$ be a geometric graph as defined in Lemma 2 having $|E_3|=m$, then \\
$EDM(G_1, G_3) + EDM(G_3, G_2)$ \\
$= \min_{\varphi \in S_m} \sum_{i=1}^m ([((\varTheta_{i}-\varTheta^{}_{k})\frac{\pi}{180^{\circ}})^2]^{1/2} + [(d_{i}-D^{}_{k})^2]^{1/2} + (\sqrt{(a_i-u_{k})^2 + (b_i-v_{k})}^2 + \sqrt{(a_{p}-u_{r})^2 + (b_{p}-v_{r})}^2)/2 )$ \\
 $\text{ }\text{ } + \min_{\varphi \in S_m} \sum_{k=1}^m ([((\varTheta^{}_{k}-\varTheta^{}_{j})\frac{\pi}{180^{\circ}})^2]^{1/2} + [(D^{}_{k}-D_{j})^2]^{1/2} + (\sqrt{(u_k-x_{j})^2 + (v_k-y_{j})}^2 + \sqrt{(u_{r}-x_{q})^2 + (v_{r}-y_{q})}^2)/2 )$ \\ 
$= \min_{\varphi \in S_m} \sum_{i=1}^m (E_{ik}^{A} + E_{ik}^{L} + E_{ik}^{P}) + \min_{\varphi \in S_m} (\sum_{k=1}^m  E_{kj}^{A} + E_{kj}^{L} + E_{kj}^{P} )$ using Lemma 2 and the fact that each $E_{ij}^{P}$ is itself a Euclidean distance function, we have \\
$\min_{\varphi \in S_m} \sum_{i,k=1}^m [ (E_{ik}^{A} +E_{kj}^{A}) + (E_{ik}^{L}+ E_{kj}^{L}) + (E_{ik}^{P} + E_{kj}^{P})]$
$\geq  \sum_{i=1}^m (E_{ij}^{A} + E_{ij}^{L} + E_{ij}^{P}) = EDM(G_1, G_2)$
\end{proof}
For the purpose of showing the triangle inequality  property of $VD$, $ED$ or $EDM$; we can use the properties of their cost matrix. %we can formulate these problems in terms of weighted bipartite matching or network flows problem.
We can observe that $EDM(G_1, G_2)$ is the minimum cost of assignment of edges from $G_1$ to $G_2$ using its cost function. Similarly $EDM(G_1, G_3)$ and $EDM(G_3, G_2)$ are the minimum cost of selecting edges from $G_1$ to $G_3$ and $G_3$ to $G_2$ to respectively. So, $EDM(G_1, G_3) + EDM(G_3, G_2)$ is a valid assignment from $G_1$ to $G_2$. Since, $EDM(G_1, G_2)$ pertains to assignment of edges from $G_1$ to $G_2$ with minimum cost,  we have $EDM(G_1, G_2) \leq EDM(G_1, G_3) + EDM(G_3, G_2)$. 

Graph distance between two geometric graphs is defined as a linear combination of both $VD$ and $ED$ for the same pair of graphs.
\begin{defn}
 %\textbf{Definition 4.}
 Let $G_1= (V_1,E_1,c_1)$ and $G_2= (V_2,E_2,c_2)$ be two geometric graphs with $|V_1|=|V_2|=n$ and $|E_1|=|E_2|=m$. Then the graph distance between the two geometric graphs $G_1$ and $G_2$ is defined as 
 \begin{equation}
  GD(G_1, G_2) = VD(G_1, G_2) + ED(G_1, G_2). 
 \end{equation}
\end{defn}
 $GD$ combines both $VD$ and $ED$. We call the first component in the definition of $GD$ as $VD$-term and the second component as $ED$-term. 
% theorem 1
 \begin{thm}
 $GD(G_1, G_2)$ satisfy the following properties: \\
 (i) $GD(G_1, G_2) \geq 0$, if $G_1 = G_2$ then $GD(G_1, G_2) = 0$ \\
 (ii) $GD(G_1, G_2) = GD(G_2, G_1)$ \\
 (iii) $GD(G_1, G_2) \leq GD(G_1, G_3) + GD(G_3, G_2)$
\end{thm}
The above follows from Lemma \ref{lem:lemma1} and Lemma \ref{lem:lemma2}. Since both $VD$-term and $ED$-term separately satisfy the above properties. \\

%When we use $EDM$ instead of $ED$ in the definition of $GD$, we call it as Graph Distance Modified ($GDM$). Therefore, $GDM(G_1, G_2) = VD(G_1, G_2) + EDM(G_1, G_2)$.

\begin{defn}
 %\textbf{Definition 5.}
 Let $G_1= (V_1,E_1,c_1)$ and $G_2= (V_2,E_2,c_2)$ be two geometric graphs with $|V_1|=|V_2|=n$ and $|E_1|=|E_2|=m$. Then the Graph Distance Modified  or Graph Distance Metric ($GDM$) between the two geometric graphs $G_1$ and $G_2$ is defined as 
 \begin{equation}
  GDM(G_1, G_2) = VD(G_1, G_2) + EDM(G_1, G_2). 
 \end{equation}
\end{defn}

In the above definition of graph distance modified we use $EDM$ instead of $ED$ for the computation of edge distance between two graphs.

% theorem 2 
\begin{thm}
 $GDM(G_1, G_2)$ is a metric over set of all geometric graphs $\mathcal{G}$.
\end{thm}

\begin{proof}
 From Lemma \ref{lem:lemma1} and Lemma \ref{lem:lemma3}, we observe that both $VD$-term and $EDM$-term of $GDM$ satisfy properties (ii), (iii) of metric. Now, we consider property (i): \\ 
 Here certainly, $GDM(G_1, G_2) \geq 0$ since $VD(G_1, G_2) \geq 0$ and $EDM(G_1, G_2) \geq 0$. \\
 If $G_1 = G_2$ then $VD(G_1, G_2) = 0$ and $EDM(G_1, G_2) = 0$ and therefore $GDM(G_1, G_2) = 0$. \\
 Now if $GDM(G_1, G_2) = 0$ then both $VD$-term and $EDM$-term are 0. When $VD$-term is 0, i.e., $VD(G_1, G_2) = 0$, then from Lemma 1, $G_1 = G_2$ without considering the links, in other words positions of all matched vertices between $G_1$ and $G_2$ coincide. When $EDM$-term is 0, then from Lemma 3, $E_{ij}^{A} = E_{ij}^{L} = E_{ij}^{P}=0$ and therefore, angular orientation, length and positions of each matched edges are exactly the same. By combining the above two facts we get $G_1 = G_2$, whenever $GDM(G_1, G_2) = 0$.
\end{proof}

\section{Geometric Graph Matching}

\subsection{Exact Geometric Graph Matching}

In this section, we describe an algorithm to check whether two geometric graphs are isomorphic. Before checking the isomorphism between two geometric graphs, their reference coordinate must be identical. Therefore, first, we carry out graph alignment of two input graphs so that their reference coordinates become identical. Steps to perform graph alignment of $G_2$ with respect to $G_1$ is described in Algorithm \ref{algorithm:graph-allignment}. The input to this algorithm is two geometric graphs $G_1$ and $G_2$ and output is the geometric transform of $G_2$ having the maximum number of edges having similar orientation to $G_1$. Line 1 of this algorithm selects an edge $e$ of $E_1$ to be a reference axis. The \textit{for} loop in line 2 of the algorithm choose an edge $f$ of $E_2$ to perform geometric-transform on $G_2$. The function geometric-transform process the coordinates of $G_2$ using a transformation matrix to make edge $f$ to have same orientation to reference axis $e$ with the left coordinate of edge $f$ coincident to left coordinate of $e$. This particular geometric configuration of $G_2$ will be translation and rotation invariant to $G1$. To make this geometric configuration scaling invariant also, we make the edges $e$ and $f$ of unit length and uniformly scale the other edges of $G_1$ and $G_2$. Line 4 of the algorithm computes $ED$ between $G_1$ and $G_2'$ and in line 6, the algorithm updates $d_{min}$ to find the minimum $ED$ over all the geometric configuration of $G_2$. Line 10 of the algorithm returns the graph $G$ corresponding to minimum $ED$ value.

\begin{algorithm}
\caption{\bf:Graph-Alignment$(G_1,G_2)$} \label{algorithm:graph-allignment}
\begin{algorithmic}[1]
\Require Two undirected geometric graphs $G_1$, $G_2$, where $G_1 = (V_1, E_1, c_1), G_2 = (V_2, E_2, c_2)$ with $|V_1|=|V_2|=n, |E_1|=|E_2|=|m|$ 
\Ensure Geometric transform of $G_2$ with maximum edges aligned to $G_1$ 

     \State Select an edge $e \in E_1$ as a reference axis
     \For {(each edge $f \in E_2$)}
         \State $G_2' \leftarrow$ geometric-transform$(G_2)$ 
         \State $d \leftarrow ED(G_1, G_2')$
          \If {$(d < d_{min})$}
           \State $d_{min} \leftarrow d$
           \State $G \leftarrow G_2'$
          \EndIf 
      \EndFor  \\
    \Return $(G)$
   \end{algorithmic}
\end{algorithm}

Test to perform geometric graph isomorphism between two graphs is presented in Algorithm \ref{algorithm:geometric-graph-iso}. The input to this algorithm is two geometric graphs $G_1$ and $G_2$. Output to this algorithm is isomorphic graphs when $G_1$ is isomorphic to $G_2$; $t$-tolerant isomorphic graphs when $G_1$ is $t$-tolerant isomorphic to $G_2$; and graph distance $GD$ when $G_1$ is neither isomorphic to $G_2$ nor $t$-tolerant isomorphic to $G_2$. Line 1 of Geometric-Graph-Isomorphism algorithm calls Algorithm \ref{algorithm:graph-allignment} to perform graph alignment of $G_2$ with respect to $G_1$. Vertex assignments from $G_1$ to $G_2$ using the definition of $VD$ is given in line 2. We can use the Hungarian algorithm for optimal assignment of vertices. Based on the assignment $VD$ is calculated in line 3. Angle and length component of $ED$ is initialized in lines 4--5. Line 7 computes $ED$ based on edge assignments from $E_1$ to $E_2$. Final $GD$ value is calculated in line 8. The \textit{if} loop in line 9 checks whether $GD$ is 0, and line 10 also check uniqueness of edge assignments which ensures that each assigned edge from $G_1$ to $G_2$ connects the same set of coordinate points there by implying a valid geometric graph isomorphism. The algorithm then returns in line 11, indicating that $G_1$ is isomorphic to $G_2$. The \textit{for} loop in the lines 13--15 of algorithm checks if all the assigned coordinates from $G_1$ to $G_2$ are within distance $t$, and again it checks uniqueness of edge assignments, it then returns in line 17 showing that $G_1$ is $t$-tolerant isomorphic to $G_2$. When $GD$ is neither 0 nor the assigned coordinates form $G_1$ to $G_2$ are within specified distance $t$, the algorithm outputs the $GD$ value in line 19. 

\begin{algorithm}
\caption{\bf:Geometric-Graph-Isomorphism$(G_1,G_2)$} \label{algorithm:geometric-graph-iso}
\begin{algorithmic}[1]
\Require Two undirected geometric graphs $G_1$, $G_2$, where $G_1 = (V_1, E_1, c_1), G_2 = (V_2, E_2, c_2)$ with $|V_1|=|V_2|=n, |E_1|=|E_2|=|m|$ 
\Ensure Isomorphic graphs or $t$-tolerant isomorphic graphs or graph distance between $G_1$ and $G_2$ 
    
    \State $G_2 \leftarrow$ Graph-Alignment$(G_1,G_2)$ %\Comment{preprocessing step}
    \State Compute vertex assignment from $V_1$ to $V_2$
    \State $VD \leftarrow  \sum_{i=1}^n \sqrt{(a_i-x_{\varphi(i)})^2 + (b_i-y_{\varphi(i)})^2}$
    \State $E_{ij}^{A} \leftarrow \sqrt{((\varTheta_{ij}-\varTheta^{'}_{ij})\frac{\pi}{180^{\circ}})^2}$ 
    \State $E_{ij}^{L} \leftarrow \sqrt{(d_{ij}-D_{ij})^2}$
    \State Compute edge assignment from $E_1$ to $E_2$
    \State $ED \leftarrow  \sum_{i=1}^n ( E_{i \varphi(i)}^{A} + E_{i\varphi(i)}^{L} ) $
    \State $GD \leftarrow  VD + ED$ 
    \If {$(GD==0)$ 
     \State Check uniqueness of edge assignments} \\
     \Return $G_1$ is isomorphic to $G_2$
      \ElsIf { \For {each $i=1 \text{ to } n$}
       \State $(a_i-x_i)< t$ \textbf{and} $(b_i-y_i)<t$ 
       \EndFor 
       \State Check uniqueness of edge assignments} \\
     \Return $G_1$ is $t$-tolerant isomorphic to $G_2$
     \Else \\
    \Return $(GD)$
    \EndIf 
   \end{algorithmic}
\end{algorithm}

\begin{prop}
 Graph-Alignment algorithm executes in $O(m.n^3)$ time.
\end{prop}
The geometric-transform operation in line 3 of Algorithm \ref{algorithm:geometric-graph-iso} takes $O(n^2)$ time, once a transformation matrix is constructed. The $EDM$ operation in line 4 is performed in $O(n^3)$ time, hence total time taken by this algorithm is $O(m.n^3)$.  

\begin{prop}
 Geometric-Graph-Isomorphism algorithm executes in $O(n^3 + m^3)$ time exclusive of Algorithm \ref{algorithm:graph-allignment} in line 1.
\end{prop}
 The assignment of vertices in line 2 can be performed in $O(n^3)$ using Munkres or Hungarian algorithm. Similarly, the assignment of edges in line 6 can be achieved in $O(n^3)$. Remaining steps can be computed in $O(n^2)$ time. Therefore overall execution time becomes $O(n^3 + m^3)$ time.

\subsection{Error-Tolerant Geometric Graph Matching}
%\subsection{Geometric Graph Distance Algorithm}
The computation of graph distance between two geometric graphs $G_1$ and $G_2$ is described in Algorithm \ref{algorithm:geometric-gd}. The input to the algorithm is two undirected geometric graphs $G_1$ and $G_2$ and four weighting parameters $w_1,w_2,w_3$ and $w_4$, which are application dependent. $G_1$ has $n_1$ vertices and $m_1$ edges, whereas $G_2$ has $n_2$ vertices and $m_2$ edges. For simplicity, we have assumed that $n_1 \geq n_2$. The output of the algorithm is the geometric graph distance between $G_1$ and $G_2$.  %If one graph is identical to other by performing geometric transformation like translation, rotation, and scaling, then the input graphs are processed to make their coordinate reference frame aligned.

When the number of vertices in $|V_1|=n_1$ is greater than that of $|V_2|=n_2$ then line 1 of the Geometric-Graph-Distance algorithm computes the mean values of $x$ and $y$ coordinate of graph $G_2$ as $x=(\sum_{i=1}^{n_2} x_i)/n_2$, $y=(\sum_{i=1}^{n_2} y_i)/n_2$ and $n_1-n_2$ vertices with these coordinate positions are appended to $V_2$ in line 4 of the \textit{for} loop in lines 3--5 to make $|V_1|=|V_2|=n$. Similarly, \textit{if} loop in lines 7--11 check the number of edges in $G_1$ and $G_2$ and it makes them equal by appending empty edges with length and angle value both 0 to $E_2$ or $E_1$ depending on whether $m_1 > m_2$ or $m_1 < m_2$ respectively. 

\begin{algorithm}
\caption{\bf:Geometric-Graph-Distance$(G_1,G_2)$} \label{algorithm:geometric-gd}
\begin{algorithmic}[1]
\Require Two undirected geometric graphs $G_1$, $G_2$, where $G_1 = (V_1, E_1, c_1), |V_1|=n_1, |E_1|=m_1, G_2 = (V_2, E_2, c_2), |V_2|=n_2, |E_2|=|m_2|$ and weighting factors $w_i$ for $i=1$ to $4$; assume $n_1 \geq n_2$
\Ensure Geometric graph distance between $G_1$ and $G_2$ 
    
    \State $(x,y) \leftarrow ((\sum_{i=1}^{n_2} x_i)/n_2, (\sum_{i=1}^{n_2} y_i)/n_2)$
    \If {$(n_1 > n_2)$}
     \For {$i=0 \text{ to } (n_1-n_2)$}
       \State $V_2 \leftarrow V_2 \cup \{(x, y)\}$ 
     \EndFor
    \EndIf 
    \If {$(m_1 > m_2)$}
     \State Append $(m_1-m_2)$ empty edges to $E_2$ %of length 0 and angle 0
     \Else
     \State Append $(m_2-m_1)$ empty edges to $E_1$%of length 0 and angle 0
    \EndIf 
    \State $G_2 \leftarrow$ Graph-Alignment-Modified$(G_1,G_2)$ \Comment{preprocessing step}
    \State Compute vertex assignment $\varphi$ from $V_1$ to $V_2$
    \State $VD \leftarrow  \sum_{i=1}^n \sqrt{(a_i-x_{\varphi(i)})^2 + (b_i-y_{\varphi(i)})^2}$
    \State Compute edge assignment $\varphi$ from $E_1$ to $E_2$
    \State $E_{ij}^{A} \leftarrow \sqrt{((\varTheta_{i}-\varTheta^{}_{j})\frac{\pi}{180^{\circ}})^2}$; $E_{ij}^{L} \leftarrow \sqrt{(d_{i}-D_{j})^2}$ 
    \State $E_{ij}^{P} \leftarrow \sqrt{(a_i-x_{j})^2 + (b_i-y_{j})}^2 + \sqrt{(a_{p}-x_{q})^2 + (b_{p}-y_{q})}^2$
    \State $E_{ij}^{P} \leftarrow (E_{ij}^{P})/2$
    \State $EDM \leftarrow  \sum_{i=1}^n (w_2 \cdot E_{i \varphi(i)}^{A} + w_3 \cdot E_{i\varphi(i)}^{L} + w_4 \cdot E_{i\varphi(i)}^{P}) $
    \State $GDM \leftarrow w_1 \cdot VD + EDM$ \\
    \Return $(GDM)$
   \end{algorithmic}
\end{algorithm}

One optional step of this algorithm in line 12 is the preprocessing phase for graph alignment to make their reference coordinates identical. 
Geometric-Graph-Distance algorithm computes vertex assignment from $V_1$ to $V_2$ based on the cost function given in the definition of $VD$ in line 13. We can use Munkres or Hungarian algorithm for optimal assignment of vertices. Based on the assignment $VD$ is calculated in line 14. Similarly, line 15 computes edge assignment from $E_1$ to $E_2$ using the cost function given in the definition of $EDM$. Line 19 computes $EDM$ by multiplying weighting factor $w_2$ to orientation term $E_{ij}^{A}$, $w_3$ to length term $E_{ij}^{L}$ and $w_4$ to position term $E_{ij}^{P}$. Finally, line 20 computes $GDM$ by multiplying $w_1$ to $VD$ and adding it to $EDM$.     

Steps to perform graph alignment of $G_2$ with respect to $G_1$ is described in Graph-Alignment-Modified algorithm. This algorithm is same as Graph-Alignment algorithm except that it calls $EDM(G_1, G_2')$ instead of $ED(G_1, G_2')$ in line 4 to find the minimum value of $d$.

\begin{algorithm}
\caption{\bf:Graph-Alignment-Modified$(G_1,G_2)$} \label{algorithm:graph-align-mod}
\begin{algorithmic}[1]
\Require Two undirected geometric graphs $G_1$, $G_2$, where $G_1 = (V_1, E_1, c_1), |V_1|=n_1, |E_1|=m_1, G_2 = (V_2, E_2, c_2), |V_2|=n_2, |E_2|=|m_2|$ 
\Ensure Geometric transform of $G_2$ having maximum edges and vertices aligned to $G_1$ 

     \State Select an edge $e \in E_1$ as a reference axis
     \For {(each edge $f \in E_2$)}
         \State $G_2' \leftarrow$ geometric-transform$(G_2)$ 
         \State $d \leftarrow EDM(G_1, G_2')$
          \If {$(d < d_{min})$}
           \State $d_{min} \leftarrow d$
           \State $G \leftarrow G_2'$
          \EndIf 
      \EndFor  \\
    \Return $(G)$
   \end{algorithmic}
\end{algorithm}

\begin{prop}
 Geometric-Graph-Distance algorithm executes in $O(n^3 + m^3)$ time.
\end{prop}
 The assignment of vertices in line 13 can be performed in $O(max\{n_1, n_2 \}^3)$ using Munkres or Hungarian algorithm. Similarly, the assignment of edges in line 15 can be achieved in $O(max\{m_1, m_2 \}^3)$. Remaining steps can be computed in $O(n^2)$ time. Let $n=max(n_1, n_2)$ and $m=max(m_1, m_2)$. Therefore overall execution time becomes $O(n^3 + m^3)$ time.
%We can observe that the assignment of vertices and edges in lines 1--2 can be performed in $O(n^3)$ by Munkres algorithm and the remaining steps can be computed in $O(n^2)$, therefore overall execution time remains $O(n^3)$.

\begin{prop}
 Graph-Alignment-Modified algorithm executes in $O(m_2.n^3)$ time.
\end{prop}
The \textit{for} loop in line 2 of Algorithm \ref{algorithm:graph-align-mod} executes $m_2$ times, where $|E_2|=|m_2|$. The geometric-transform operation in line 3 takes $O(n_2^2)$ time; once a transformation matrix is constructed, where $|V_2|=n_2$. The $EDM$ operation in line 4 is performed in $O(n^3)$ time, where $n=max(n_1, n_2)$. Hence total time taken by this algorithm is $O(m_2.n^3)$.

\section{Experimental Evaluation}

\subsection{Exact Geometric Graph Matching}
To assess the effectiveness Geometric-Graph-Isomorphism algorithm, we apply it for exact graph matching. We use letter dataset of IAM graph database repository for the demonstration of exact graph matching \cite{RiesenBunke2008}. 
%Letter dataset contains capital letters of the English alphabet, which can be written using straight lines only. It consists of 15 classes of letters such as A,E,F,H,J,K,L,M,N,T,V,W,X,Y, and Z. For each class, various graphs are generated by applying different distortion level of high, medium, and low magnitude. 
Each node in letter graphs is labeled with an $(x, y)$ coordinate to represent its location in the two-dimensional plane.

% Make equal edges
Depending on three \textit{if} loop conditions in the Geometric-Graph-Isomorphism algorithm, there are three cases to be considered. In the first case, we take several geometric graphs, apply composite geometric transformation using the translation, rotation, scaling and their combination to these sample graphs and verify whether the concerned pair is geometrically isomorphic using Geometric-Graph-Isomorphism algorithm. In the second case, we apply random distortion within a specified range to the vertices of the sample graphs and check whether they are $t$-tolerant isomorphic with respect to each other. Finally, in the third case, we apply the proposed algorithm to low, medium and high distortions of letter dataset to compute the graph matching, and compare their relative graph distances.

In the first set of experiments, we take letter graphs from each of low, medium and high class and apply an arbitrary transformation to the coordinate of vertices. We then apply the Geometric-Graph-Isomorphism algorithm to each pair of the original and transformed graph to verify the isomorphism. Since the Graph-Alignment algorithm makes the reference coordinate of two graphs identical, it undoes the transformations applied on the processed letter graph to output the two graphs to be isomorphic in each test. 

In the second set of experiments, we select letter graphs from each of low, medium, and high class, and apply random distortions from $0$ to $t$ distance on each coordinate of vertices of the graph. We check $t$-tolerant isomorphism for each pair of graphs for four different values of $t$ which are $t/2, t, 3t/2, 2t$. Comparison of accuracy of $t$-tolerant isomorphism for different values of $t$, when Geometric-Graph-Isomorphism algorithm is applied to letter graphs of low, medium and high distortion classes is shown in Figure ~\ref{fig:tolerant-geometric-graph-somorphism}. Here we observe that for the value of $2t$ every pair of graphs is correctly identified to be isomorphic.

% Fig.4
\begin{figure}[!t]
\centering
\includegraphics[scale=.6]{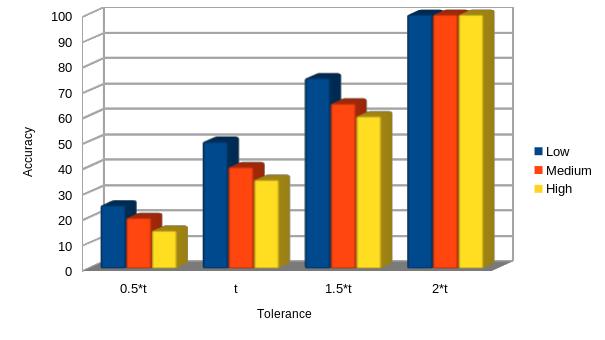}
\caption{$t$-tolerant geometric graph isomorphism}
\label{fig:tolerant-geometric-graph-somorphism}
\end{figure} 
 
In the third set of experiments, we apply Geometric-Graph-Isomorphism algorithm on each letter dataset to perform exact graph matching. For this experiment, we select graphs of equal vertices and edges. Based on the $GD$ value returned by the algorithm, we compute the classification accuracy of letter dataset using the nearest neighborhood classifier. Classification accuracy of each of the fifteen letters of letter graphs of low distortion class is shown in Figure \ref{fig:accuracy-low}.  
 
% Fig.5
\begin{figure}[!t]
\centering
\includegraphics[scale=.5]{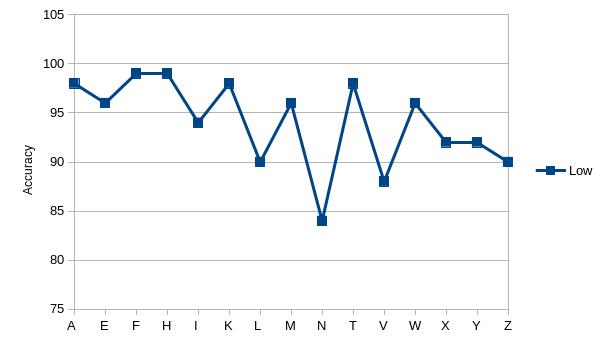}
\caption{Accuracy Low}
\label{fig:accuracy-low}
\end{figure}  

Similarly, classification accuracy of each letters for medium and high distortion class of letter dataset is shown in Figure \ref{fig:accuracy-medium} and Figure \ref{fig:accuracy-high} respectively.

% Fig.6
\begin{figure}[!t]
\centering
\includegraphics[scale=.5]{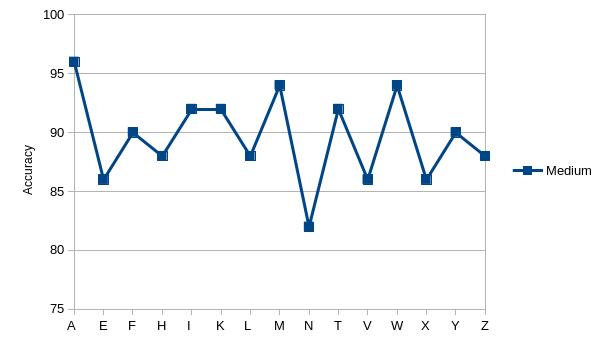}
\caption{Accuracy Medium}
\label{fig:accuracy-medium}
\end{figure}  
 
% Fig.7
\begin{figure}[!t]
\centering
\includegraphics[scale=.5]{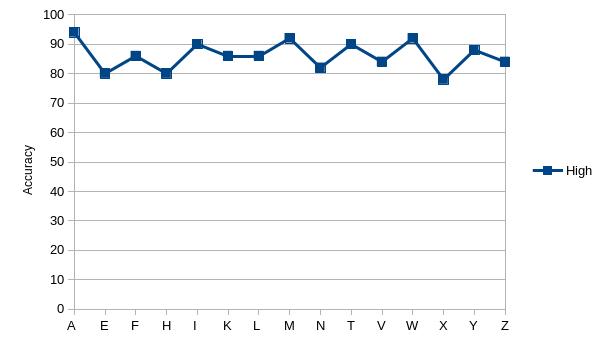}
\caption{Accuracy High}
\label{fig:accuracy-high}
\end{figure}  

The comparison of classification accuracy of each of the low, medium, and high class of letter dataset is shown in Figure \ref{fig:accuracy-ggi}. Here we observe that the accuracy of letter dataset of the low class is higher than that of medium class, which in turn is higher than that of letter dataset of high distortion.

% Fig.8
\begin{figure}[!t]
\centering
\includegraphics[scale=.5]{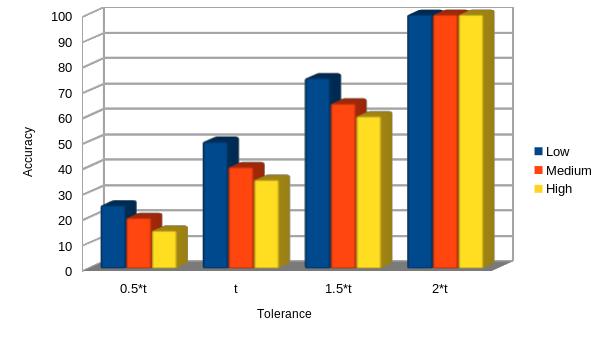}
\caption{Accuracy}
\label{fig:accuracy-ggi}
\end{figure}

\subsection{Error-Tolerant Geometric Graph Matching}

To evaluate the usefulness of the geometric graph similarity framework, we apply it for inexact graph matching task. In this section, we use Geometric-Graph-Distance algorithm for geometric graph matching. We use the letter dataset and AIDS dataset of IAM graph database repository \citep{RiesenBunke2008} for comparison of execution time and accuracy of geometric graph matching with other important graph matching algorithms. Each node in letter graphs is labeled with an $(x, y)$ coordinate to represent its location in the two-dimensional plane. Every node of AIDS graphs also has its associated coordinate position.

\subsubsection{Execution Time Comparison}
Results in this section are computed using a system having 9.8 GB of memory and running the processor at 3.40 GHz. For comparison purpose, we use $A^{*}$ based graph edit distance, beamsearch \citep{Neuhausetal2006} with beam width $s=10$, and bipartite graph matching \citep{RiesenBunke2009}. Comparison of the average running time of graph matching methods in milliseconds for geometric graph matching, beamsearch having beam width $w=10$, and GED is shown in Table \ref{table:comp-avg-time}. 

% Table 1
\begin{table}[!t]
\renewcommand{\arraystretch}{1.0}
\caption{Comparison of average computation time in $ms$}\label{table:comp-avg-time}
\label{table }
\begin{center}
\begin{tabular}{|c| c| c| c|} 
\hline
Algorithm & $T_{Letter}$ & Algorithm & $T_{AIDS}$  \\ [1ex] \hline\hline 
Geometric & 5.7 & Geometric & 1.39 \\ \hline
Beam & 4.5 & Beam & 6.48  \\ \hline
GED & 8.1 & Bipartite & 12.11 \\ [1ex] \hline

\end{tabular} 
\end{center}
\end{table} 

We may not use $A^{*}$ based graph edit distance on AIDS dataset, as it usually becomes infeasible beyond 10 to 20 nodes. Comparison of the average execution time of graph matching in milliseconds for geometric graph matching, beamsearch and bipartite graph matching is shown in Table \ref{table:comp-avg-time}. On an average geometric graph matching takes less execution time than either beamsearch optimization or bipartite graph matching heuristic. 

\subsubsection{Accuracy Comparison}
To compare the $GDM$ computed by Algorithm \ref{algorithm:geometric-gd} to the GED for the same pair of graphs, we compute these values for the first few letter graphs from each of high, medium and low distortion letter graphs as shown in Figure \ref{fig:gdm-vs-ged-high}, Figure \ref{fig:gdm-vs-ged-medium}, and Figure \ref{fig:gdm-vs-ged-low} respectively. We note that $GDM$ and GED values are more similar in low distortion letter graphs differing only by a shift as compared to high distortion letter graphs. It implies that in lower distortion, the two distance measures seems to correlate more than that with higher distortion levels. Also, the accuracy for low distortion letter dataset is higher than the accuracy for high distortion dataset. It shows that the proposed model is more accurate for unmodified graphs.
% Fig.3
\begin{figure}[!t]
\centering
\includegraphics[scale=.15]{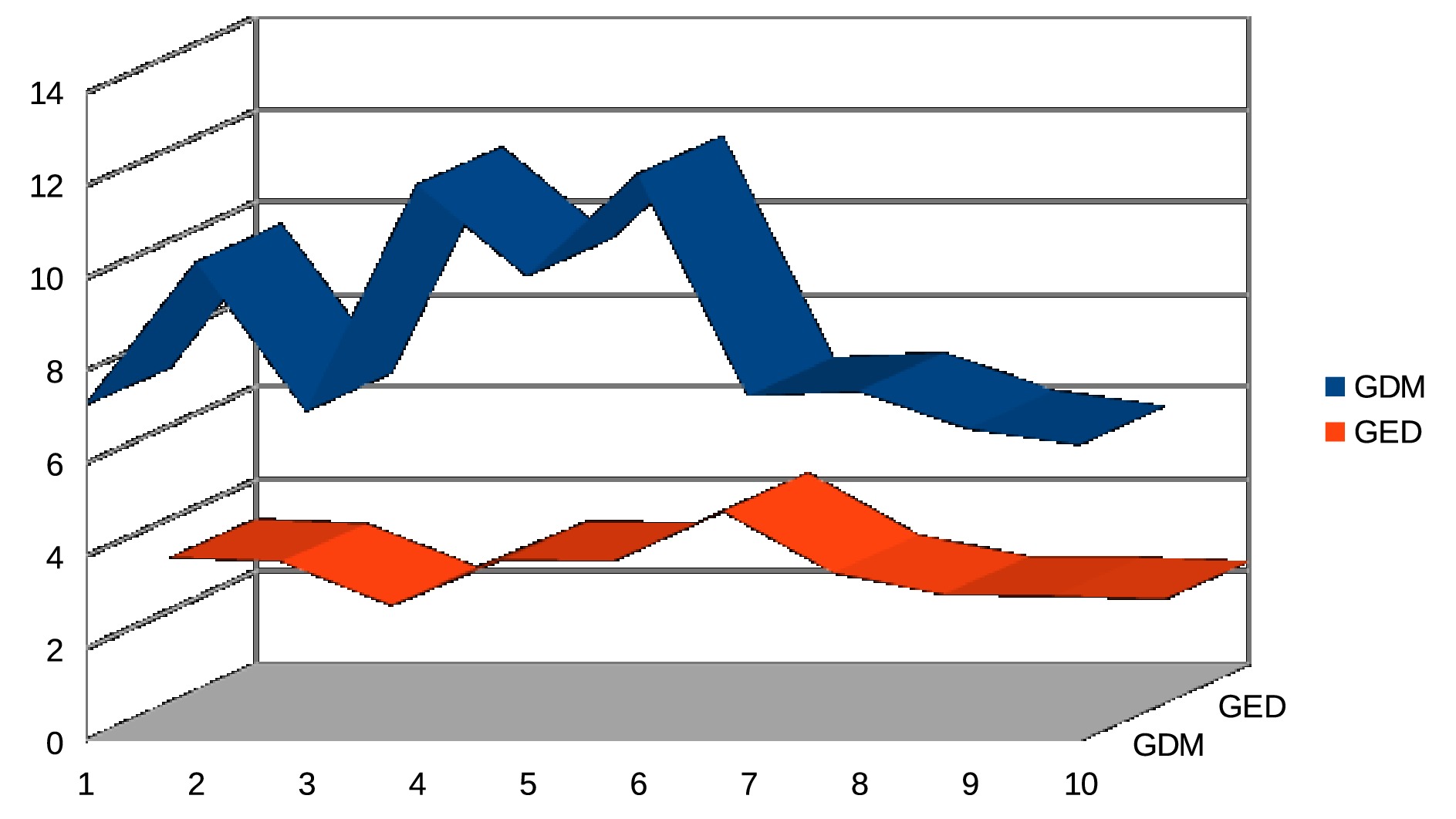}
\caption{GDM vs GED for high distortion letter graphs}
\label{fig:gdm-vs-ged-high}
\end{figure}
% Fig.4
\begin{figure}[!t]
\centering
\includegraphics[scale=.15]{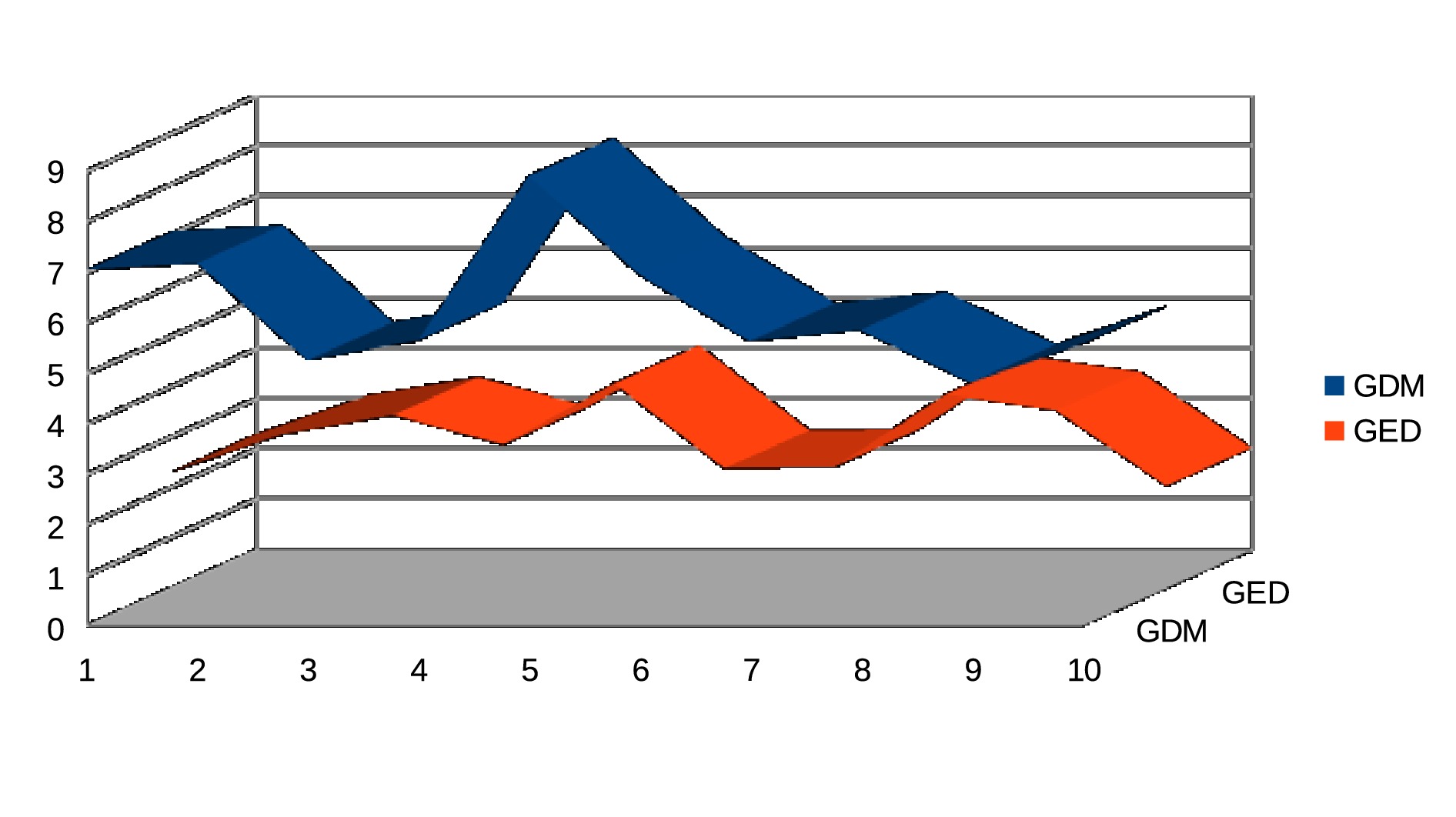}
\caption{GDM vs GED for medium distortion letter graphs}
\label{fig:gdm-vs-ged-medium}
\end{figure}
% Fig.5
\begin{figure}[!t]
\centering
\includegraphics[scale=.15]{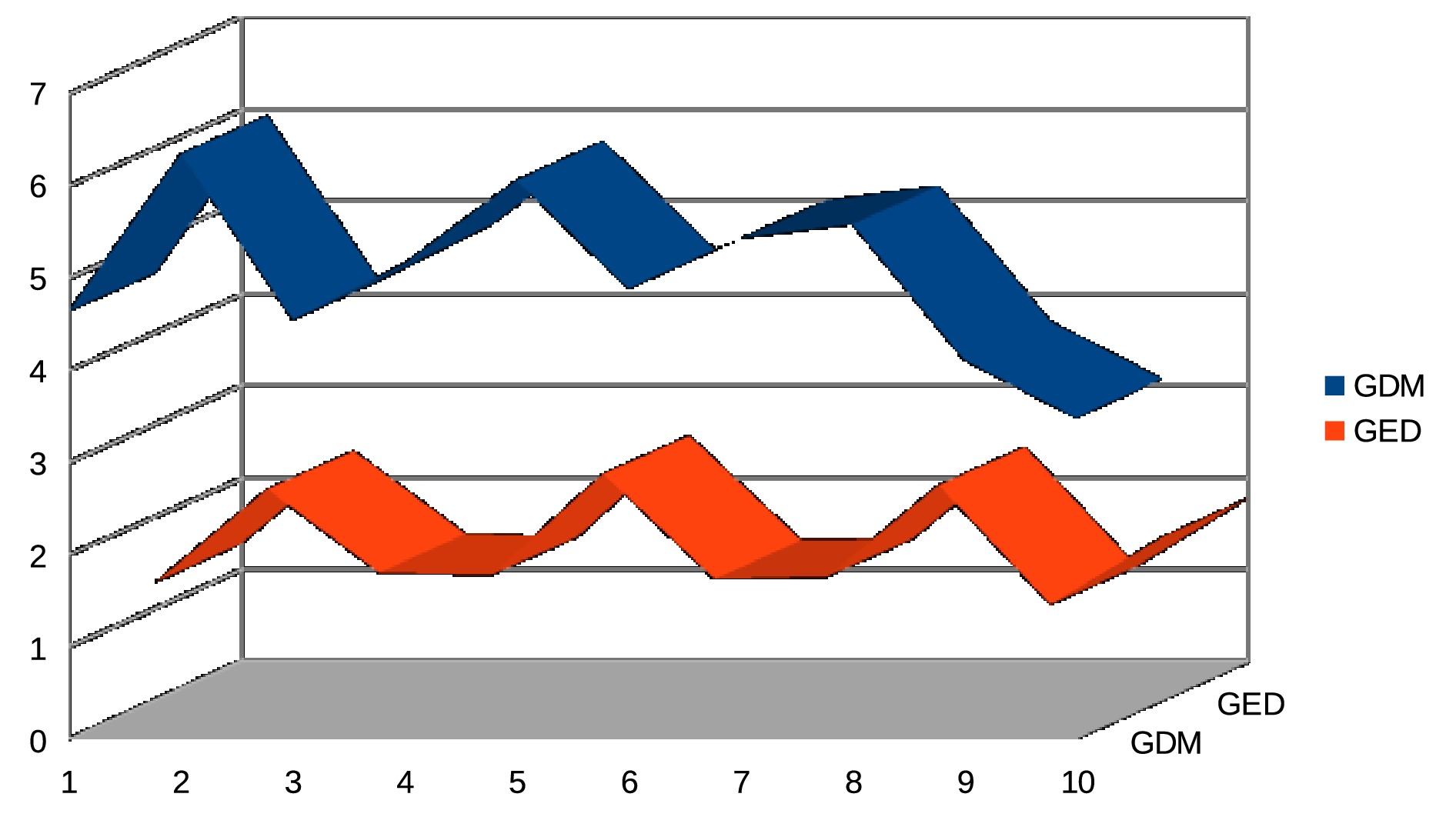}
\caption{GDM vs GED for low distortion letter graphs}
\label{fig:gdm-vs-ged-low}
\end{figure}

For accuracy evaluation of geometric graph matching, we consider the problem of classification of letter graphs and AIDS graphs using nearest neighbor classifier. For accuracy evaluation of letter dataset, we use high distortion letter graphs consisting of 50 training, validation and test graphs for each of the 15 classes of letters. We have trained the proposed model using the training set, optimized the weight parameters using the validation set and computed the classification accuracy using the test set. We use steepest-ascent search in the parameter space to find the weight parameters that are as well performing on the validation set as possible. The best weight parameter obtained are $w_1=0.35$, $w_2=0.23$, $w_3=0.11$, and $w_4=0.31$.

Table \ref{table:accuracy-on-letter-using-vd-gd} shows the accuracy ratio for each of the 15 letters of high distortion letter graphs using Algorithm \ref{algorithm:geometric-gd}. For comparison we have shown accuracy using $VD$, $ED$ and $EDM$ separately along with $GD$ and $GDM$ computed using Algorithm \ref{algorithm:geometric-gd}. We observe that accuracy using $GDM$ is always higher than that of $VD$ and $EDM$ individually; similarly, the accuracy using $GD$ is more than that of $VD$ and $ED$ separately. We also note that the average accuracy of $GDM$ is higher than that of $GD$ for the computation of graph distance in Algorithm \ref{algorithm:geometric-gd}.

To compare the accuracy of geometric graph matching with other important graph matching techniques, we apply these algorithms to letter dataset with high distortion. The average classification accuracy achieved using geometric $GDM$, beamsearch and GED are 89.06, 84.40 and 90 respectively. Comparison of the accuracy of letter dataset using geometric $GDM$, beamsearch and, GED for every individual letter is shown in Figure \ref{fig:comparison-accuracy-letter}.
%In this table, we notice that accuracy using $GDM$ instead of $GD$ in Algorithm 1 is slightly lower for some letter graphs because the third term $E_{ij}^{P}$ may not add more advantage to the definition of $EDM$ as it is similar to $VD$ and position of vertices are already being considered in $VD$. Comparison of the accuracy of letter dataset using geometric $GD$, beamsearch and, GED are shown in Fig.6. %Here, we observe that the accuracy ratio using geometric $GM$ is usually more than beamsearch and GED.
% Table 2
\begin{table}[!t]
\renewcommand{\arraystretch}{1.0}
\caption{Accuracy on letter dataset using $VD$, $ED$, $EDM$, $GD$ and $GDM$}\label{table:accuracy-on-letter-using-vd-gd}
\label{table }
\begin{center}
\begin{tabular}{|c| c| c| c| c| c|} 
\hline
 Class & $VD$ & $ED$ & $EDM$ & $GD$ & $GDM$ \\ [1ex] \hline\hline 
A & 84 & 80 & 84 & 98 & 98 \\ \hline
E & 36 & 78 & 80 & 80 & 82 \\ \hline
F & 72 & 64 & 70 & 86 & 88 \\ \hline
H & 38 & 72 & 74 & 82 & 84 \\ \hline
I & 96 & 76 & 78 & 96 & 94 \\ \hline
K & 42 & 74 & 82 & 92 & 94 \\ \hline
L & 84 & 52 & 60 & 88 & 90 \\ \hline
M & 38 & 88 & 88 & 96 & 96 \\ \hline
N & 34 & 64 & 68 & 84 & 84 \\ \hline
T & 84 & 62 & 74 & 90 & 92 \\ \hline
V & 74 & 50 & 58 & 88 & 88 \\ \hline
W & 38 & 82 & 82 & 94 & 96 \\ \hline
X & 30 & 52 & 66 & 70 & 72 \\ \hline
Y & 46 & 56 & 64 & 84 & 88 \\ \hline
Z & 40 & 68 & 72 & 88 & 90 \\ [1ex] \hline

\end{tabular} 
\end{center}
\end{table} 

% Fig.6
\begin{figure}[!t]
\centering
\includegraphics[scale=.15]{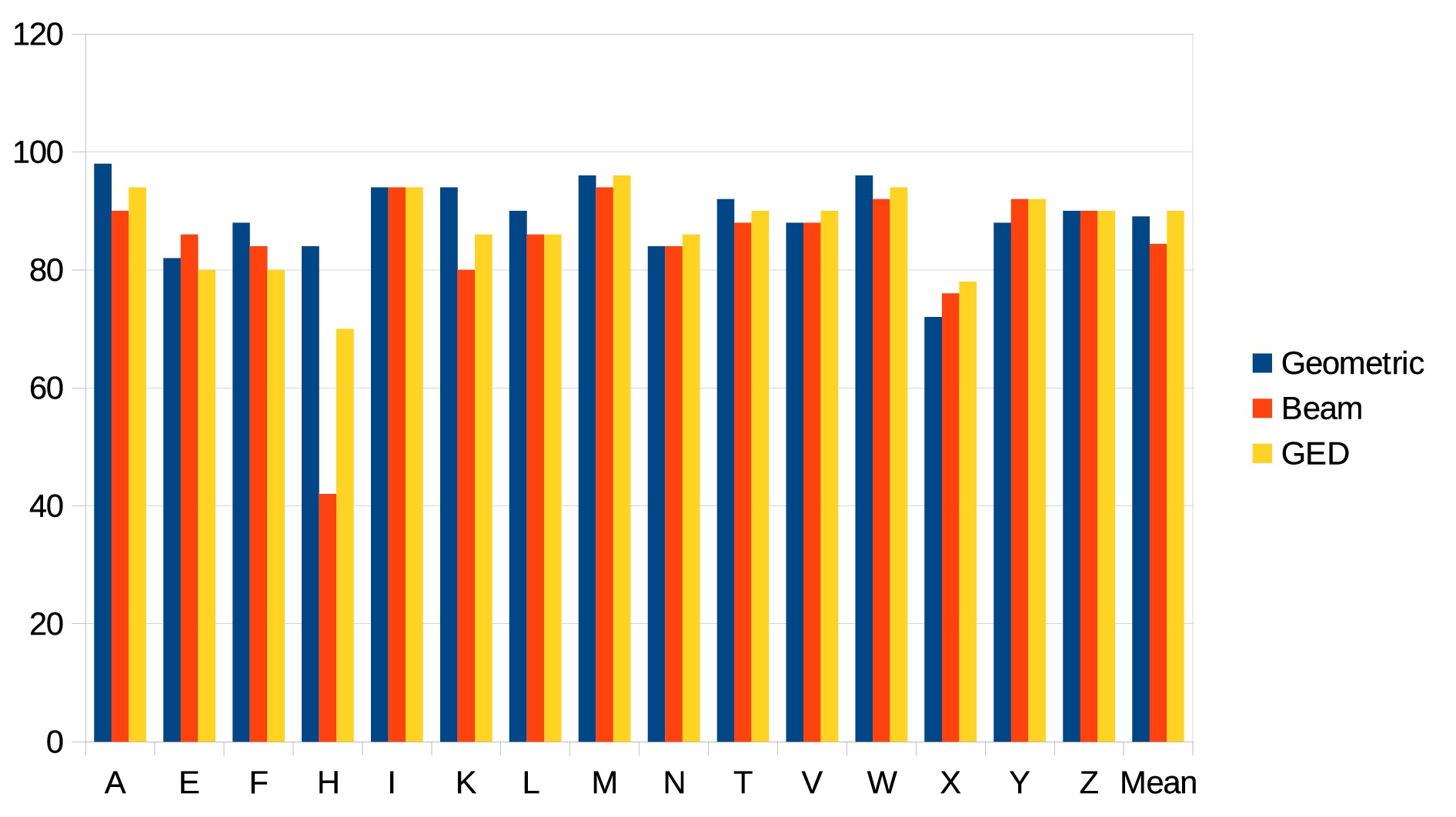}
\caption{Comparison of accuracy on letter dataset}
\label{fig:comparison-accuracy-letter}
\end{figure}

AIDS training dataset consists of 250 graphs, out of which 50 graphs are from active class and the remaining 200 from inactive class. Test dataset includes 1500 graphs, having 300 graphs from active and rest 1200 graph are from inactive class. The classification accuracy of AIDS dataset using geometric GM, beamsearch and bipartite GM is shown in Table \ref{table:acuracy-on-aids}. We can observe that recognition rate of confirmed active molecules using geometric GM is more than both of beamsearch and bipartite GM, whereas classification rate of inactive molecules is less than that of beamsearch and bipartite GM.   

% Table 3
\begin{table}[!t]
\renewcommand{\arraystretch}{1.0}
\caption{Accuracy on AIDS dataset}\label{table:acuracy-on-aids}
\label{table }
\begin{center}
\begin{tabular}{|c| c| c|} 
\hline
Algorithm used & Active & Inactive   \\ [1ex] \hline\hline 
Geometric & 99.33 & 98.50 \\ \hline
Beam & 98.33 & 99.91 \\ \hline
Bipartite & 98.66 & 99.91 \\ [1ex] \hline

\end{tabular} 
\end{center}
\end{table}

The proposed graph distance measure can be used to compare the structural similarity between different graphs. It can be particularly useful in real-world applications, where the graph data is large and can be modified by noise. Depending on application requirement, we can select weighting factors such that $\sum_{i=1}^n w_i=1$. When the position of vertices is more dominant, we can select $w_1$ to be higher, if angular structures are more important then $w_2$ can be prominent. Otherwise, if edge differences are more essential, we can select $w_3$ to be higher. We can either use equal weight factors or train them to optimize the classification ratio. In this section, we have trained the weight parameters for the computation of $GD$ and $GDM$ in the Geometric-Graph-Distance algorithm.  

\section{Summary}
In this chapter, we described an intuitive approach to compute similarity between two geometric graphs and used it to perform exact and error-tolerant graph matching. In a geometric graph, every vertex has an associated coordinate, which specify its distinct position in the plane. We use this fact to define the \textit{VD} between the two graphs. We defined \textit{EDM} between the two graphs using the angular, edge length, and position components between corresponding edges of two graphs. Finally, we combined \textit{VD} and \textit{EDM} to form \textit{GDM} and used it to compute error-tolerant graph matching between two graphs. The experimental results show that this graph matching approach can be promising to graph dataset in which every node has a coordinate position in a two-dimensional plane. 
 
% Chapter Template

\chapter{Conclusions and Future Work} % Main chapter title

\label{Chapter6} % Change X to a consecutive number; for referencing this chapter elsewhere, use \ref{ChapterX}

\lhead{Chapter 6. \emph{Conclusions and Future Work}} % Change X to a consecutive number; this is for the header on each page - perhaps a shortened title

%----------------------------------------------------------------------------------------
%	SECTION 1
%----------------------------------------------------------------------------------------

\section{Conclusions}
This thesis addresses the problem of graph matching, which is the process of finding the similarity between the two graphs. Depending on the requirements of matching, graph matching is classified into two categories. When a strict matching between two graphs is needed, exact matching is used. Whereas inexact graph matching is used, when some flexibility and tolerance to error is allowed.  Graph matching has a wide range of applications in structural pattern recognition, biometric identification, computer vision, biological and chemical applications, etc. The main advantage of using graphs as compared to vectors in object recognition is its representational power and flexibility to link to any number of objects. However, with the advantage of using graphs, it brings some limitations. For example, one of the key issues of graph matching techniques is that they are computationally much expensive. Due to the non-availability of an efficient solution, various approximation and suboptimal algorithms have been proposed. Another key issue in graph matching is the lack of standard similarity measures to find similarity or dissimilarity between two graphs. This thesis considers the above key issues and presents algorithms for exact, approximate and error-tolerant graph matching.  

An extensive survey of various exact and inexact graph matching techniques was provided. Graph edit distance is one of the most flexible technique to perform graph matching. It is the minimum number of edit operations needed to transform one graph into another one. A common set of edit operations includes insertion, deletion and substitution of nodes and edges. A variation to graph edit distance using the concept of graph homeomorphism is introduced. The homeomorphic graph edit distance between two graphs is equal to graph edit distance between two graphs after applying path contraction on both input graphs. The path contraction operation on a graph removes nodes of degree two of all simple paths of the graph such that all intermediate nodes except first and last is of degree two. It leads to a reduction in search space and substantial saving in computation time.

A class of graph matching algorithms is presented, which diminishes the graph size by deleting the less important nodes using some measure of relevance. Node contraction approach reduces the graph size by deleting the nodes with least degree provided these nodes are not cut vertex. The condition for not being cut vertex is put to avoid unrestricted deletion of nodes, which may lead to many disconnected components of the graph. The extended graph edit distance between two graphs is defined, which is equal to graph edit distance between two graphs after applying the appropriate node contraction operations on both input graphs. Computational results show that this technique can be used for a trade-off between execution time and classification accuracy.

One limitation of the above node contraction approach is that the number of nodes of a given degree in different graphs can differ from each other, so the number of nodes removed may not be uniform. An extension to the above approach is proposed, which reduces the graph size by removing a fraction of nodes from both graphs based on a given centrality measure. Four different centrality measures, namely, degree, betweenness, eigenvector and PageRank centrality measures are used to contract the graphs. Experiments show that different centrality criteria lead to a different saving in computation time and classification ratio. Depending on the application requirements, a suitable centrality measure can be selected to achieve the best performance. One advantage of this proposed method is that it can be used to perform graph matching under given time constraints. For large graphs, an early estimate of graph matching can be obtained based on a fixed fraction of nodes of both graphs, which are most important according to a given centrality measure. One can note that this approach, along with the above approaches of node contraction and homeomorphism, can be used on top of any graph matching technique as an optimization criterion.   

A novel metric for measuring the similarity between two geometric graphs is introduced. The proposed similarity measure is promising yet straightforward. Vertex distance between two geometric graphs is defined as a linear sum assignment problem formulation for the assignment of the vertex set of the first graph to the vertex set of the second graph so that their overall distance is minimized. Edge distance between two graphs is defined using three feature of edges, which are orientation, length and position of edges. Finally, the graph distance metric is defined by combining the vertex distance and the three components of edge distance. The proposed graph distance can be considered as a step towards introducing a standard similarity measure for finding similarity between two graphs embedded in a plane.

The proposed geometric graph similarity framework is applied to exact and error-tolerant graph matching. An algorithm for geometric graph isomorphism is presented, which checks whether two graphs are geometrically isomorphic. In case there is a chance of slight distortion in the coordinates of vertices, this algorithm can also check that the two graphs are $t$-tolerant isomorphic when the distance between two corresponding coordinates is within the distance $t$. Before checking for geometric graph isomorphism, graph alignment algorithm is used to ensure that the reference coordinates of two graphs as identical as possible. The proposed similarity scheme is also used for error-tolerant graph matching. A set of weight parameters are used to combine the vertex distance and the angular, length and position components of edge distance. This framework is particularly useful for application involving large graphs, where data can be altered by distortions. Depending on the application need, suitable weight factors can be selected to optimize the performance. Computational results show that this framework is promising to graph dataset in which each vertex has its associated coordinate point.

\section{Future Directions}
% Edge contraction, other centrality measures, Other dataset, Parallelization, Graph kernel and embedding, weight parametes

Graph matching using node contraction is based on the concept of ignoring nodes having less importance, so as reduce the total search space. One of the future work can be to explore the performance of removing edges rather than nodes during the preprocessing of the graph using some relevant criteria. The proposed work of error-tolerant graph matching using centrality information considers four centrality measures. Future work can be to consider evaluating the performance of other centrality measures on the different dataset. The experimental results are performed mainly on letter and molecules graph dataset; future work can be to evaluate the performance of the algorithms on the other graph dataset, especially chemical and biological graph dataset. Another direction can be to implement the parallel version of the algorithm and compare its performance with the classical one. In the proposed work of error-tolerant geometric graph matching framework, before the computation of vertex distance, size of both graphs are made equal by inserting nodes in the smaller graph with coordinate equal to the average of all its existing coordinates. Other techniques can be explored to insert the nodes so that they are more representative concerning initial graphs. Similarly, other approaches can be investigated to add the edges in one graph to make the edge size of both graphs equal. Weight parameters are used to combine the vertex distance and the other three components of edge distance. Future work can be to perform experiments for the different combination of weight parameters and compare their performance. Future work can also be to apply the proposed geometric graph similarity framework for the design of graph kernel and embedding.

%----------------------------------------------------------------------------------------
%	THESIS CONTENT - APPENDICES
%----------------------------------------------------------------------------------------

\addtocontents{toc}{\vspace{2em}} % Add a gap in the Contents, for aesthetics

\appendix % Cue to tell LaTeX that the following 'chapters' are Appendices

% Appendix Template

\chapter{Graph Matching Tools} % Main appendix title

\label{AppendixA} % Change X to a consecutive letter; for referencing this appendix elsewhere, use \ref{AppendixX}

\lhead{Appendix A. \emph{Graph Matching Tools}} % Change X to a consecutive letter; this is for the header on each page - perhaps a shortened title

This appendix describes some of the important software toolkit and links for graph matching. Technical committee \#15\footnote{\url{https://iapr-tc15.greyc.fr}} of the International Association for Pattern Recognition\footnote{\url{www.iapr.org}} (IAPR) is devoted to encouraging the research and innovation in graph-based representation in the pattern recognition field. It provides details of various software toolkits, links and documentation useful for working in graph matching and graph-based representations. It also provides links for various benchmarking and graph datasets\footnote{\url{https://iapr-tc15.greyc.fr/links.html}}.

Various data structures and graph types are proposed to represent different types of graphs. Holt \textit{et al.} \citep{Holtetal2000} in 2000 describe one of the important structures known as Graph eXchange Language (GXL)\footnote{\url{http://www.gupro.de/GXL/}} format. GXL is designed to provide a standard exchange format for graphs \citep{Holtetal2006}. It is an XML-based standard exchange format to share data between tools and it can also be extended to represent hypergraphs and hierarchical graphs. Other XML based graph file formats includes GraphML, DGML and XGMML. GraphML\footnote{\url{http://graphml.graphdrawing.org/}} is an XML-based comprehensive file format for graphs. It can represent different types of graphs, including directed, undirected, hypergraphs, and hierarchical graphs. Directed Graph Markup Language (DGML) is used for directed graphs. XGMML (eXtensible Graph Markup and Modeling Language) is an XML application based on GML, which is used for graph representation. Graph Modelling Language (GML) is a hierarchical text-based file format for describing graphs. Other text-based file formats for graphs are DOT and TGF. DOT is a graph description language. The DOT language only defines a graph; it does not provide support for rendering the graph. Trivial Graph Format (TGF) is a simple text-based file format for describing graphs using adjacency list data structure. 

Graphviz\footnote{\url{https://graphviz.org/}}  is an open-source graph visualization tool. It provides different graphical interfaces and support for the conversion of various graph file formats. Other graph visualization and analysis tools includes iGraph\footnote{\url{https://igraph.org/}}, Jgraph \footnote{\url{https://sourceforge.net/projects/jgraph/}}.

There are several publicly available graph datasets and benchmarks\footnote{\url{https://iapr-tc15.greyc.fr/links.html}} for evaluation and implementation of graph matching algorithms. IAM\footnote{\url{https://fki.tic.heia-fr.ch/databases/iam-graph-database}} Graph Database Repository is one of the publicly available graph databases for graph-based pattern recognition and machine learning. IAM graph database uses GXL file formats for graph representation and manipulation \citep{RiesenBunke2008}. The ARG Database\footnote{\url{https://mivia.unisa.it/datasets/graph-database/arg-database/}} is a large collection of labeled and unlabeled graphs provided by the MIVIA Group for benchmarking of graph matching algorithms. Graph Data Repository For Graph Edit Distance (GDR4GED)\footnote{\url{http://www.rfai.li.univ-tours.fr/PublicData/GDR4GED/home.html}} provide datasets for testing graph edit distance algorithms. Benchmark Datasets\footnote{\url{https://ls11-www.cs.tu-dortmund.de/staff/morris/graphkerneldatasets}} for Graph Kernels are provided evaluation and testing of graph kernels.

A novel Graph Matching Toolkit (GMT)\footnote{\url{http://www.fhnw.ch/wirtschaft/iwi/gmt}} is a publicly available software toolkit, and it introduces a flexible software package for various implementation of graph edit distance computation \citep{Riesenetal2013}. GMT toolkit implements different algorithms of graph edit distance computation in JAVA. The VFLib\footnote{\url{https://mivia.unisa.it/vflib/}} is a graph matching library which provides various algorithms for graph isomorphism, subgraph isomorphism \citep{Cordellaetal2004}. The library is an efficient implementation of the VF2 graph matching algorithm and is written in C++. LAD\footnote{\url{http://liris.cnrs.fr/csolnon/LAD.html}} is a software toolkit for computing the subgraph isomorphism problem, and it is implemented in C. GEDLIB\footnote{\url{https://github.com/dbblumenthal/gedlib}} is a library for finding edit distances between graphs using different algorithms, and it is written in C++. Another Graph Matching Toolkit (GMT)\footnote{\url{https://www3.cs.stonybrook.edu/~algorith/implement/gmt/implement.shtml}}, provides C++ implementation of various graph matching algorithms. Graph-based tools for data mining and machine learning is provided in \citep{Bunke2003}. Overview of existing software tools for graph matching is described in \citep{Lasinger2013}.

% Appendix Template

\chapter{List of Publications} % Main appendix title

\label{AppendixB} % Change X to a consecutive letter; for referencing this appendix elsewhere, use \ref{AppendixX}

\lhead{Appendix B. \emph{List of Publications}} % Change X to a consecutive letter; this is for the header on each page - perhaps a shortened title

\begin{itemize}
 \item Shri Prakash Dwivedi, Ravi Shankar Singh. ``Error-Tolerant Geometric Graph Similarity and Matching'', \textbf{Pattern Recognition Letters}, Elsevier, vol. 125, pp. 625-631, 2019. %[An official publication of the IAPR] 
 \item Shri Prakash Dwivedi, Ravi Shankar Singh. ``Error-Tolerant Graph Matching using Node Contraction'', \textbf{Pattern Recognition Letters}, Elsevier, vol. 116, pp. 58-64, 2018. %[An official publication of the IAPR]
 \item Shri Prakash Dwivedi, Ravi Shankar Singh. ``Error-Tolerant Geometric Graph Similarity'', In IAPR Joint Intl.Workshop on Structural, Syntactic, and Statistical Pattern Recognition (\textbf{S+SSPR}), \textbf{Lecture Notes in Computer Science}, vol 11004, pp. 337-344, Springer 2018.
 \item Shri Prakash Dwivedi, Ravi Shankar Singh. ``Error-Tolerant Graph Matching using Homeomorphism'', In Intl. Conference on Advances in Computing, Communication and Informatics (\textbf{ICACCI}), pp. 1762-1766, IEEE 2017.
 \item Shri Prakash Dwivedi, Ravi Shankar Singh. ``Error-Tolerant Approximate Graph Matching utilizing Node Centrality Information'', \textbf{Pattern Recognition Letters}, Elsevier. (Submitted) %[An official publication of the IAPR] 
  
\end{itemize}

\addtocontents{toc}{\vspace{2em}} % Add a gap in the Contents, for aesthetics

\backmatter

%----------------------------------------------------------------------------------------
%	BIBLIOGRAPHY
%----------------------------------------------------------------------------------------
\cleardoublepage                                                                                                                                                                                                                                                                                                                                                                                                                                                                                                                                                                                                                                                                                                                                                                                                                                                                      
\nocite{*}
\label{Bibliography}

\lhead{\emph{Bibliography}} % Change the page header to say "Bibliography"

\bibliographystyle{unsrt}
\refstepcounter{chapter}
\bibliography{Bibliography}

\end{document}